\documentclass[iop,apj, floatfix]{emulateapj}
\usepackage{amsmath,amssymb,amstext}
\usepackage{graphicx}

\usepackage[breaklinks,colorlinks,citecolor=blue,linkcolor=magenta]{hyperref} 
\usepackage[all]{hypcap}
\usepackage{aas_macros}
\usepackage{natbib}
\usepackage{color,soul}
\newcommand{\acounits}{\mbox{M$_\odot$ pc$^{-2}$ (K km s$^{-1}$)$^{-1}$}}

\shorttitle{Gallagher et al.}
\shortauthors{Gallagher et al.}

\begin{document}

\title{Dense Gas, Dynamical Equilibrium Pressure, and Star Formation in Nearby Star-Forming Galaxies}
\author{
Molly J. Gallagher\altaffilmark{1}, 
Adam K. Leroy\altaffilmark{1}, 
Frank Bigiel\altaffilmark{2}, 
Diane Cormier\altaffilmark{3,2}, 
Mar\'{i}a J. Jim\'{e}nez-Donaire\altaffilmark{4}, 
Eve Ostriker\altaffilmark{5}, 
Antonio Usero\altaffilmark{6}, 
Alberto D. Bolatto\altaffilmark{7}, 
Santiago Garc\'{i}a-Burillo\altaffilmark{6}, 
Annie Hughes\altaffilmark{8,9}, 
Amanda A. Kepley\altaffilmark{10},
Mark Krumholz\altaffilmark{11},
Sharon E. Meidt\altaffilmark{12}, 
David S. Meier\altaffilmark{13, 14},
Eric J. Murphy\altaffilmark{10},
J\'{e}r\^{o}me Pety\altaffilmark{15,16}, 
Erik Rosolowsky\altaffilmark{17}, 
Eva Schinnerer\altaffilmark{12},
Andreas Schruba\altaffilmark{18},
Fabian Walter\altaffilmark{12}
}
\altaffiltext{1}{Department of Astronomy, The Ohio State University, 140 West 18th Avenue, Columbus, OH 43210, email: {\tt gallagher.674@osu.edu}.}
\altaffiltext{2}{Institute f\"ur theoretische Astrophysik, Zentrum f\"ur Astronomie der Universit\"at Heidelberg, Albert-Ueberle Str. 2, 69120 Heidelberg, Germany.}
\altaffiltext{3}{Laboratoire AIM, CEA/DSM - CNRS - Universit\'e Paris Diderot, Irfu/Service \ d'Astrophysique, CEA Saclay, 91191
  Gif-sur-Yvette, France}
\altaffiltext{4}{Harvard-Smithsonian Center for Astrophysics, 60 Garden St., Cambridge, MA 02138, USA}
\altaffiltext{5}{Department of Astrophysical Sciences, Princeton University, Princeton, NJ 08544}
\altaffiltext{6}{Observatorio Astron\'omico Nacional (IGN), C/ Alfonso XII, 3, 28014 Madrid, Spain}
\altaffiltext{7}{Department of Astronomy, Laboratory for Millimeter-wave Astronomy, and Joint Space Institute, University of Maryland, College Park, Maryland 20742, USA}
\altaffiltext{8}{CNRS, IRAP, 9 av. du Colonel Roche, BP 44346, F-31028 Toulouse cedex 4, France}
\altaffiltext{9}{Universit\'{e} de Toulouse, UPS-OMP, IRAP, F-31028 Toulouse cedex 4, France}
\altaffiltext{10}{National Radio Astronomy Observatory, 520 Edgemont Road, Charlottesville, VA 22903, USA}
\altaffiltext{11}{Research School of Astronomy \& Astrophysics, Australian National University, Canberra, ACT 2611, Australia}
\altaffiltext{12}{Max Planck Institute f\"ur Astronomie, K\"onigstuhl 17, 69117, Heidelberg, Germany}
\altaffiltext{13}{Department of Physics, New Mexico Institute of Mining and Technology, 801 Leroy Place, Soccoro, NM 87801, USA}
\altaffiltext{14}{National Radio Astronomy Observatory, P. O. Box O, 1003 Lopezville Road, Socorro, NM, 87801, USA}
\altaffiltext{15}{Institut de Radioastronomie Millim\`{e}trique (IRAM), 300 Rue de la Piscine, F-38406 Saint Martin d'H\`{e}res, France}
\altaffiltext{16}{Observatoire de Paris, 61 Avenue de l'Observatoire, F-75014 Paris, France}
\altaffiltext{17}{Department of Physics, University of Alberta, Edmonton, AB T6G 2E1, Canada}
\altaffiltext{18}{Max-Planck-Institut f\"ur extraterrestrische Physik, Giessenbachstra{\ss}e 1, 85748 Garching, Germany}

\begin{abstract}
We use new ALMA observations to investigate the connection between dense gas fraction, star formation rate, and local environment across the inner region of four local galaxies showing a wide range of molecular gas depletion times. We map HCN (1-0), HCO$^+$ (1-0), CS (2-1), $^{13}$CO (1-0), and C$^{18}$O (1-0) across the inner few kpc of each target. We combine these data with short spacing information from the IRAM large program EMPIRE, archival CO maps, tracers of stellar structure and recent star formation, and recent HCN surveys by Bigiel et al. and Usero et al. We test the degree to which changes in the dense gas fraction drive changes in the SFR. $I_{HCN}/I_{CO}$ (tracing the dense gas fraction) correlates strongly with $I_{CO}$ (tracing molecular gas surface density), stellar surface density, and dynamical equilibrium pressure, $P_{DE}$. Therefore, $I_{HCN}/I_{CO}$ becomes very low and HCN becomes very faint at large galactocentric radii, where ratios as low as $I_{HCN}/I_{CO} \sim 0.01$ become common. The apparent ability of dense gas to form stars,  $\Sigma_{SFR}/\Sigma_{dense}$ (where $\Sigma_{dense}$ is traced by the HCN intensity and the star formation rate is traced by a combination of H$\alpha$ and 24$\mu$m emission), also depends on environment. $\Sigma_{SFR}/\Sigma_{dense}$ decreases in regions of high gas surface density, high stellar surface density, and high $P_{DE}$. Statistically, these correlations between environment and both $\Sigma_{SFR}/\Sigma_{dense}$ and $I_{HCN}/I_{CO}$ are stronger than that between apparent dense gas fraction ($I_{HCN}/I_{CO}$) and the apparent molecular gas star formation efficiency $\Sigma_{SFR}/\Sigma_{mol}$. We show that these results are not specific to HCN.
\end{abstract}

\keywords{molecular cloud, gas density, star formation, ALMA}
\maketitle

\section{Introduction}
\label{sec:Introduction}

Observations of star forming regions in the Milky Way (e.g. \citealt{LADA03}; \citealt{KAINULAINEN09}; \citealt{HEIDERMAN10}; \citealt{LADA10}; \citealt{ANDRE14}) indicate that stars form mainly in dense substructures. These studies have inspired a large body of literature investigating how the amount of dense gas relates to the star formation rate of a cloud or galaxy. \cite{LADA12} found the star formation rate (SFR) surface density ($\Sigma_{SFR}$) in individual Milky Way clouds to be proportional to the fraction of the gas that is dense (where here ''dense'' is n $\ge 3\times 10^{4}$cm$^{-3}$). Similarly, both \cite{EVANS14} and \cite{LADA10} found that the SFR in individual clouds relates linearly to the molecular gas mass above an extinction threshold chosen to select only dense gas: A$_{V}\approx$8~mag and A$_{K}\approx$0.8~mag, respectively \citep[see also][]{KONYVES15}. In a pioneering study, \cite{GAO04} showed that galaxy-integrated SFR, traced by IR emission, relates linearly to the total mass of dense molecular gas, traced by HCN emission. Based on this result, they argued for a similar picture for whole galaxies, with the SFR set by the amount of dense, HCN-emitting gas.

Spectroscopic studies of other galaxies indicate a more complex relationship between dense gas and SFR. \cite{USERO15} surveyed $\sim 60$ regions across $30$ star-forming galaxies with a $\sim 1{-}2$ kpc-sized beam. They found that the dense gas star formation efficiency ($SFE_{dense}$, defined as $\Sigma_{SFR}/\Sigma_{dense}$) inferred from HCN, H$\alpha$, and 24$\mu$m emission, anti-correlates with both the stellar surface density ($\Sigma_*$) and the fraction of gas in the molecular phase ($f_{ mol}$). Using one of the first whole-galaxy HCN galaxy maps, \cite{BIGIEL16} found similar trends across the disk of NGC~5194 \citep[consistent with][]{CHEN15}: at $\sim$kpc resolution, $SFE_{dense}$ drops with increasing $\Sigma_*$ and increasing $f_{ mol}$. The sense of these results agrees with recent work studying the Milky Way, which finds the rate of star formation per unit dense gas to be lower in the Galactic Center than in Solar Neighborhood molecular clouds \citep[see][]{LONGMORE13}.

The abundance of dense gas is also of interest and appears to depend strongly on environment. \citet{GAO04} observed large variations in the dense gas fraction ($f_{dense} \equiv \Sigma_{dense}/\Sigma_{mol}$), traced by the $I_{HCN}/I_{CO}$ ratio. \citet{USERO15} and \citet{BIGIEL16} found that $f_{dense}$ varies significantly, correlating with $\Sigma_*$ and $f_{mol}$. Furthermore, the Milky Way center appears far richer in dense gas \citep{LONGMORE13} than the local clouds studied by \cite{LADA10} and \cite{EVANS14}.

\citet{USERO15} and \citet{BIGIEL16} found that $SFE_{dense}$ and $f_{dense}$ vary as a function of $\Sigma_*$ and $f_{ mol}$. Both $\Sigma_*$ and $f_{mol}$ relate closely to the interstellar pressure needed to support a gas disk in vertical dynamic equilibrium, $P_{DE}$ \citep[][]{ELMEGREEN89}. We know $P_{DE}$ correlates closely with $f_{mol}$ \citep{WONG02,BLITZ06,LEROY08} and with the internal pressure of molecular clouds \citep{HUGHES13}. We also expect a correlation between $P_{DE}$ and the density of the interstellar gas. Indeed, \citet{HELFER97} suggested this correlation to help explain the paucity of HCN emission outside galaxy centers. 

This central role of gas pressure and ISM weight appears to be consistent with star formation self-regulation theory (see e.g. \citealt{OSTRIKER10}; \citealt{OSTRIKER11}; \citealt{KIM11}) and simulations (see e.g. \citealt{KIM11}; \citealt{KIM13}, \citealt{KIM15}), which predict that $\Sigma_{SFR}$ will be proportional to the total pressure. However, $P_{DE}$ should not be the only factor at play. The turbulent Mach number \citep[e.g.,][]{KRUMHOLZ07,GARCIABURILLO12,USERO15}, virial parameter \citep{KRUIJSSEN14}, and large scale kinematics \citep[][]{MEIDT18} may also influence $SFE_{dense}$.

Testing these ideas and advancing this field requires more resolved multi-line mapping of galaxies, in the style of \citet{USERO15}, \cite{CHEN15}, and \citet{BIGIEL16}. In this approach, one observes a suite of lines whose emissivity peaks at different densities, $n_{eff}$\footnote{$n_{eff}$ refers to the lowest density for which a line achieves 95\% of its maximum emissivity at a given $T_{\rm kin}$ and $\tau$. It is closely related, though not identical, to the effective critical density, see \citet{SHIRLEY15}, \citet{LEROY17}}. To first order, molecular line emission will be proportional the mass of gas above $n_{eff}$, as long as all other physical conditions remain fixed. Thus, changes in the ratio of intensities between two emission lines with different $n_{eff}$ can indicate a changing ratio of masses above each density \citep[though there are crucial subtleties, e.g., see][]{KRUMHOLZ07,LEROY17}. Because this method gives access to changes in the density distribution without the need to resolve molecular cloud substructure, it can be deployed to study changing gas density distributions across whole galaxies or large parts of galaxies. This, in turn, gives access to a much wider range of physical conditions in which we can study the origin and role of dense gas.
 
Until recently, interferometers have lacked the surface brightness sensitivity to survey high $n_{eff}$ lines like HCN~(1-0) across the disks of nearby normal galaxies\citep[for an early attempt limited by sensitivity, see][]{HELFER97}. HCN~(1-0), often the brightest dense gas tracer, can be $\sim 30$ or more times fainter than CO~(1-0) \citep{USERO15}. The Atacama Large Millimeter/submillimeter Array (ALMA) changes this. ALMA makes it practical to map entire nearby galaxies with multi-line, density-sensitive spectroscopy. The high sensitivity of ALMA allows us to reach noise levels comparable to those of previous single dish maps \citep[e.g.,][]{BIGIEL16} in less than an hour. The high resolution of ALMA helps to bridge the gap in scale between the individual molecular clouds and cores studied in the Milky Way and galaxy averages used in previous seminal studies \citep[e.g.,][]{GAO04,GARCIABURILLO12}. ALMA's resolution also makes it possible to distinguish distinct dynamical regions, for example disentangling the nuclear gas structures in galaxies (analogs to the Milky Way's ``central molecular zone,'' CMZ) from extended emission associated with the disk and spiral arms.

Here, we employ ALMA multi-line spectroscopy to study the origin and role of dense molecular gas in nearby galaxies. Our main goals are to measure any variations in the dense gas fraction and the star formation efficiency of dense gas, and to understand the physical drivers of such variations. We present new ALMA observations of NGC~3351, NGC~3627, NGC~4254, and NGC~4321. These observations cover high $n_{eff}$ transitions --- HCN~(1-0), HCO$^{+}$~(1-0), and CS~(2-1) --- and two CO isotopologues  --- $^{13}$CO~(1-0), and C$^{18}$O~(1-0) --- across the inner $\approx 1'$ ($\approx3-5$kpc) of each galaxy. We chose these four targets because together they show a wide range of apparent SFR per unit molecular gas ($SFE_{mol}$) over their inner few kpc \citep[][]{LEROY13}. By observing the dense gas and contrasting it with SFR tracers and CO imaging, we aim to understand if these apparent variations in the $SFE_{mol}$ within and among our targets are driven by changes in the dense gas fraction. We  also aim to understand the physical drivers of the dense gas fraction.

In this paper, we combine these new observations with the data from \citet{USERO15} and \citet{BIGIEL16} to test the hypotheses that 1) the dense gas fraction alone drives the star formation rate and 2) that the mean midplane pressure drives the density distribution and the role of dense gas in star formation.

In \S \ref{sec:data}, we describe our ALMA observations (\S \ref{subsec:MLO}), previous CO observations (\S \ref{subsec:COO}), and data processing (\S \ref{subsec:DCP}). We also summarize the multiwavelength data used in our analysis (\S \ref{subsec:SFR_lit}, \S \ref{subsec:SM_lit}, \S \ref{subsec:GM_lit}, \S \ref{subsec:Ph}). Then, in \S \ref{sec:Results}, we describe some qualitative properties of our data (\S \ref{subsec:RPR}) and explore the quantitative relationship between dense gas and star formation rate in our sample (\S \ref{subsec:DGFASFE}). We discuss validity of using $I_{HCN}/I_{CO}$ as a tracer of gas density (\S \ref{subsec:density}), and then explore what sets the star formation efficiency of dense gas (\S \ref{subsec:sfedense}) and the dense gas fraction (\S \ref{subsec:DG}), highlighting the possible role of ISM pressure (\S \ref{subsec:pressure_disc}). In \S \ref{sec:discussion} we discuss the implications of our results, including theoretical implications for the link between gas density, star formation, and galactic environment (\ref{subsec:theory}). Finally, we lay out key caveats and next steps (\S \ref{subsec:caveats}), and then summarize our findings in \S \ref{sec:summary}.
 
\section{Data}
\label{sec:data} 

We mapped tracers of dense gas over the inner regions of NGC~3351, NGC~3627, NGC~4254, and NGC~4321. We chose these targets from the sample of nearby galaxies of \citet{LEROY13}, which had the best available supporting multiwavelength data (CO, {\sc Hi}, infrared (IR), H$\alpha$, etc. mapping) at the time of the proposal. Based on the measured IR surface brightness of the targets in that study, we estimated the likely HCN emission of each target \citep[following][]{GAO04,USERO15}. Out of the subset of \citet{LEROY13} targets that could feasibly be detected by ALMA, we chose these four because together they spanned a large range of CO-to-$24\mu$m ratios over their inner few kpc (see plots below). This implies variations in the efficiency with which the total molecular gas reservoir forms stars, making this small sample ideal to test the hypothesis that variations in the molecular gas depletion time are driven by variations in the dense gas fraction.

\subsection{ALMA Molecular Line Observations}
\label{subsec:MLO}

\capstartfalse
\begin{deluxetable}{lcccc}
\tabletypesize{\scriptsize}
\tablecaption{Lines Observed \label{tab:line_data}}
\tablewidth{0pt}
\tablehead{
\colhead{Line} & 
\colhead{$\nu_{\rm rest}$\tablenotemark{a}} &
\colhead{Fiducial $\tau$\tablenotemark{b}} &
\colhead{ $n_{eff}$\tablenotemark{c}} \\[2px]
& \multicolumn{1}{c}{(GHz)} & & \multicolumn{1}{c}{(cm$^{-3}$)}}
\startdata
$^{12}$CO (1-0) & 115.27& 10 &$1\times10^2$ \\
$^{13}$CO (1-0) & 110.20& 0.1 & $8\times10^2$ \\
C$^{18}$O (1-0)\tablenotemark{d} & 109.78& 0.1 & $8\times10^2$\\
CS (2-1)\tablenotemark{d} & 97.98& 1 &$7\times10^4$ \\
HCO$^+$ (1-0)& 89.19 & 1 & $4\times10^4$ \\
HCN (1-0)& 88.63& 1 & $2\times 10^5$
\enddata
\tablenotetext{a}{From Splatalogue ({\tt http://splatalogue.net/}).}
\tablenotetext{b}{Typical optical depth. See \cite{JIMENEZDONAIRE17,JIMENEZDONAIRE17B} for more details.}
\tablenotetext{c}{Density at which the emissivity reaches 95\% of its maximum given our fiducial $\tau$ and taking $T_{\rm kin} = 25$~K. From \citet{LEROY17}.}
\tablenotetext{d}{CS~(2-1) was only observed by ALMA, not the IRAM 30-m, and so not covered in NGC~5194 and not short-spacing corrected in the other targets.}
\end{deluxetable}
\capstarttrue

\capstartfalse
\begin{deluxetable*}{lcccccccc}[ht!]
\tabletypesize{\scriptsize}
\tablecaption{Targets Observed \label{tab:galaxy_data}}
\tablewidth{0pt}
\tablehead{
\colhead{Galaxy} & 
\colhead{R.A.} &
\colhead{Dec.} &
\colhead{Morphology} &
\colhead{Inclination} &
\colhead{Position Angle} &
\colhead{Distance} &
\colhead{Linear Beam} & 
\colhead{F.o.V.} \\[2px]
& \multicolumn{1}{c}{(J2000)} & \multicolumn{1}{c}{(J2000)}  & \multicolumn{1}{c}{}  & \multicolumn{1}{c}{(deg)} & \multicolumn{1}{c}{(deg)} & \multicolumn{1}{c}{(Mpc)} & \multicolumn{1}{c}{(pc)} & \multicolumn{1}{c}{(kpc)}
}
\startdata
NGC 3351 & 160.990417 & 11.703806 & SB(r)b & 41.0 & 192.0 & 9.33 & 392 & 3.67 \\ 
NGC 3627 & 170.0623508 & 12.9915378 & SAB(s)b & 62.0 & 173.0  & 9.38 & 380 & 3.56 \\ 
NGC 4254 & 184.706682 & 14.416509 & SA(s)c & 32.0 & 55.0 & 14.4 & 605 & 5.67 \\ 
NGC 4321 & 185.728463 & 15.821818 & SAB(s)bc & 30.0 & 153.0 & 14.3 & 636 & 5.96 \\ 
NGC 5194\tablenotemark{30m} & 202.4667 & 47.1947 & SA(s)bc pec& 22.0 & 7.5 & 7.6 & 295 & 9.22 x 12.55
\enddata
\tablenotetext{30m}{Observed with the IRAM 30-m. Data from \citet{BIGIEL16}, see that paper for more details. The other targets were all observed with ALMA with short and zero spacing correction from the IRAM 30-m. See \S \ref{sec:data}.}
\tablecomments{{\bf R.A., Dec.}: adopted center of the galaxy. {\bf Inclination}: inclination used to construct the radial profiles and correct surface densities for projection. {\bf Position Angle}: position angle measured north through east used to construct the radial profiles. {\bf Distance}: adopted distance to the galaxy in Mpc. {\bf Linear Beam}: linear scale in pc of the FWHM of the beam used in this analysis. This is $8\arcsec$ at the distance of the galaxy for the first four targets and $30\arcsec = 1.1$~kpc for NGC 5194 \citep{BIGIEL16}. {\bf F.o.V.} --- Field of view across the dense gas maps at the distance of the galaxy without accounting for inclination. {\bf References:} Centers and morphologies adopted from NASA Extragalactic Database, which draws key information from RC3 \citep{RC3}. Source for orientation parameters: NGC~3351, \citet{DICAIRE08}, NGC~3627, \citet{DEBLOK08}, NGC~4254, \citet{BOISSIER03}, NGC~4321, \citet{MUNOZMATEOS09}, NGC~5194 \citet{COLOMBO14B}. Distances adopted from \citet{KENNICUTT11}.}
\end{deluxetable*}
\capstarttrue

As part of ALMA's Cycle 2 campaign, we observed HCN~(1-0), HCO$^{+}$~(1-0), CS~(2-1), $^{13}$CO~(1-0), and C$^{18}$O~(1-0) in four galaxies using ALMA's main array of 12m antennas. For the remainder of the paper, we will refer to these lines as HCN, HCO$^{+}$, CS, $^{13}$CO, and C$^{18}$O. HCN, HCO$^{+}$, and CS all have $n_{eff}$ $\sim 10^4{-}10^5$~cm$^{-3}$ (see Table \ref{tab:line_data}), and so are expected to trace mainly dense gas \citep[though in the absence of such gas they can still emit, e.g.,][]{SHIRLEY15,LEROY17}. The CO isotopologues, $^{13}$CO and C$^{18}$O, trace lower density gas, $n_{eff} \sim 10^{3}$~cm$^{-3}$. The contrast between the optically thin isotopologues and the optically thick $^{12}$CO emission constrains the optical depth and physical conditions in the bulk of the molecular gas. We make limited use of $^{13}$CO and C$^{18}$O in this paper. These data are analyzed in detail by \citet{JIMENEZDONAIRE17}. Fainter lines in the bandpass were analyzed by \citet{JIMENEZDONAIRE17B}.

Table \ref{tab:galaxy_data} gives our adopted position, morphology, orientation, distance, beam size, and field of view for each target. We observed seven fields in a hexagonally packed mosaic pointed towards the center of each galaxy. The mosaic pattern used the default Nyquist-spacing set by the ALMA observing tool.

We observed each galaxy with two spectral setups. The first covered lines tracing the dense gas: HCN, HCO$^{+}$, and CS. The four spectral windows covered 85.4--87.2 GHz, 87.2--89.0 GHz, 97.2--99.0 GHz, and 99.0--100.8 GHz. The second spectral setup covered lines tracing the overall distribution of molecular gas: $^{13}$CO and C$^{18}$O. Those four spectral windows covered 98.2--100.0 GHz, 96.6--98.4 GHz, 108.5--110.3 GHz, and 110.3--112.1 GHz. For both setups, we observed using a channel width of 976.6 kHz ($\sim 3$~km~s$^{-1}$ at $\nu = 100$~GHz) and bandwidth 1.875~GHz, sufficient to cover the full velocity extent of each line in question.

We observed in a compact configuration in order to emphasize flux recovery and surface brightness sensitivity, reflecting the faint nature of the dense gas tracers. After calibration, the HCN observations had 703 (NGC 3351), 630 (NGC 3627), 561 (NGC 4254), and 561 (NGC 4321) baselines, with minimum and maximum unprojected lengths of 15m and 348m, median baseline length of 90{-}100m, and 20-90\% range of typically 50 to 195m. For reference at the $\nu \sim 89.5$~GHz of the HCN and HCO$^+$ lines, 50m, 100m, and 200m correspond to $\sim 13.8\arcsec$, $6.9\arcsec$, and $3.5\arcsec$.

We processed the data using the the CASA software package (\citealt{MCMULLIN07}) and the observatory-provided calibration scripts. Most of the calibration occurred in CASA version 4.2.2, with one data set calibrated in CASA version 4.3.1. The calibration scripts were a mixture of the observatory-produced CASA scripts and calls to the formal ALMA pipeline. In all cases they are available, along with the data, from the ALMA archive. After inspecting the pipeline calibrated data, we imaged each line separately. For the final version of the imaging, we used CASA version 4.6.0.

We first subtracted continuum emission using the CASA task {\tt uvcontsub} and avoiding the frequencies of known bright lines. We then imaged each cube using natural weighting, averaging in frequency to produce 10~km~s$^{-1}$ wide channels, and applying a small $u-v$ taper ($2{-}3\arcsec$ depending on the line and target). The taper further emphasizes surface brightness sensitivity. The small loss of resolution is irrelevant to the science in this paper because the comparison to tracers of recent star formation already limits our work to $\gtrsim 5\arcsec$ resolution. 

With the taper, but before any other processing the synthesized beams in the deconvolved HCN images were: $4.2\arcsec \times 3.6\arcsec$ (NGC~3351), $4.4\arcsec \times 3.7\arcsec$ (NGC~3627), $6.0\arcsec \times 3.5\arcsec$ (NGC~4254), and $4.6\arcsec \times 3.2\arcsec$ (NGC~4321). The pixel size adopted during imaging was always chosen to heavily oversample the beam. After imaging, we convolved each cube using the CASA task {\tt imsmooth} to have a round $8\arcsec \times 8\arcsec$ Gaussian beam. This allowed us to beam match the poorer resolution CO and infrared data that are crucial to the analysis. The final images used in this analysis all have $8\arcsec$ (FWHM) beams, 10~km~s$^{-1}$ channel width, and $0.5\arcsec$ pixels that heavily oversample the beam.

Given the $u-v$ coverage mentioned above, we expect structures larger than $\sim 14\arcsec$ in a single channel to suffer from spatial filtering in the ALMA main array data. To account for this, we combined our 12-m HCN, HCO$^+$, $^{13}$CO, and C$^{18}$O data with single dish maps obtained as part of the IRAM EMPIRE Survey \citep[][M. Jimenez Donaire et al. in preparation]{BIGIEL16,CORMIER18}. To do this, we aligned the IRAM maps to the grid of the ALMA data, applied the primary beam response of the ALMA images to the IRAM data, and converted the IRAM data to have units of Jy~beam$^{-1}$. Then we combined the two data sets using the CASA task {\tt feather}. After the combination, we verified that the feathered cube indeed matched the spectral profile of the IRAM 30-m cube when both were convolved to a common $30\arcsec$ resolution. Short spacing data were not available for the CS, so those data are from ALMA's main 12-m array only in this paper.

Table \ref{tab:flux_recov} reports the total flux recovered for each line from each galaxy both with and without the addition of the IRAM 30-m data. We calculate the total flux by summing the pixels in the original data cubes. The Table shows that ALMA recovers $\gtrsim 95$\% of the flux for all lines in NGC 3351, which is dominated by a bright, compact nuclear source. In fact, for the two faintest lines, ALMA finds slightly more flux than the IRAM-30m, indicating modest ($\sim 10$\%) discrepancies in the flux calibration scale. ALMA recovers a lower fraction of the flux for galaxies with more extended, low S/N emission. On average, ALMA recovers $\sim 60{-}80$\% of the flux found by the IRAM 30-m. Our worst case is NGC 4254, where the faint line emission appears extended and ALMA recovers only $\sim 30{-}50$\% of the flux seen by the IRAM 30-m. This likely results from two factors: the large extent of the CO emission in the galaxy and the lack of a compact, bright nuclear source (in contrast to the other three targets). This variable, sometimes poor flux recovery emphasizes the importance of including short spacing data.

Throughout the paper, we work with intensity in units of brightness temperature, $T_B$. We convert our final data cubes from their native units of Jy/beam to $T_B$ via the standard Rayleigh-Jeans formula:

\begin{equation}
\label{eq:tb}
T_B {\rm \left[ K \right]} = \frac{c^2}{2\times10^{23}~k_B~\nu^2}~I_\nu~\left[{\rm Jy/beam} \right]
\end{equation}

\noindent where $\nu$ is the frequency of the line, $k_{B}$ is Boltzmann's constant, and $I_\nu$ is the original intensity in Jy~beam$^{-1}$.

Finally, we measure the rms noise from the signal free region of each cube. At $8\arcsec$ resolution, the statistical noise in each $10$~km~s$^{-1}$ channel is $\sim 5{-}10$~mK for HCN, HCO$^+$, and CS and $\sim 10$~mK for $^{13}$CO, C$^{18}$O. These vary somewhat from cube to cube, and we use the correct local value to construct uncertainty maps for the integrated intensity. The nominal flux calibration accuracy of ALMA in Band 3 during Cycle 2 was 5\% according to the ALMA Cycle 2 Technical Handbook, though this may be somewhat optimistic. The IRAM 30-m intensity calibration for EMPIRE observations is internally consistent to $\approx 5$\% from night to night, but the absolute calibrations scale is uncertain at the 10{-}15\% level (M. Jimenez Donaire et al. in preparation).

\capstartfalse
\begin{deluxetable}{lccccccccccc}
\tabletypesize{\scriptsize}
\tablecaption{Flux Recovery \label{tab:flux_recov}}
\tablewidth{0pt}
\tablehead{
\colhead{Data Type} & 
\colhead{$^{13}$CO} &
\colhead{C$^{18}$O} &
\colhead{HCO$^+$} &
\colhead{HCN} \\
\colhead{} &
\multicolumn{4}{c}{($10^{3}$~K~km~s$^{-1}$~arcsec$^2$)}
}
\startdata
\sidehead{\bf{NGC3351}}
ALMA+IRAM~30-m  & 3.32 & 0.41 & 1.21 & 2.24\\
ALMA only &  3.20 & 0.45 & 1.24 & 2.14\\
Fraction recovered & 0.96 & 1.11 & 1.02 & 0.96\\
\sidehead{\bf{NGC3627}}
ALMA+IRAM~30-m & 16.98 & 1.55 & 4.96 & 6.47 \\
ALMA only &  11.98 & 1.40 & 3.77 & 4.39 \\
Fraction recovered & 0.71 & 0.91 & 0.76 & 0.68 \\
\sidehead{\bf{NGC4254}}
ALMA+IRAM~30-m & 15.98 & 3.31 & 3.98 & 4.34\\
ALMA only & 6.84 & 0.67 & 1.11 & 1.42\\
Fraction recovered & 0.43 & 0.20 & 0.28 & 0.33\\
\sidehead{\bf{NGC4321}}
ALMA+IRAM~30-m & 14.52 & 2.61 & 3.86 & 5.10\\
ALMA only & 8.79 & 1.81 & 2.20 & 3.55\\
Fraction recovered & 0.61 & 0.70 & 0.57 & 0.70\\
\enddata
\end{deluxetable}
\capstarttrue

\subsection{CO Observations}
\label{subsec:COO}

We use $^{12}$CO~(1-0) emission, hereafter referred to as CO, to trace the overall molecular gas reservoir. We draw these maps from literature data. In each case, our CO data includes both interferometric and single dish data, allowing us to reach our working $\sim 8\arcsec$ resolution, but also recover zero- and short-spacing information. All of the CO data have a larger field of view than our ALMA maps, bandwidth that covers the entire CO line for the galaxy, and pixels that oversample the beam by more than the Nyquist rate.
    
For NGC 3351 and NGC 3627, we use CO observations from BIMA SONG (\citealt{HELFER03}). These cubes include data from both the BIMA interferometer and the NRAO 12m single dish telescope on Kitt Peak. The combined BIMA$+$12m cubes have native resolutions of $7.4\arcsec \times 5.2\arcsec \times 10$ km s$^{-1}$ (NGC 3351) and $7.3\arcsec \times 5.8\arcsec \times 10$ km s$^{-1}$ (NGC 3627). \citet{HELFER03} quote a flux calibration uncertainty of 15\%.

Nor NGC 4254, we use interferometric CO observations from CARMA STING (\citealt{RAHMAN11}) and single dish data from the CO extension to the IRAM EMPIRE survey \citep[][M. Jim\'enez-Donaire et al. in prepation]{CORMIER18}. Before convolution, the native resolution of the CARMA data are $3.3\arcsec \times 2.7\arcsec \times 5$km s$^{-1}$. \citet{RAHMAN11} do not quote an amplitude calibration uncertainty, but discuss $\sim$10\% as a typical value. We combine the CARMA and IRAM data using the CASA task {\tt feather}, which carries out a Fourier plane combination of the two cubes.
    
For NGC 4321, we use CO data provided as part of the ALMA science verification program. These include both main 12-m array and Atacama Compact Array (ACA) short spacing and total power data. We use the feathered ``reference image'' provided as part of the science verification release, which has resolution $3.9\arcsec \times 2.5\arcsec \times 5$~km~s$^{-1}$. As above, the nominal amplitude calibration accuracy for ALMA at Band 3 is $\pm 5$\%, though in practice this seems optimistic. 
We convert all CO data to units of brightness temperature following Equation \ref{eq:tb}, convolve them to our working $8\arcsec$ resolution, and align them to the astrometric and velocity grid of the ALMA dense gas data.

\subsection{CO and Dense Gas Conversion Factors}
\label{subsec:CF}

Whenever possible, we report our results in terms of observable quantities, e.g., $I_{CO}$, $I_{HCN}$, etc. We also express our results in terms of physical quantities, e.g., molecular gas mass and dense gas mass. The translation from observed to physical quantities carries substantial uncertainty, and remains a topic of active research \citep[e.g.,][]{SANDSTROM13,USERO15,LEROY17}. However, these physical quantities --- e.g., $\Sigma_{H2}$ and $\Sigma_{dense}$ --- are of considerable interest, and are central to our science goals. Therefore, following \citet{USERO15}, we will also express our results as best-estimate physical terms. Given a choice, we prefer the simplest possible translation from observed to physical quantities, and will plot both axes whenever we can.

By default, we quote total molecular gas mass assuming a CO~(1-0) to molecular gas mass conversion factor of $\alpha_{CO} \approx 4.3~\acounits$. In order to derive a fiducial dense gas mass, we convert from HCN intensity to dense gas mass surface density assuming $\alpha_{HCN} \approx 10~\acounits$. Both factors include helium.

Both of these conversion factors carry substantial uncertainty, though for different reasons. Our $\alpha_{CO} \approx 4.3~\acounits$ is often taken as the Milky Way conversion factor, and applied as a default to Solar metallicity massive galaxies \citep[see][]{BOLATTO13}. However, dust-based studies that include our current targets suggest that the gas-rich regions in the central parts of some galaxies have lower $\alpha_{CO}$ \citep{SANDSTROM13}. This presumably reflects dynamical broadening of the CO line width, leading to more CO emission per unit mass.

Meanwhile, we consider $\alpha_{HCN}$ uncertain because the abundance and opacity of HCN as a function of density and environment remain poorly known. \cite{GAO04} argue for $\alpha_{HCN} \approx 10~\acounits$ to convert from HCN intensity to surface density of gas above $n_{H2} \sim 3 \times 10^4$~cm$^{-3}$ based on LVG modeling and invoking the virial theorem. However, if the inputs to these calculations, e.g., the abundance of HCN, the opacity of HCN, or the dynamical state of dense gas, vary from their assumed values, then the absolute value of $\alpha_{HCN}$ also changes \citep[e.g., see][]{LEROY17}. Constraints on these quantities in other galaxies remain very weak \citep[see][]{MARTIN06,JIMENEZDONAIRE17}, but observations of the Milky Way do demonstrate important variations \citep[e.g.,][]{PETY17}.

In this paper, we focus on the dense gas fraction. Many environmental factors that affect CO might also affect HCN emission. Because we lack a useful prescription for $\alpha_{HCN}$, it is unclear how we should implement variations in $\alpha_{CO}$ \citep[for an in depth discussion of how these quantities may relate see][]{USERO15}. Moreover, we aim to clearly present our results in terms of observed line ratios, e.g., $I_{HCN}/I_{CO}$. This requires a simple $\alpha_{CO}$ and $\alpha_{HCN}$.

We do consider how our multi-line observations support the idea that HCN traces the dense gas. For a complete discussion of the variation in $\alpha_{CO}$, $\alpha_{HCN}$, and their effects on $f_{dense}$ and $SFE_{dense}$, we refer the reader \cite{USERO15}.

\subsection{Creation of Integrated Intensity Maps}
\label{subsec:DCP}

For each galaxy, we create a position-position-velocity mask using the CO data. We first identify pixels with signal to noise (S/N) $>$5 in at least two adjacent velocity channels. We then remove contiguous S/N$>$5 regions that are small compared to the size of the beam. Next, we grow these regions to include adjoining pixels where S/N$>$2. The resulting mask does a good job of capturing the region of bright CO emission that one would identify by eye.

CO emission tends to be brighter and easier to excite than emission from rarer isotopologues or dense gas tracers. Therefore, we take this region of bright CO emission to also represent the position-position-velocity region where we might find these fainter lines. We verify by eye that no clear dense gas tracer emission extends beyond this mask. Thus regions outside this mask are taken to be signal free and used to estimate the rms noise in each cube.

We sum the masked line data cubes along each line of sight to produce maps of integrated intensity, in K~km~s$^{-1}$. In this sum, masked pixels have a value of zero, and lines of sight with no identified signal have an integrated intensity of zero. Thus, we also produce a two dimensional mask indicating which lines of sight include any bright signal in the CO cube.

We calculate the statistical uncertainty in the integrated intensity from the rms noise in an individual channel map (in K) times the channel width (in km~s$^{-1}$) times the square root of unmasked voxels along a line of sight.  Typical rms uncertainties due to statistical errors in the integrated lines are $\sim 0.2$~K~km~s$^{-1}$ for the CO isotopologues and $\sim 0.1$~K~km~s$^{-1}$ for the dense gas tracers. The CO data tend to have poorer sensitivity, with typical uncertainty $\sim 2$~K~km~s$^{-1}$ in the integrated intensity per line of sight. However, because the CO is brighter, the CO data still have higher S/N.

\subsection{Star Formation Rates}
\label{subsec:SFR_lit}

The deep multiwavelength data available for our sample allows the prospect to estimate the surface density of recent star formation, $\Sigma_{\rm SFR}$ in several ways. In our table of radial profiles, we provide the measurements necessary to reconstruct most popular monochromatic or ``hybrid'' tracers \citep[e.g., see review in][]{KENNICUTT12}.

In the plots accompanying the main text, we present $\Sigma_{\rm SFR}$ estimated from a combination of H$\alpha$ and $24\mu$m emission following \citet{CALZETTI07} and \citet{LEROY12}. We adopt this choice because for nearby non-starburst galaxies it has emerged as a widely accepted estimator of $\Sigma_{\rm SFR}$ \citep[see][]{KENNICUTT12}.

Although H$\alpha$+24$\mu$m is widely accepted, its calibration remains fundamentally empirical, with limited fundamental work on resolved targets \citep[][]{KENNICUTT07,CALZETTI07,MURPHY11}. Much of the literature surrounding dense gas remains focused on infrared luminosity, considering the estimated bolometric luminosity (reprocessed by dust) of the young stellar population a more physical tracer of recent star formation activity. Therefore, as a complement to the H$\alpha$+24$\mu$m estimates shown in the main plots, the appendix investigates estimates of the total infrared (TIR) surface brightness. We convolve multiband {\em Herschel} and {\em Spitzer} imaging to a common, coarse resolution and apply the prescriptions of \citet{GALAMETZ13} for each galaxy. Based on this, we calibrate galaxy-by-galaxy conversions from the higher resolution $70\mu$m data to $\Sigma_{\rm TIR}$. These data are provided in the table of radial profiles and can be used to carry out a parallel analysis to the one presented in the main text.

The numerical results from these two main approaches to $\Sigma_{\rm SFR}$ differ at the $\sim 30\%$ level. However, the choice of SFR tracer does not affect our qualitative conclusions. For either case, dense gas fraction appears to be a poor predictor of the bulk star formation efficiency, and the SFR per unit dense gas drops in the high surface density, inner parts of several of our targets.

\subsubsection{H$\alpha+24\mu$m}

We calculate the surface density of recent star formation using the prescription of \cite{CALZETTI07} recast into surface brightness units. Their prescription (their Equation 7), in luminosity units is:

\begin{equation}
\label{eq:calzetti}
{\rm SFR}~\left[ {\rm M_\odot~yr^{-1}} \right] = 5.3 \times 10^{-42} \left( L_{\rm H\alpha} + 0.031~L_{24} \right)~.
\end{equation}

\noindent Here $L_{\rm H\alpha}$ is the line integrated H$\alpha$ luminosity, with units of erg~s$^{-1}$ and no correction for internal extinction. $L_{24} \equiv \nu L_\nu (24\mu{\rm m})$ is the 24$\mu$m luminosity, also in units of erg~s$^{-1}$. This luminosity comes from multiplying the specific luminosity, $L_\nu$, by the frequency at $\lambda = 24\mu$m.

For a measured line-integrated intensity, $I_{\rm H\alpha}$, in erg~s$^{-1}$~cm$^{-2}$~sr$^{-1}$, across a beam with angular area $\Omega$ at distance $d$, the corresponding SFR surface density, $\Sigma_{\rm SFR}$ is:

\begin{eqnarray}
\Sigma_{\rm SFR} &=& \frac{{\rm SFR}}{\Omega~d^2} \\
\nonumber &=& \frac{5.3 \times 10^{-42} I_{\rm H\alpha}~\Omega~4~\pi~d^2}{\Omega~d^2} \\
\end{eqnarray}

\noindent Adopting the standard units for $\Sigma_{\rm SFR}$ of ${\rm M}_{\odot}~{\rm yr}^{-1}~{\rm kpc}^{-2}$, the $d$ in the numerator is in cm whereas the $d$ in the denominator in kpc. We correct this unit inconsistency by multiplying the numerator by $9.52 \times 10^{42}$~cm$^2$~kpc$^{-2}$. This yields $\Sigma_{SFR} = 634~I_{H\alpha}$ with $\Sigma_{\rm SFR}$ in ${\rm M}_{\odot}~{\rm yr}^{-1}~{\rm kpc}^{-2}$ and $I_{\rm H\alpha}$ in erg~s$^{-1}$~cm$^{-2}$~sr$^{-1}$. 

To place the $24\mu$m term in surface brightness units, we combine the $634$ prefactor from the previous calculation, the $0.031$ relative weighting of luminosities from Equation \ref{eq:calzetti}, the conversion of $1~{\rm MJy~sr}^{-1} = 10^{-17}$~erg~s$^{-1}$~cm$^{-2}$~sr$^{-1}$~Hz$^{-1}$, and $\nu (24\mu{\rm m}) \approx 1.25 \times 10^{13}$~Hz. Then

\begin{equation}
\label{eq:HA24-to-SFR}
\Sigma_{SFR} = 634~I_{H\alpha} + 0.0025~I_{24\mu m}~
\end{equation} 

\noindent again with $\Sigma_{\rm SFR}$ in ${\rm M}_{\odot}~{\rm yr}^{-1}~{\rm kpc}^{-2}$, $I_{\rm H\alpha}$ in erg~s$^{-1}$~cm$^{-2}$~sr$^{-1}$, and $I_{24\mu m}$ in MJy~sr$^{-1}$.

We measure 24$\mu$m intensity from the SINGS maps \citep[FWHM $\sim 6.4\arcsec$][]{KENNICUTT03}. To estimate $I_{H\alpha}$, we use the collection of literature H$\alpha$ maps compiled and processed by \citet{LEROY12}. These draw heavily on LVL and SINGS \citep{LEE09,KENNICUTT03}. This estimate assumes a Kroupa IMF. The correction to a Chabrier IMF is modest.

We convolve both the H$\alpha$ and $24\mu$m maps to the $8''$ resolution of the mm-wave data. To convolve the $24\mu$m we use the kernels of \citet{ANIANO11}. Because of the extended structure of the 24$\mu$m PSF, translating from the native PSF to an $8''$ Gaussian requires the use of an ``aggressive'' kernel. This may be another reason to prefer the $70\mu$m-oriented alternative described below.

In the appendix, we show that in our sources and target regions the $24\mu$m term dominates $\Sigma_{\rm SFR}$. H$\alpha$ itself typically contributes $\sim 30$\% to the overall estimate.

\subsubsection{Indirect Total Infrared Based SFR Estimates}
\label{sec:sfr_tir}

As an alternative to the H$\alpha$+24$\mu$m approach, the appendix works out an SFR estimate based on the total infrared (TIR) emission in our targets. We match the resolution of the {\em Spitzer} and {\em Herschel} maps at $24$, $70$, $160$, and $250\mu$m at $30\arcsec$\footnote{The common $30\arcsec$ resolution was chosen based on the ability to compare to IRAM 30-m maps for a related project. In principle this comparison could be done at $22\arcsec$ (FWHM) resolution. The {\em Herschel} 250$\mu$m PSF can be ``very safely'' convolved to a Gaussian with this resolution \citep[see][]{ANIANO11}.}. Then, we use the prescriptions of \citet{GALAMETZ13} to calculate $\Sigma_{TIR}$, the total infrared luminosity surface density. This corresponds to the expected integral under the whole IR spectral energy distribution from $8{-}1000\mu$m. If the light in the region under consideration is dominated by an embedded young stellar population, then this offers an alternative tracer of recent star formation.

The full suite of IR bands is only available at poor ($22\arcsec$) resolution. To estimate the small-scale behavior of $\Sigma_{\rm TIR}$, we use the high resolution $70\mu$m maps from {\em Herschel} \citep[][]{KENNICUTT11}. In the appendix, we show that 70$\mu$m emission correlates better with $\Sigma_{\rm TIR}$ than the 24$\mu$m does at low resolution. We calibrate a simple conversion from $70\mu$m to TIR for each galaxy in our sample. This conversion is not perfect, but assuming that it holds sub-resolution allows us to estimate $\Sigma_{\rm TIR}$ at the $8\arcsec$ common resolution of our mm-wave data.

We present these ``TIR'' based estimates in the table or radial profiles, adopting the conversion from $\Sigma_{TIR}$ to $\Sigma_{SFR}$ calculated by \citet{MURPHY11}:

\begin{equation} 
\bigg(\frac{\Sigma_{SFR, TIR}}{\rm{M_{\odot} \ yr^{-1} \ \rm{kpc}^{-2}}}\bigg) = 1.48\times10^{-10}\bigg(\frac{\Sigma_{TIR}}{L_{\odot} \ \rm{kpc}^{-2}}\bigg)~.
\label{eq:TIR-to-SFR}
\end{equation}

\noindent \citet{MURPHY11} derive this conversion assuming that the entire Balmer continuum is absorbed and re-radiated by dust and assuming a continuous star formation history with no contamination due to an old background stellar population. Based on comparison with UV emission, all of our targets appear heavily extinguished across the region of interest (so that adding a GALEX-based term would not influence $\Sigma_{\rm SFR,TIR}$ significantly). However, variations in the star formation history or heating by older stellar populations remain a concern (though these are among the most gas rich, IR-bright, actively star forming regions in the KINGFISH {\em Herschel} survey).

This TIR estimate follows a less conventional approach than the H$\alpha$+24$\mu$m, but it represents our preferred SFR estimate for comparison to the literature. We recommend adopting these values from the table of radial profiles when carrying out such a comparison. However, in the main text, we will plot results deriving SFR from H$\alpha$+24$\mu$m.

\subsubsection{Literature Star Formation Rate Estimates}

We compare our results to those of \cite{USERO15}, \cite{GAO04}, and \cite{GARCIABURILLO12}. Each of these estimates the TIR luminosity and uses this as their SFR indicator. But each study calculates TIR slightly differently. \cite{USERO15} calculates TIR from combining 24-160$\mu$m data to get a 24$\mu$m/TIR ratio at coarse resolution, and then rescales the 24$\mu$m intensity at their working resolution by that factor (similar to our approach in the appendix). \cite{GAO04} calculates TIR luminosity for their whole galaxies using the prescription in \cite{SANDERS96} and measurements from all 4 IRAS flux bands- 12, 25, 60, 100$\mu$m. \cite{GARCIABURILLO12} calculates TIR intensity using a scaling of 60-100$\mu$m from \cite{GARCIACARPIO08}. Based on the appendix, we expect these different approaches to introduce scatter into any comparison at the $\sim 30\%$ level. Homogenizing the literature SFR estimates used in dense gas studies clearly has large value but is beyond the scope of this study.

\subsection{Stellar Mass}
\label{subsec:SM_lit}

\citet{USERO15} and \citet{BIGIEL16} highlighted strong correlations with stellar structure, a possible driver of gas pressure. With this in mind, we compare our observations to 3.6$\mu$m. maps, which are dominated by light from old stars. We use maps obtained by {\em Spitzer}, mostly as part of SINGS \citep{KENNICUTT03}, and processed as part of the S$^4$G survey \citet{SHETH10}. Because {\em Spitzer}'s 3.6$\mu$m band can include emission associated with recent star formation, primarily dust emission, we use the contaminant-corrected maps of \citet{QUEREJETA15}, which are based on an algorithm developed by \citet{MEIDT12}. Globally, the correction of \citet{QUEREJETA15} and \citet{MEIDT12} attributes $10{-}30$\% of the $3.6\mu$m emission to a component associated with star formation and removes it from the maps. The PSF of the IRAC 3.6$\mu$m maps has FWHM $\sim 1.9\arcsec$ before any convolution.

We calculate the stellar surface density from the contaminant-corrected 3.6$\mu$m data (\citealt{QUEREJETA15}) assuming a mass to light ratio of $352$~M$_\odot$~pc$^{-2}$~(MJy~sr$^{-1}$) (\citealt{MEIDT14}). This roughly corresponds to 0.5 M$_{\odot}$ per L$_{\odot}$. Note that \citet{USERO15} and \citet{BIGIEL16} used median-filtered versions of the 3.6$\mu$m maps to correct for contamination, rather than the ICA-corrected versions from \citet{QUEREJETA15}. \citet{USERO15} also used a different mass to light ratio. After accounting for the different mass-to-light ratio, the contaminant corrected $3.6\mu$m map that we use here agrees reasonably with the median-filtered \citet{USERO15} and \citet{BIGIEL16} data.

\subsection{Atomic Gas Mass}
\label{subsec:GM_lit}

The distribution of atomic gas ({\sc Hi}) plays a role in calculating dynamical equilibrium pressure. We compare to the {\sc Hi} maps from THINGS \citep{WALTER08} for NGC~3351, NGC~3627, and NGC~5194, and archival VLA maps for NGC~4254 and NGC~4321 \citep[from][]{SCHRUBA11,LEROY13}. We begin with the 21-cm integrated intensity (``moment 0'') maps and convert these to hydrogen column density assuming optically thin emission \citep[see][]{WALTER08}. 

The {\sc Hi} maps have coarser resolution than our other data. The naturally weighted THINGS maps for NGC 3351 and NGC 3627 have synthesized beams of $9.9\arcsec \times 7.1\arcsec$ and $10.0\arcsec \times 8.9\arcsec$. The archival maps of NGC~4254 and NGC~4321 have synthesized beams of $16.9\arcsec \times 16.2\arcsec$ and $14.7\arcsec \times 14.1\arcsec$. In all cases, the 21-cm map covers the full spectral extent of the galaxy and the pixels heavily oversample the synthesized beam. However, we note that the VLA maps do not include zero spacing data, and can be expected to filter out emission with angular extent $\gtrsim 30'$ in an individual channel. Despite this, \citet{WALTER08} show that for the THINGS maps maps the agreement between the flux measured in the interferometer-only cube and single dish measurements is quite good. In practice, we suggest to consider the zero level of these profiles to be perhaps biased slightly low, by $\lesssim 2$~M$_\odot$~pc$^{-2}$.

In our analysis, we will assume the {\sc Hi} to be smooth sub-resolution. This agrees with analysis showing the {\sc Hi} to be only weakly clumped \citep{LEROY13B} and with the relatively flat shape of the radial profiles \citep[][and see below]{SCHRUBA11}. The main application of the {\sc Hi} profiles will be to calculate total gas surface density for use in pressure estimates, so the analysis is not terribly sensitive to this assumption.

\subsection{Dynamical Equilibrium}
\label{subsec:Ph}

\citet{USERO15} and \citet{BIGIEL16}, following \citet{HELFER97}, suggested that ISM pressure may play a central role in determining the gas density distribution and the role of gas at any particular density in star formation. We estimate the approximate midplane gas pressure needed to maintain a vertical dynamical equilibrium following \citet{ELMEGREEN89,BLITZ06,WONG02,LEROY08,OSTRIKER10}. Specifically, we take:

\begin{equation} 
\label{eq:ph}
P_{DE} \approx \frac{\pi G \Sigma_{gas}^2}{2}+ \Sigma_{gas} (2 G \rho_\star)^{1/2} ~\sigma_{gas, z}~.
\end{equation}

\noindent Here $\Sigma_{gas}$ is the total atomic plus molecular gas surface density, $\rho_\star$ is the volumetric mass density of stars and dark matter at the midplane, and $\sigma_{gas, z}$ is the vertical velocity dispersion of the gas.

$P_{DE}$ expresses the pressure needed to balance the vertical gravity on the gas in the galaxy disk. The first term in Equation \ref{eq:ph} reflects the gas self-gravity, the second term reflects the weight of the gas in the potential well of the stars. For our target regions, the stellar potential will exceed the gas self-gravity, and we expect the second term be dominant.

Only the stars and dark matter within the gas layer are relevant to the potential. We expect the density of dark matter within the gas layer to be small compared to that of the stars, and so neglect that contribution to $\rho_\star$. Thus, $\rho_\star$ rather than $\Sigma_\star$ enters Equation \ref{eq:ph}. Estimating the three dimensional structure of the disk represents the main challenge to gauging $P_{DE}$ from observations. We need estimates of the local stellar scale height to translate our inferred $\Sigma_\star$ into $\rho_\star$ via:

\begin{equation}
\label{eq:pstar}
\rho_\star \approx \frac{\Sigma_\star}{2~h_\star}
\end{equation}

\noindent following \citet{VANDERKRUIT88}.

A full treatment of the height of stellar disks as a function of morphology, mass, and galactocentric radius is beyond the scope of this paper. Observations still do not provide a perfect prescription, and more recent observations show that the topic may be more complex than suggested by earlier work \citep[e.g.,][]{COMERON12,COMERON14}. Lacking a better prescription, we follow \citet{LEROY08} and adopt the typical flattening ratio found by \citet{KREGEL02}, $l_*/h_*=7.3 \pm 2.2$, to relate the stellar scale height and the observed scale length for the disk component. 

We adopt $\sigma_{z,gas} \approx 15$~km~s$^{-1}$. This is towards the high end of the observed gas velocity dispersions in galaxy disks, appropriate for large scales and the gas-rich, high surface density inner regions of galaxy disks \citep[e.g.,][]{CALDUPRIMO13,TAMBURRO09,MOGOTSI16}.

For these assumptions, we calculate $P_{DE}$ in each ring, and report this in the online table described in Appendix \ref{sec:table}.

Finally, we provide a word on the interpretation of $P_{DE}$. $P_{DE}$ is the time-averaged pressure needed to support the gas disk against its weight due to both self-gravity and the gravity of the stellar potential well. We include all gas in this calculation, and expect Equation \ref{eq:ph} to hold averaged over time and space (several gas scale heights), with all gas participating in this equilibrium. Thus, we expect $P_{DE}$ to express the average state of a region of the ISM. Equation \ref{eq:ph} does not distinguish between a gravitationally-bound component and a diffuse component of the gas. This is consistent with recent simulations (e.g. \citealt{KIM17}) and observations that indicate self-gravitating structures to be quite transient \citep{KAWAMURA09,SHCRUBA10}. Such transient structures may be thought of as fully participating in the ISM's large-scale dynamics when averaged over several scale heights and vertical crossing times. Similarly, for a turbulence-dominated system we do not think of $P_{DE}$ as an external pressure that acts on cloud "surfaces;" rather, it is the mean pressure averaging over both high- and low-density regions.

\subsection{Convolution to a Common Resolution}
\label{subsec:align}

We convolved all data to a working resolution of $8\arcsec$, using the kernels provided by \citet{ANIANO11} when needed. After convolution, we aligned all of the broadband multi-wavelength data to the common astrometric grid shared by the HCN, HCO$^{+}$, CS, $^{13}$CO, and C$^{18}$O data. As mentioned above, the {\sc Hi} data could not be beam matched to an $8\arcsec$ beam, and we use these at their native resolution, assuming a smooth distribution within the beam. 

We use a combination of the convolved {\em Spitzer} 24$\mu$m data and H$\alpha$ data as our primary SFR tracer. Despite the modest $\sim 6\arcsec$ core of the 24$\mu$m PSF, the {\em Spitzer} PSF has significant extended structure, and an ``aggressive'' kernel is required to convolve a 24$\mu$m image to have an $8\arcsec$ Gaussian PSF \citet{ANIANO11}. As a result, mild artifacts are visible surrounding the nucleus of NGC~3351 in the convolved $24\mu$m image. The alternative $70\mu$m based tracer has a cleaner PSF. Here again, comparing the two approaches offers a useful way to assess systematic effects.

Note that we use this 8" working resolution for NGC 3351, NGC 3627, NGC 4254, and NGC 4321. We use a lower 30" working resolution for NGC 5194, set by the resolution of the IRAM-30m data. Otherwise the procedure is the same.

\subsection{Construction of Radial Profiles}
\label{subsec:RP}

The most striking contrast in our data is between the bright central regions and the surrounding disk, which includes fainter emission from gas in the spiral arms and bars. Even ALMA struggles to detect emission from the extended disk at high significance. To improve our signal-to-noise ratio, we conduct much of our analysis using radial profiles of line emission and supporting data. Although azimuthal (e.g., arm-interarm) contrast will also be interesting for future studies, our present data lack the sensitivity to recover faint interarm emission.

We set the width of each radial bin to be half the FWHM of the working beam  ($15\arcsec$ for NGC 5194 and $4\arcsec$ for all other galaxies). This should critically sample the radial structure of the galaxy. It also yields profiles in which adjacent rings are correlated, so that we have twice as many data in each plot than independent radial measurements.

In each ring, we calculate two values: (1) the mean intensity of all pixels within the two-dimensional version of the CO mask, ignoring all other values, and (2) the mean intensity of all pixels in the ring. For (2), we treat pixels outside the map as having intensity $0$ for the molecular lines, so that the ring averaged value represents the values in the mask diluted by the empty space outside the mask. For the ancillary data, (3.6$\mu$m, H$\alpha$, 24$\mu$m, etc.), we calculate the mean value only within the two-dimension CO-based mask for profile (1) and the mean value over the whole ring for profile (2). We report both values in the table described in Appendix \ref{sec:table}. Note that for these other data, we do not set any pixels to 0 when taking the mean over the whole ring, but instead use all of the data.

We calculate the uncertainty for these mean intensities by propagating the error from individual intensity measurements. In this case the uncertainty is

\begin{equation}
\Sigma_{ii} = \frac{\sqrt{\sigma_1^2 + \sigma_2^2 + \ldots + \sigma_N^2}}{N} \times \sqrt{O}~.
\end{equation}

\noindent Here $\Sigma_{ii}$ is the uncertainty in the integrated intensity averaged over the ring and $N$ is the number of pixels in the ring. $O$ is a factor (the ``oversampling factor'') designed to account for the non-independence of the pixels. If all data are independent with the same uncertainty, $\sigma_{II} = \sigma_1 / \sqrt{N}$ as expected for the uncertainty in the mean of $N$ data. Generally $O$ should be the number of pixels per beam. Here, given that our rings are thinner than a beam, we take $O$ equal to the ring thickness in pixels times the FWHM beam size in pixels. We never allow $N/O < 1$, so that $\sigma_{II}$ is at most the uncertainty in an individual point. When calculating this mean intensity for the line data, pixels outside of the mask are ignored even for profile (2). That is, our uncertainties do not reflect uncertainty in mask construction.

The intensity profiles of each line, along with profiles of the supporting multiwavelength data, are a main observational result of the paper. These are available in full as online material. All surface brightness and surface densities quoted in the paper have been corrected for the effects of inclination.

\section{Results}
\label{sec:Results}

\begin{figure*}
\plottwo{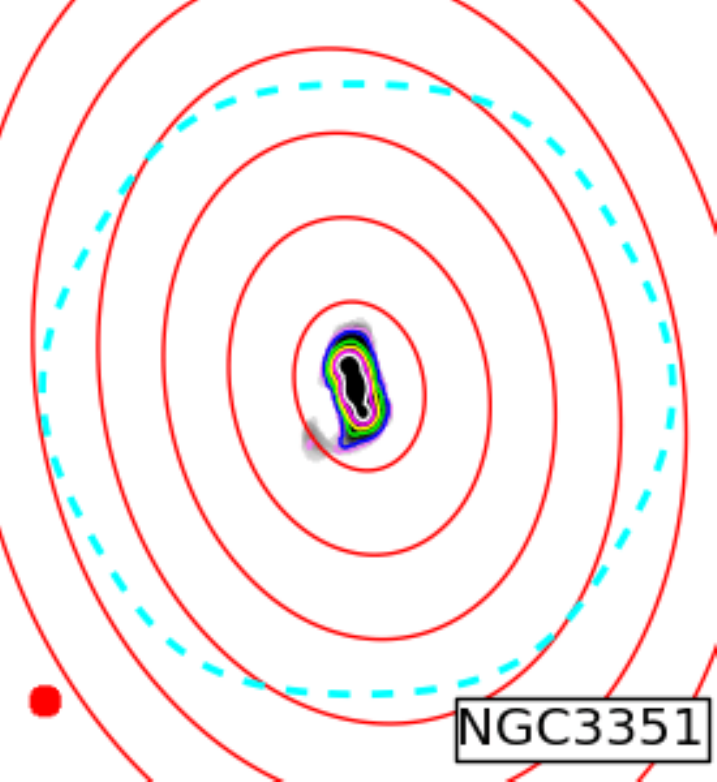}{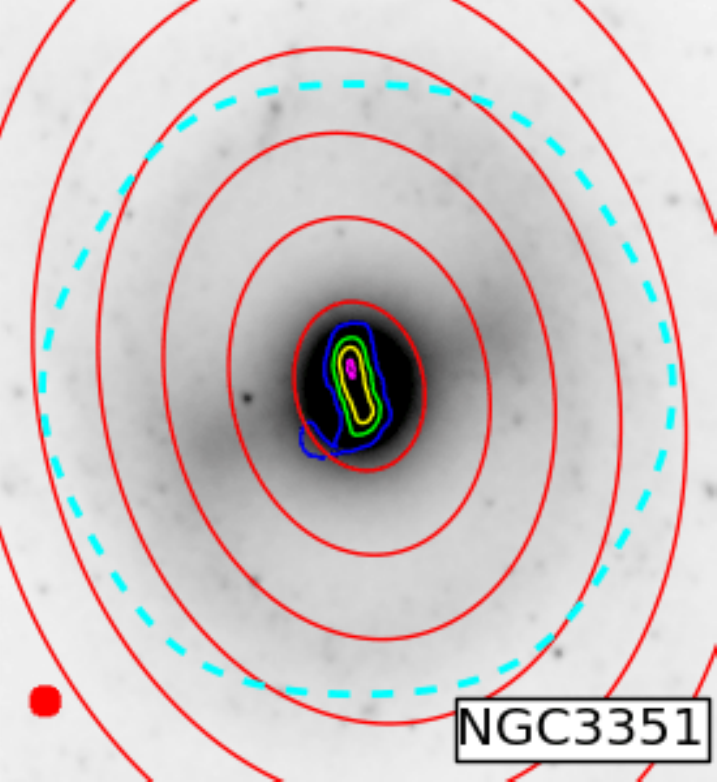}
\plottwo{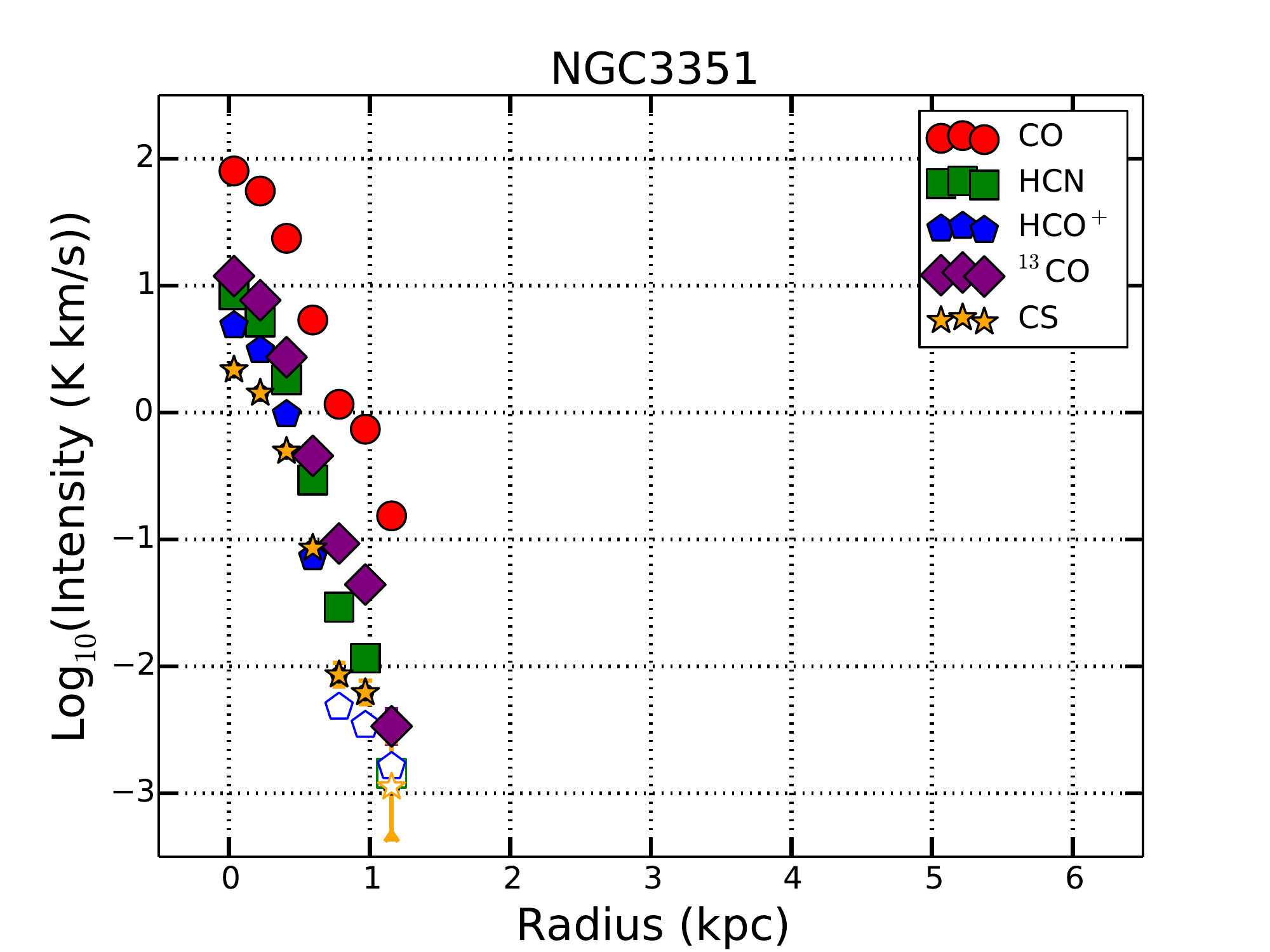}{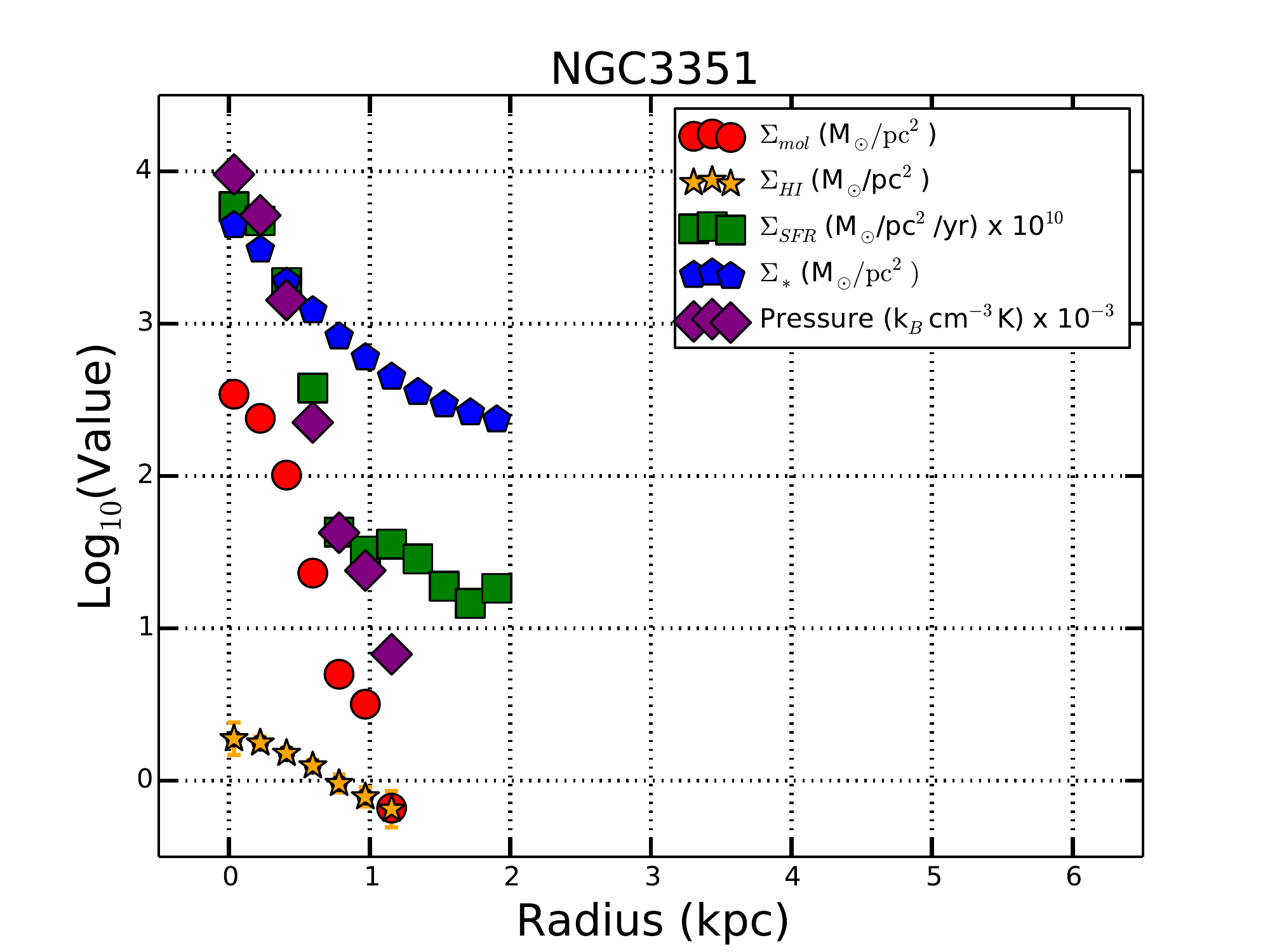}
\plottwo{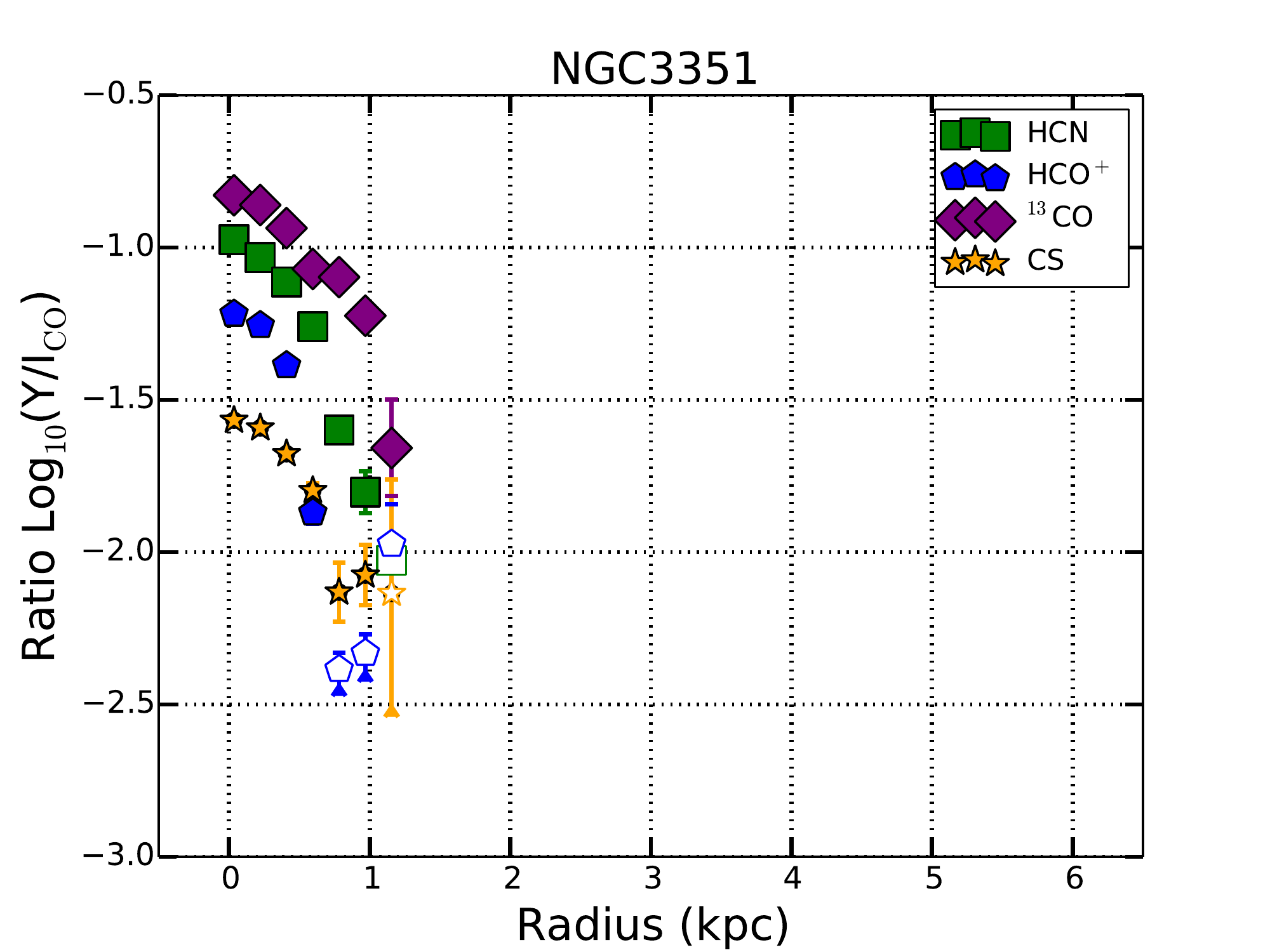}{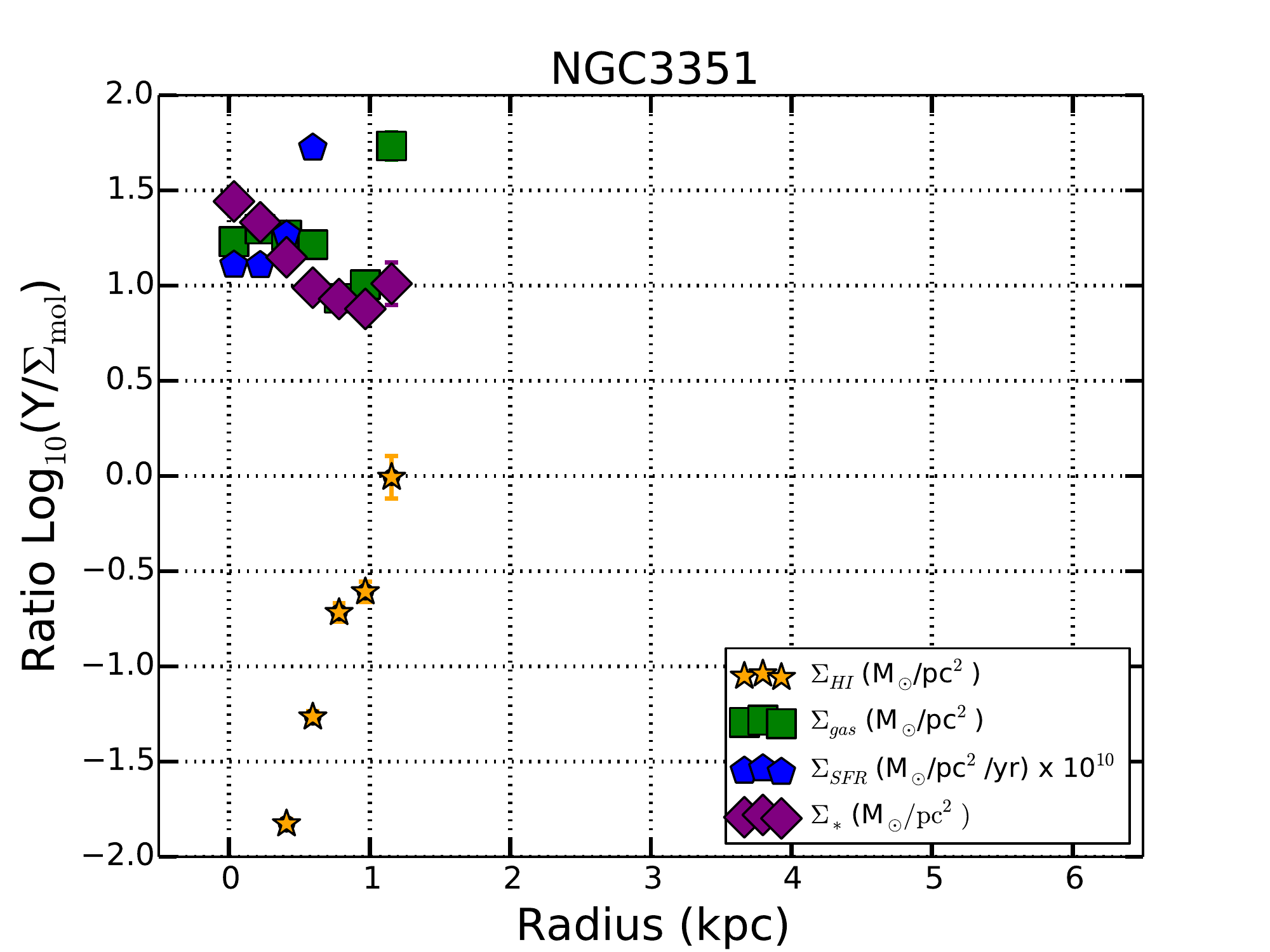}
\caption{Our ALMA observations in context for NGC~3351. ({\em Top left:}) HCN contours (color) over CO integrated intensity (grayscale). HCN contours: show 0.5 (violet), 1 (blue), 3.5 (green), 6.5 (yellow), 9.5 (magenta), and 12.5 (white)~K~km~s$^{-1}$. Red rings show fixed galactocentric radius in the plane of the galaxy, spaced by 1~kpc. The filled red circle in the lower left shows the common $\theta=8\arcsec$ (FWHM) beam used in our analysis. The dashed cyan contour shows the ALMA field of view for HCN emission.  ({\em Top right:}) 3.6$\mu$m map, tracing stellar structure, with CO integrated intensity contours at 10, 60, 110, 160, and 210~K~km~s$^{-1}$. ({\em Center left}:) Azimuthally averaged integrated intensity profiles for CO, HCN, CS, HCO$^+$, and $^{13}$CO (see \S \ref{subsec:RP}), restricting the average to regions with statistically significant CO emission. Filled points have SNR $\geq 2$. Error bars show uncertainty in the mean. Empty points indicate upper limits.  ({\em Center right:}) Intensity profiles for tracers of galaxy and ISM structure: CO (tracing molecular gas), H$\alpha$ and 24$\mu$m emission (tracing star formation), and contaminant-corrected 3.6$\mu$m emission (tracing stellar mass).({\em Bottom left:}) Ratio of other lines to CO intensity as a function of radius. ({\em Bottom right:}) Ratios among tracers of galaxy and ISM structure, along with estimated dynamical equilibrium pressure. The radial profiles are available online as described in Appendix \ref{sec:table}.}
\label{fig:3351}
\end{figure*}

\begin{figure*}
\plottwo{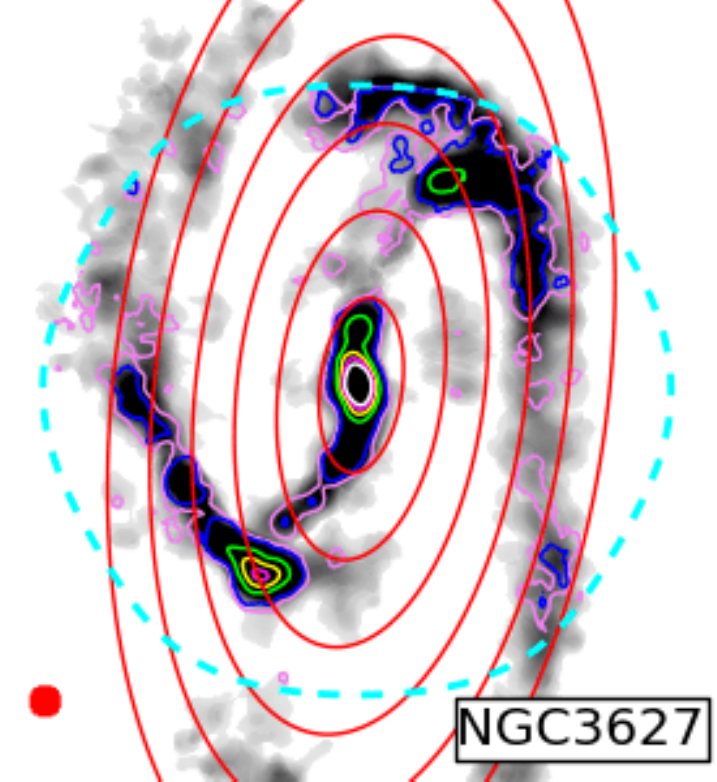}{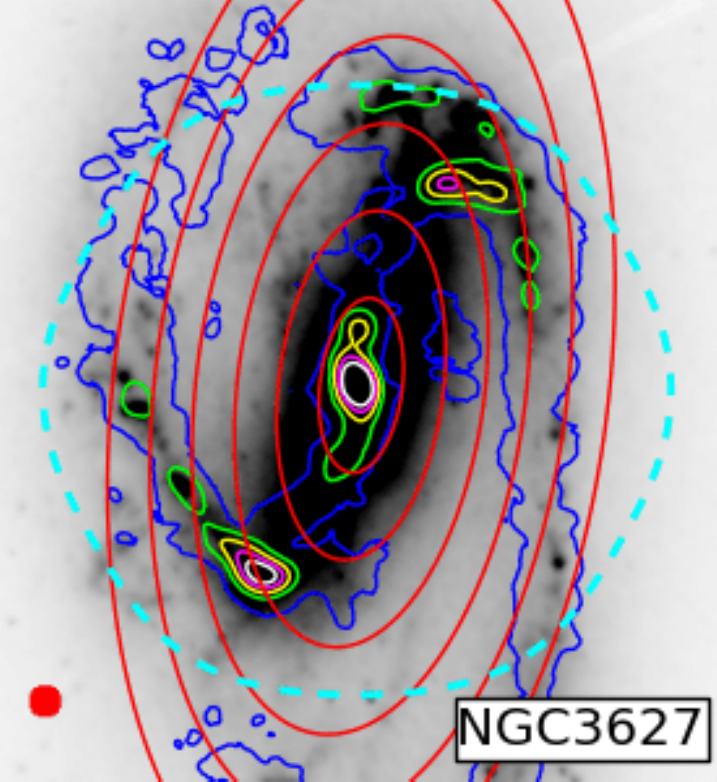}
\plottwo{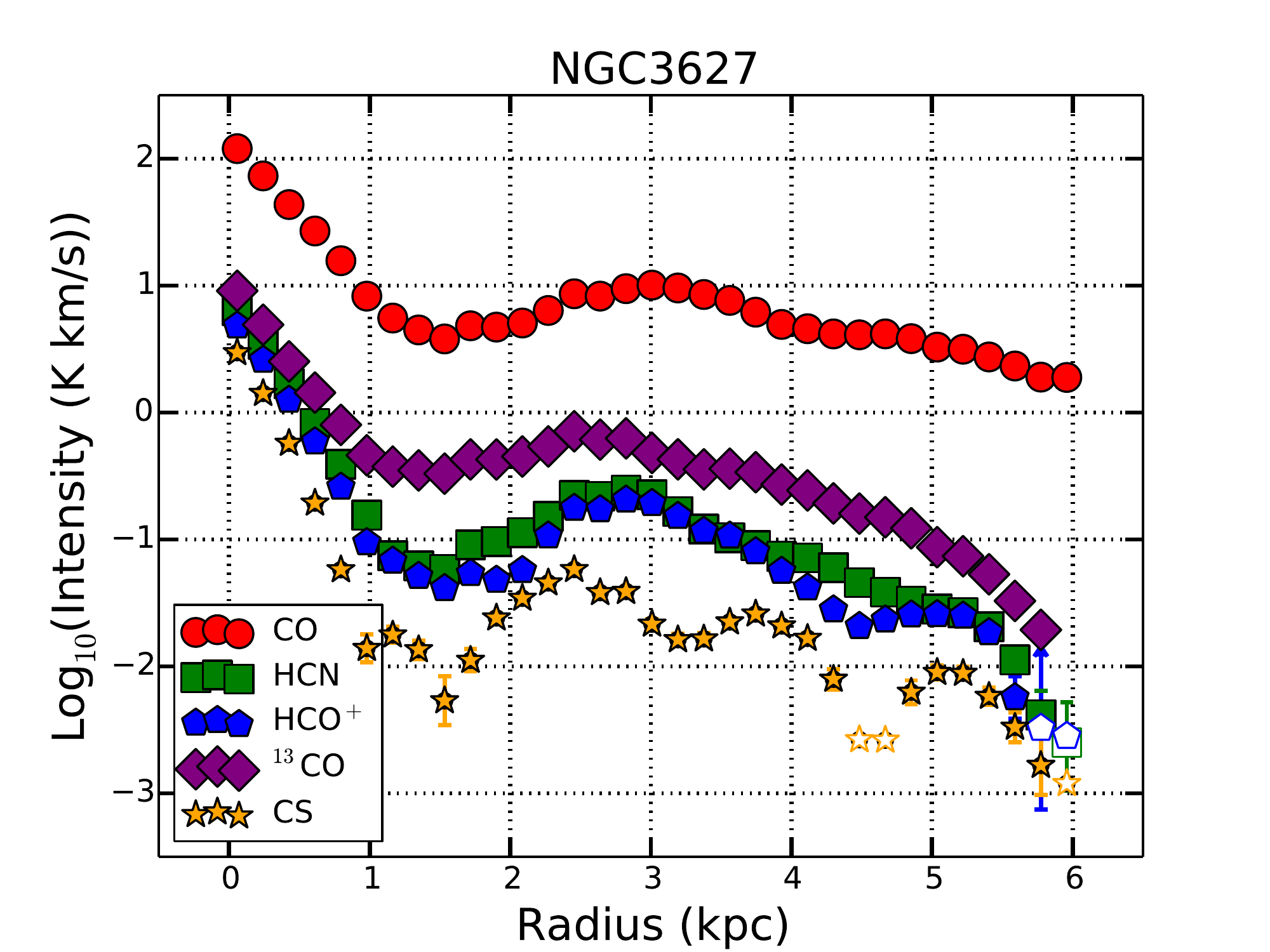}{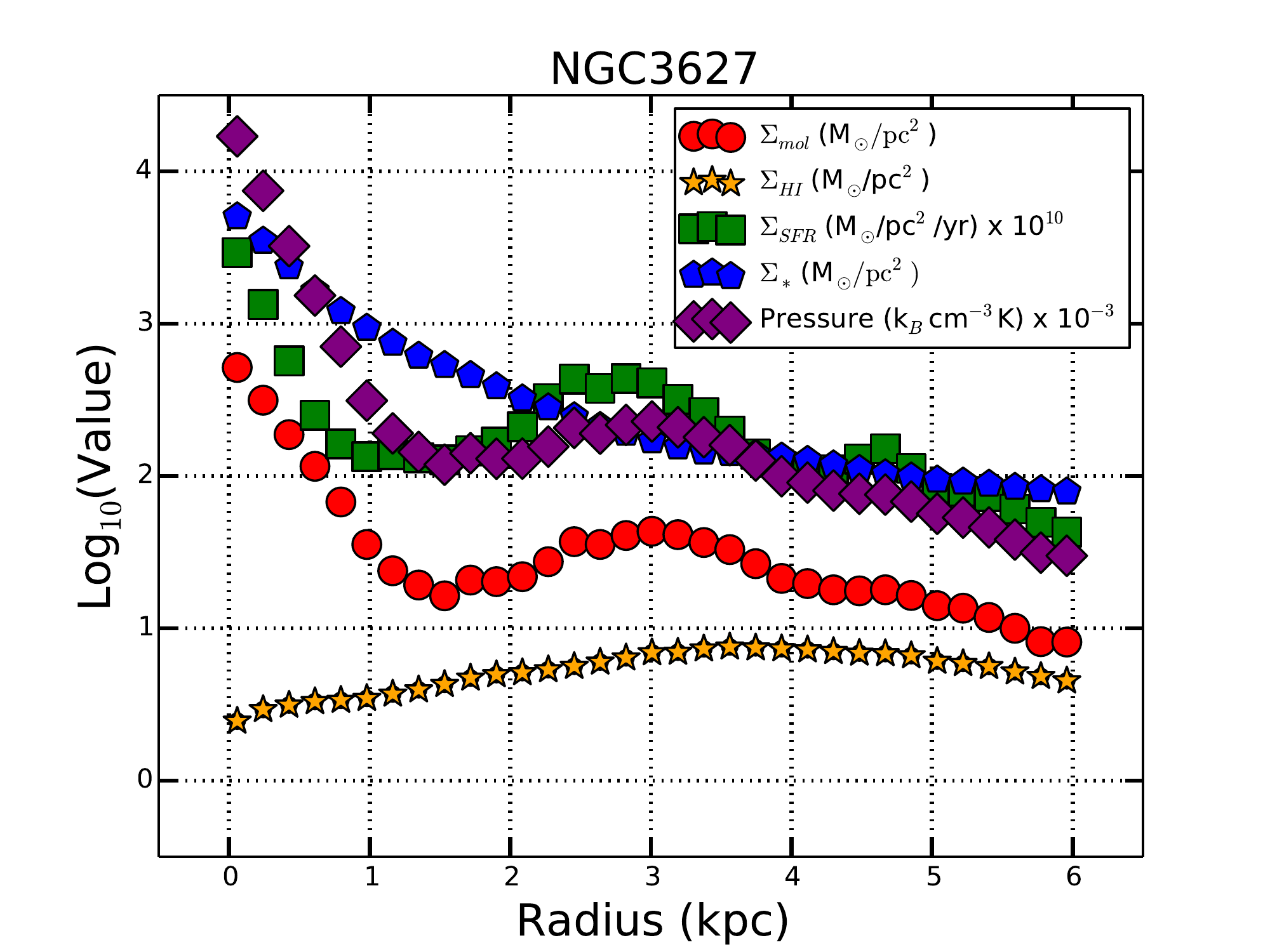}
\plottwo{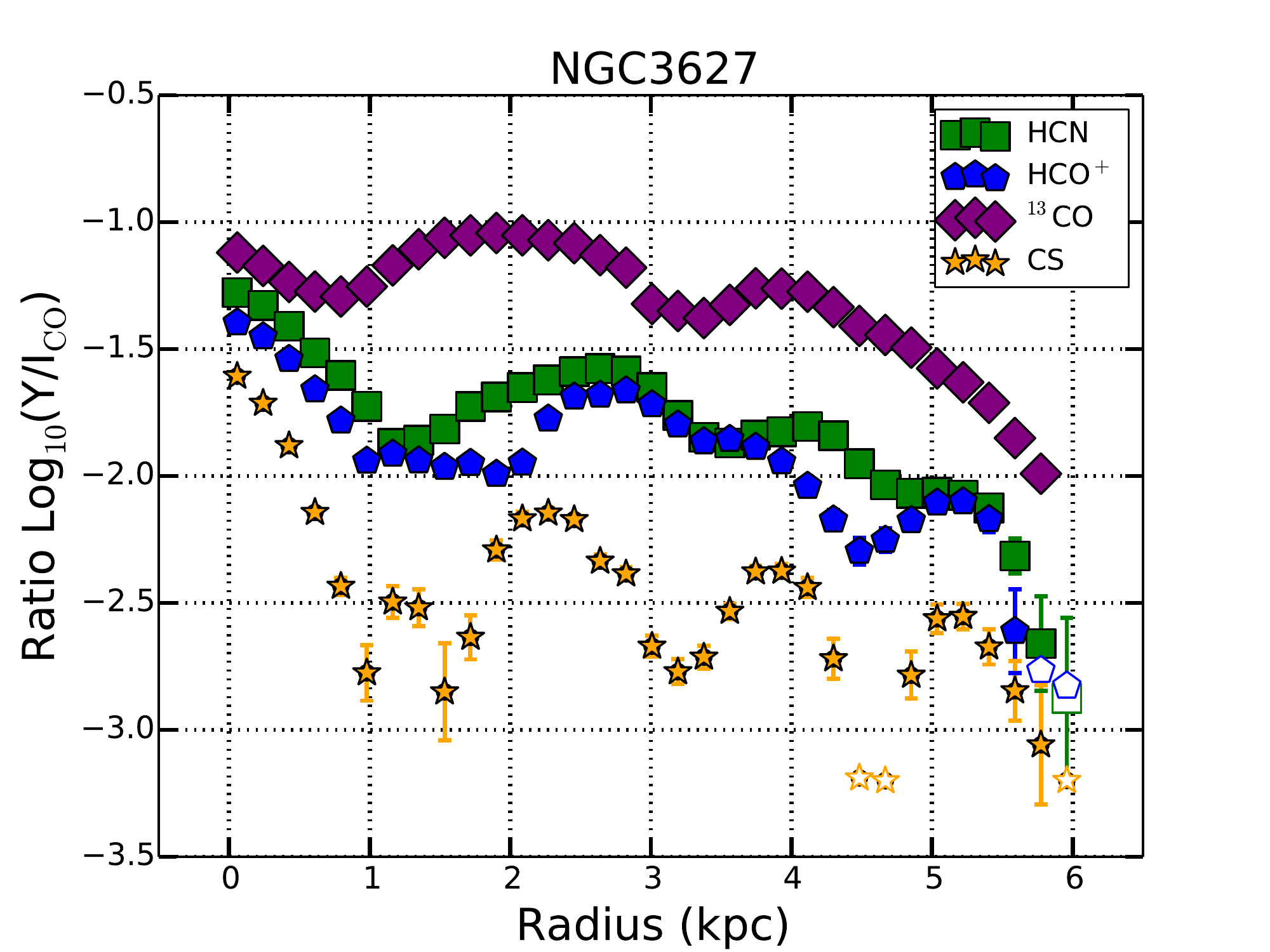}{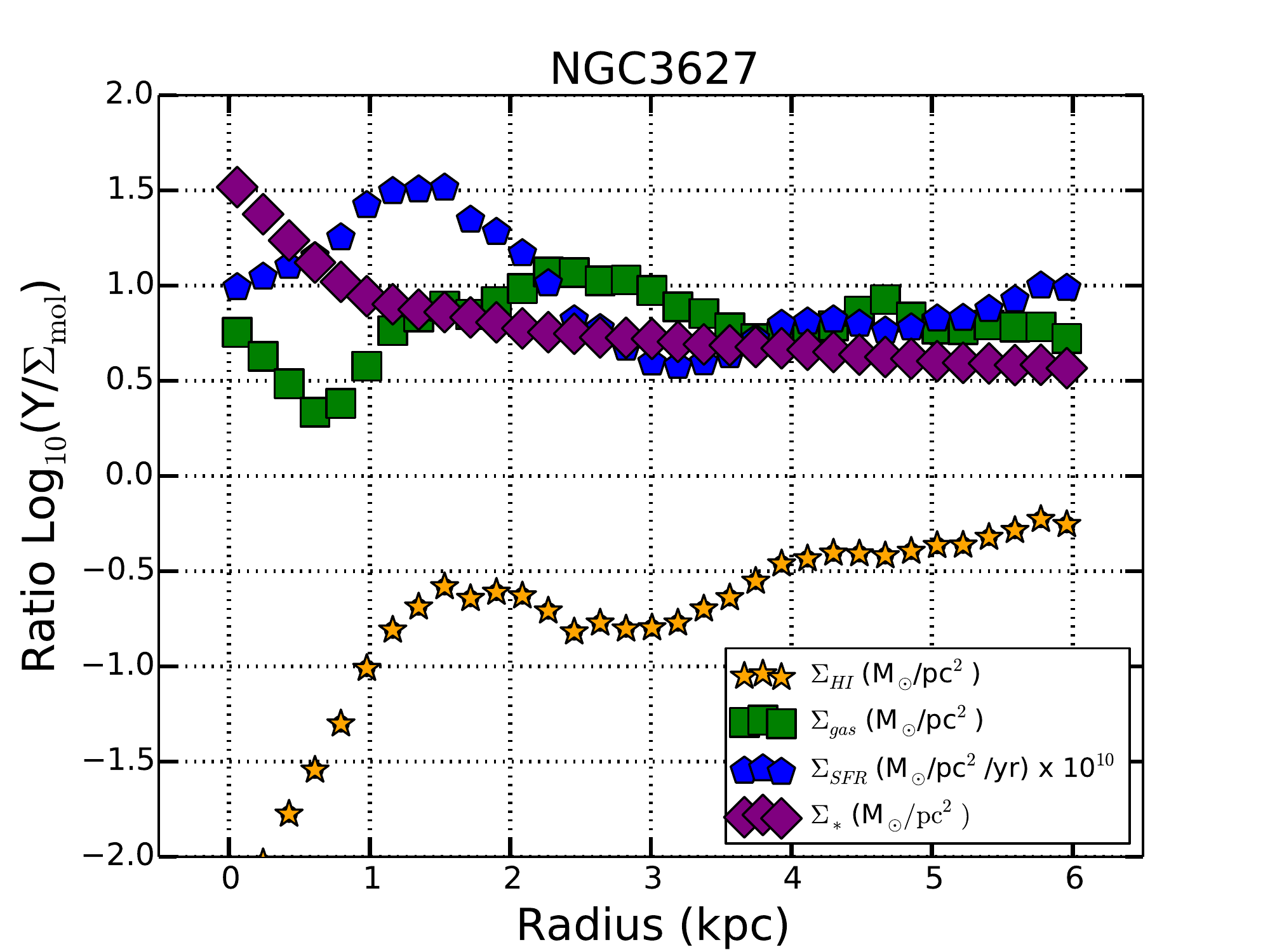}
\caption{As Figure \ref{fig:3351} but for NGC~3627.}
\label{fig:3627}
\end{figure*}

\begin{figure*}
\plottwo{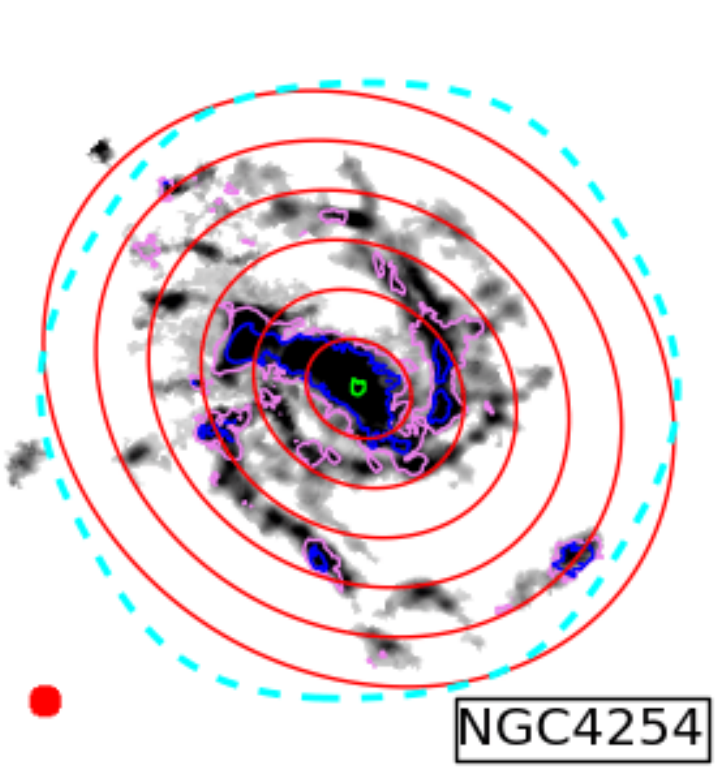}{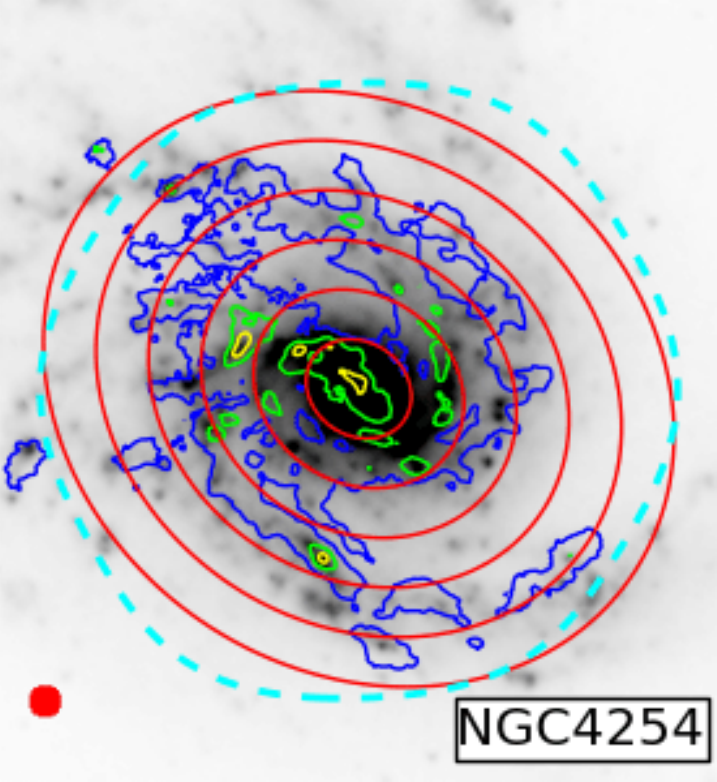}
\plottwo{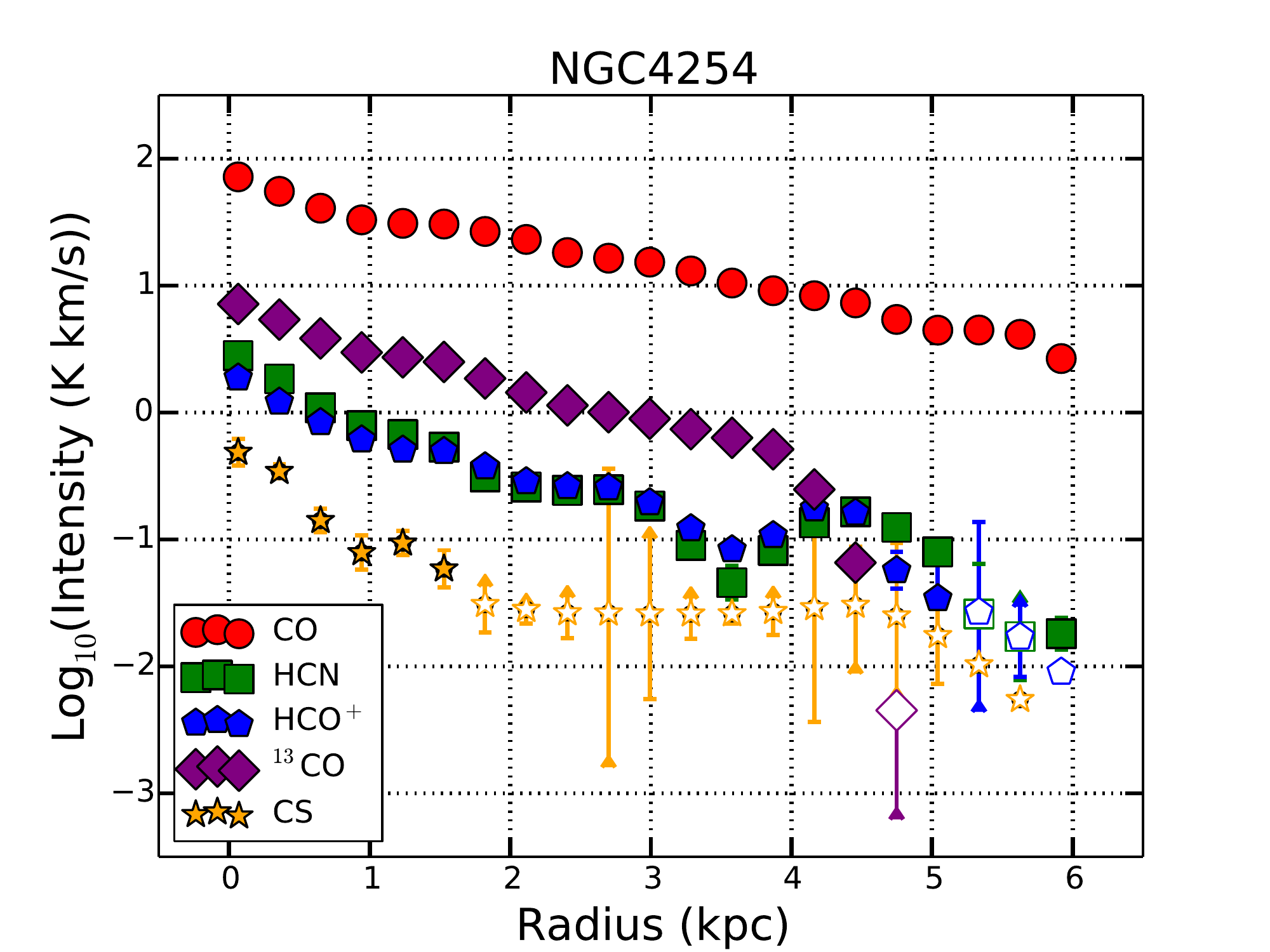}{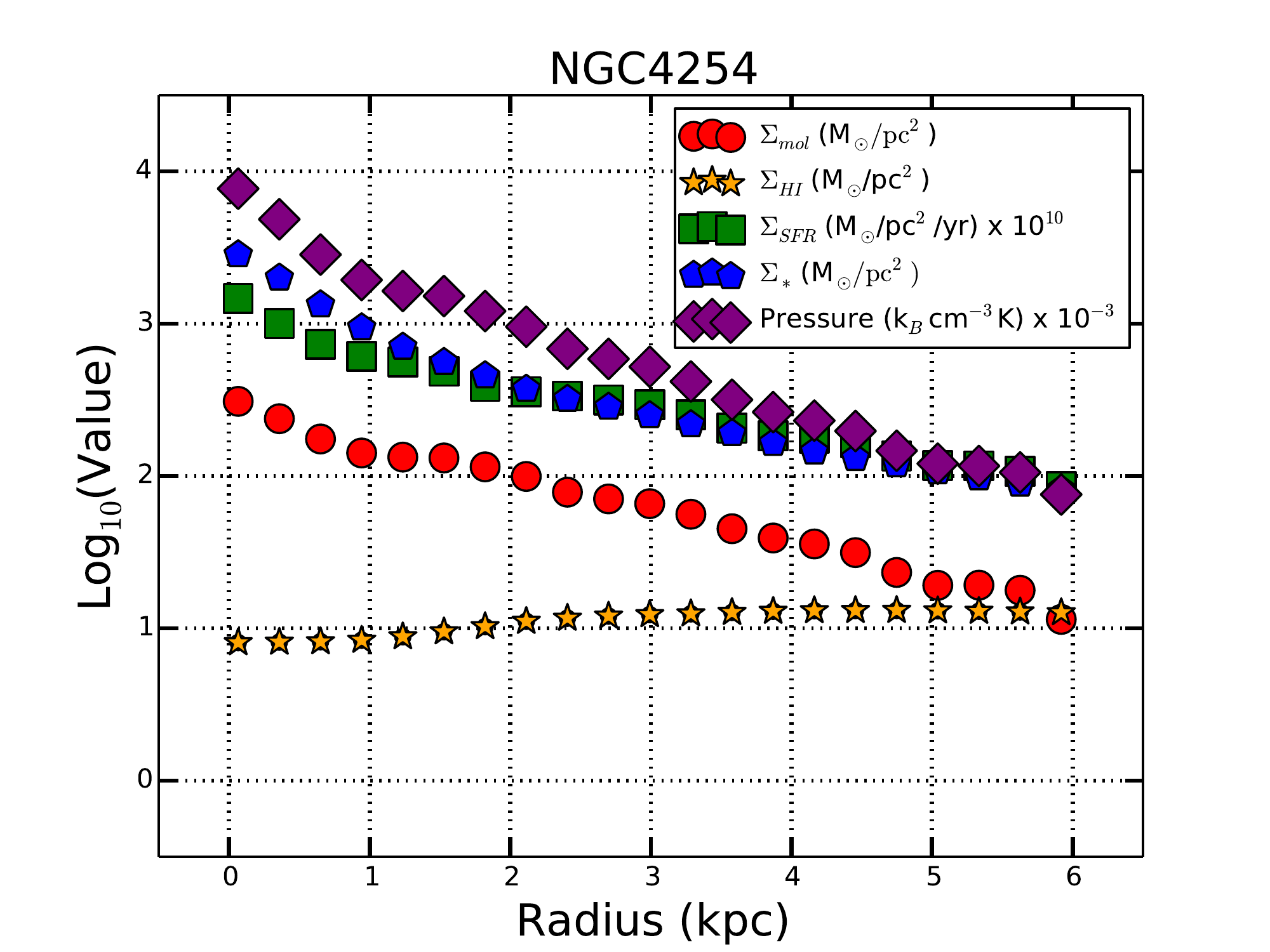}
\plottwo{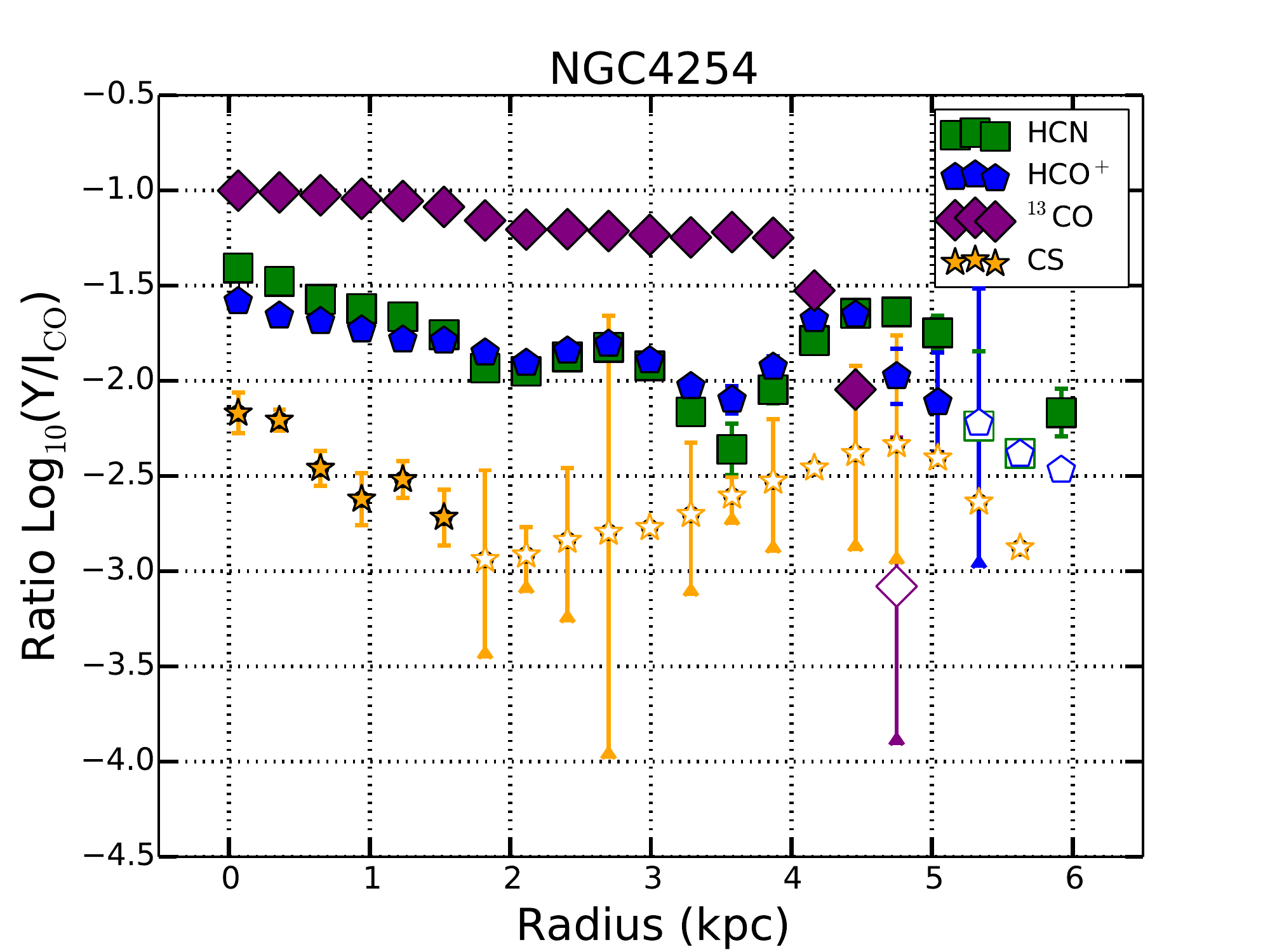}{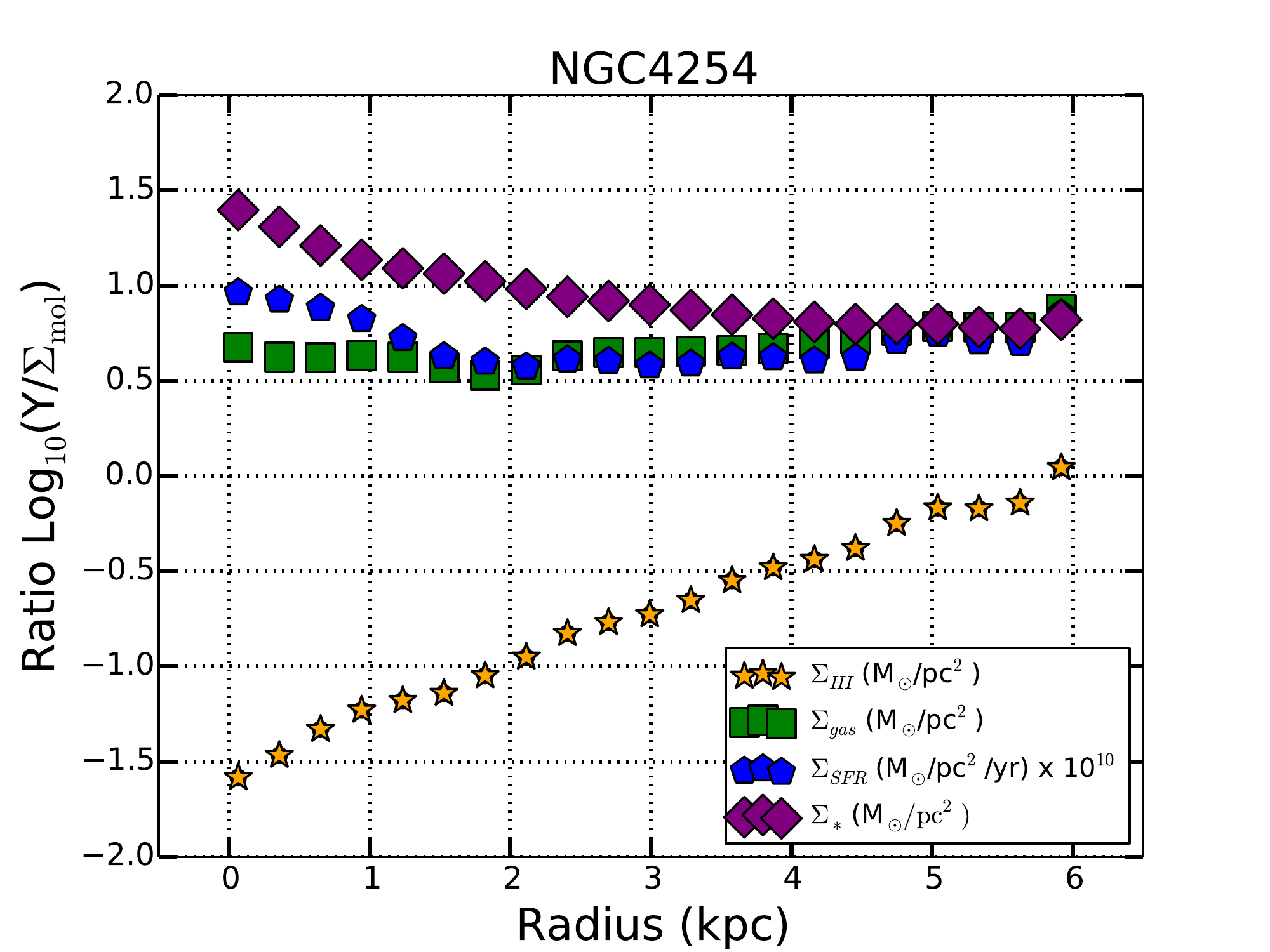}
\caption{As Figure \ref{fig:3351} but for NGC~4254.}
\label{fig:4254}
\end{figure*}

\begin{figure*}
\plottwo{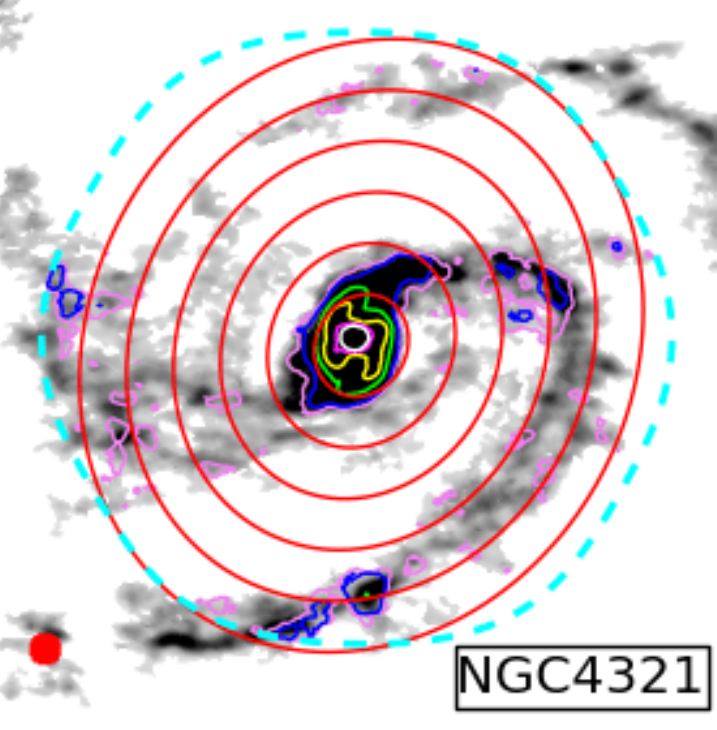}{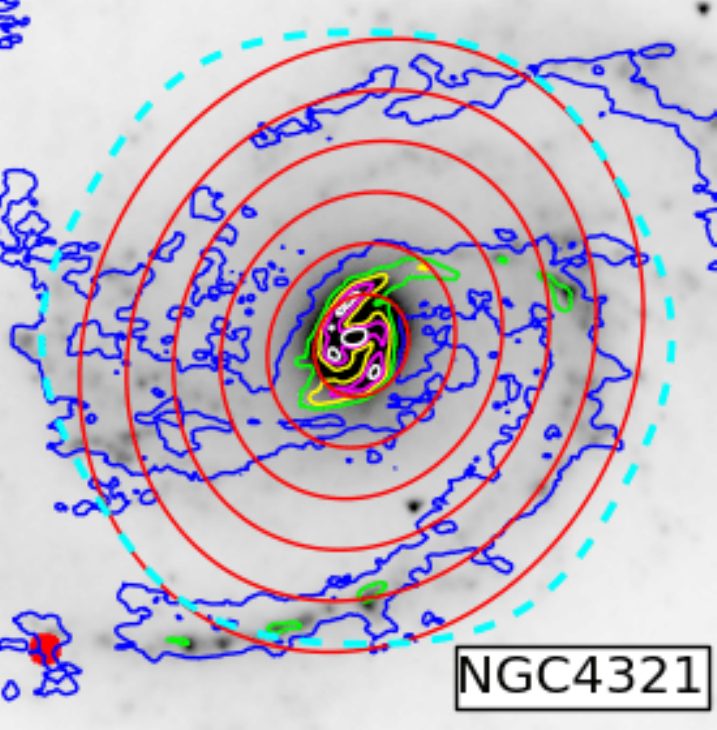}
\plottwo{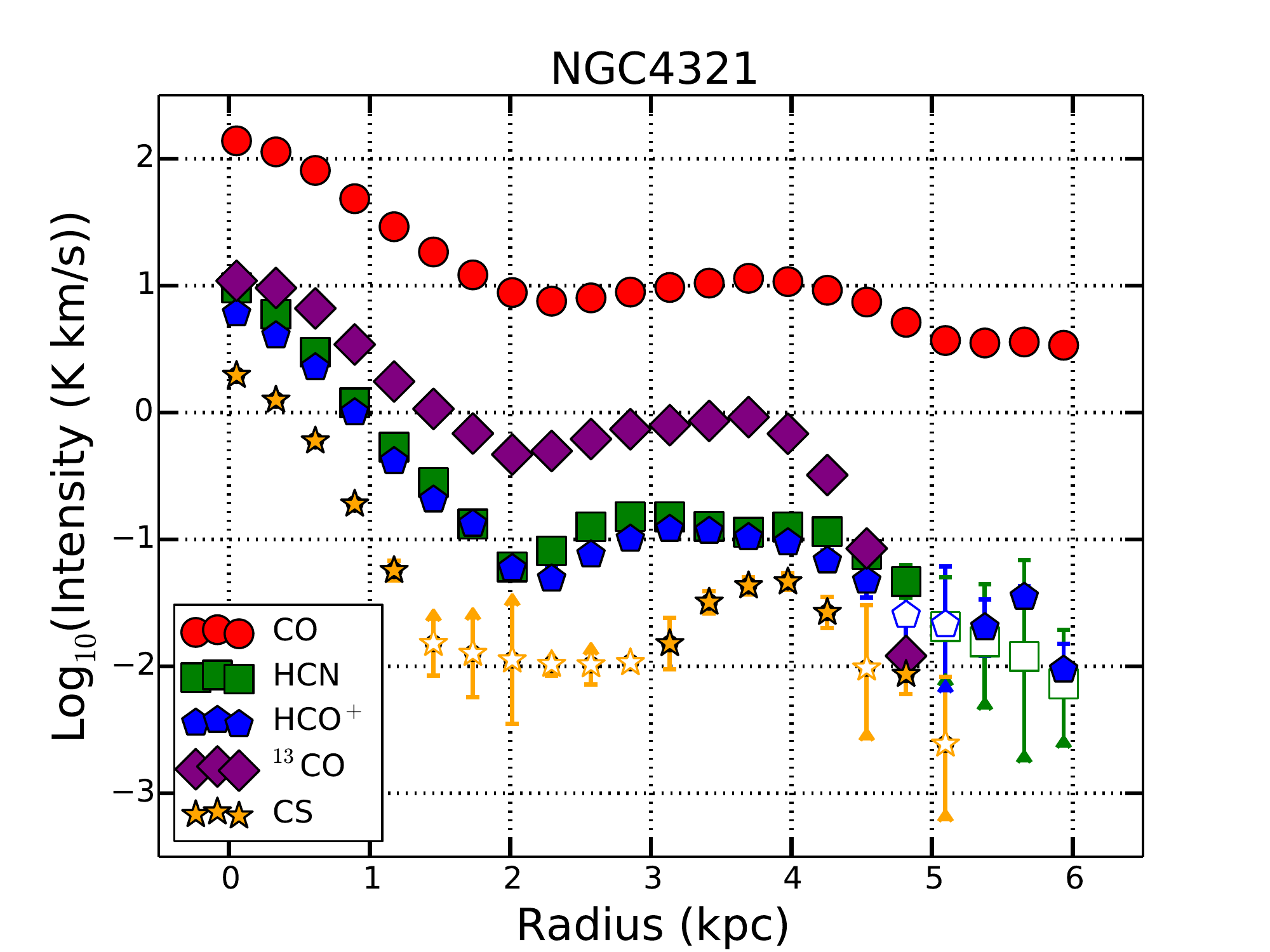}{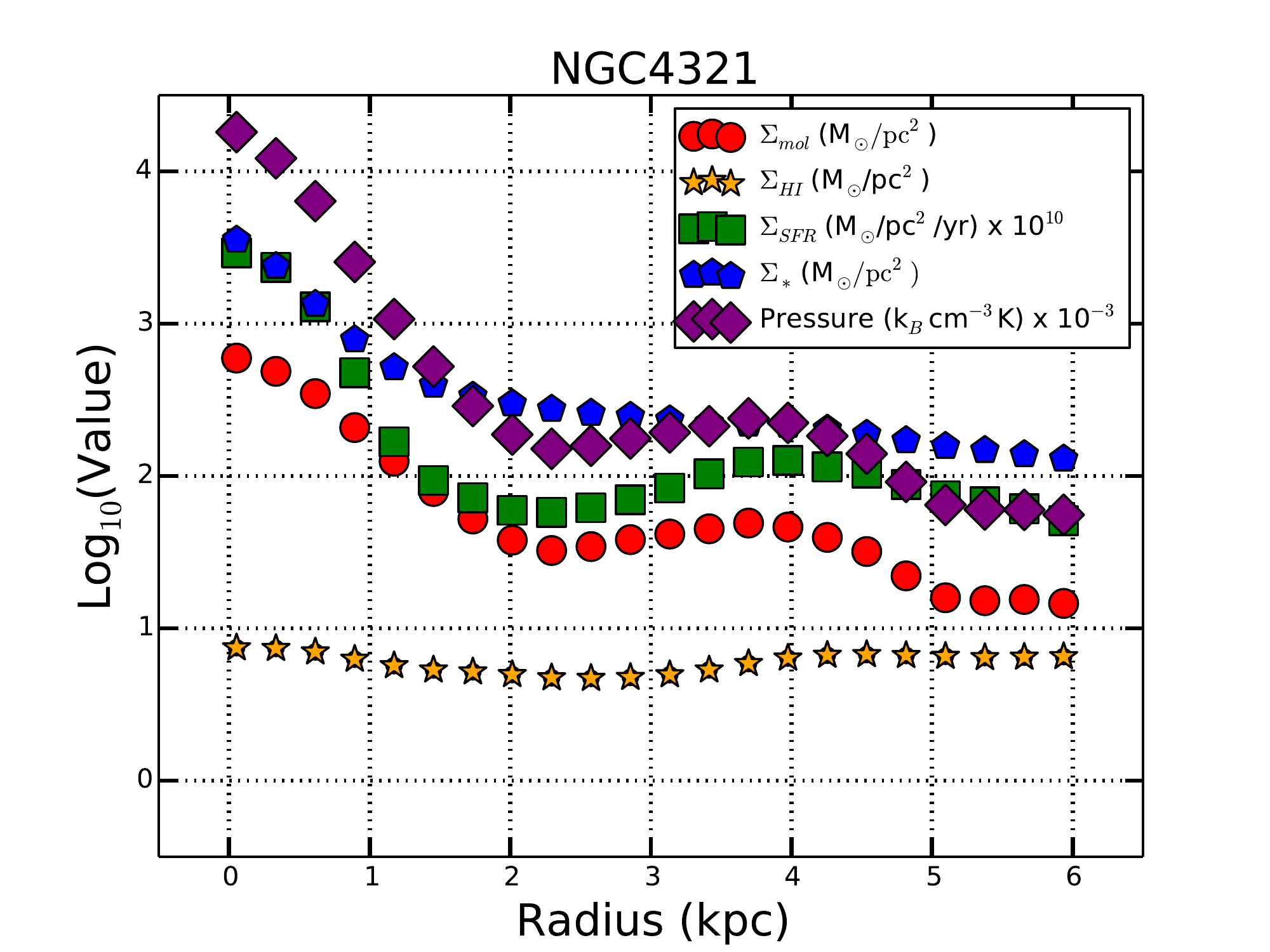}
\plottwo{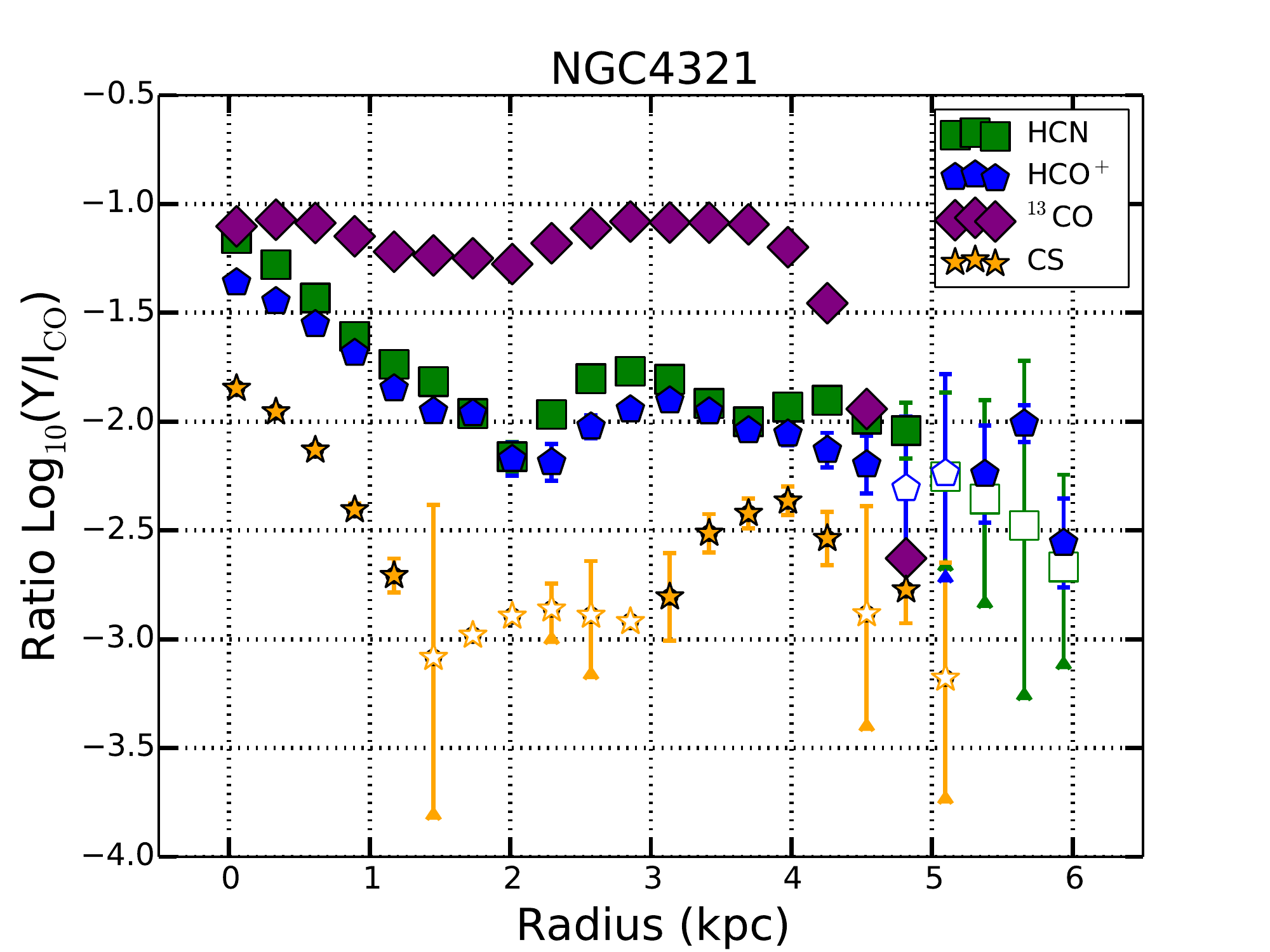}{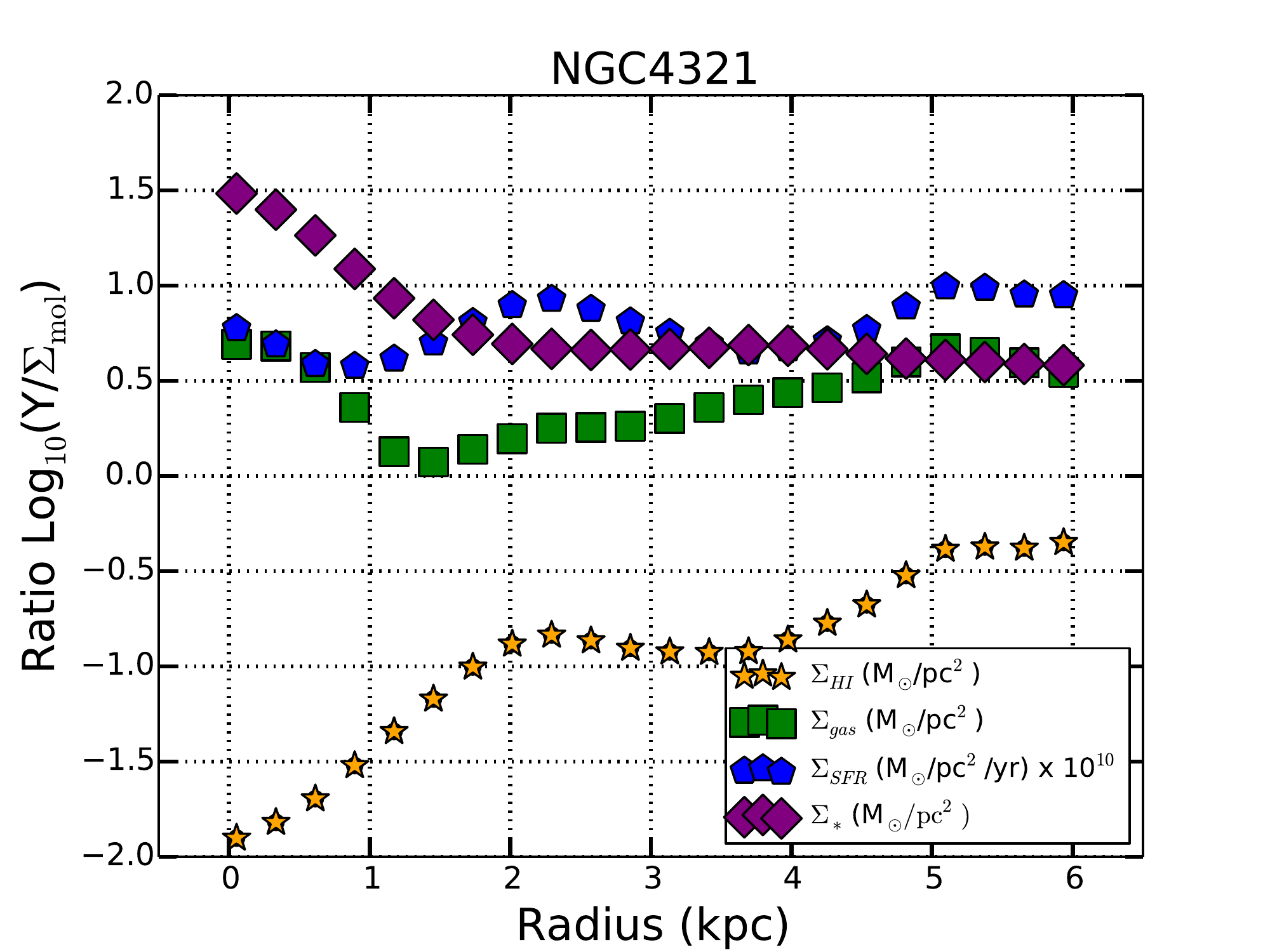}
\caption{As Figure \ref{fig:3351} but for NGC~4321.}
\label{fig:4321}
\end{figure*}

\begin{figure*}
\plottwo{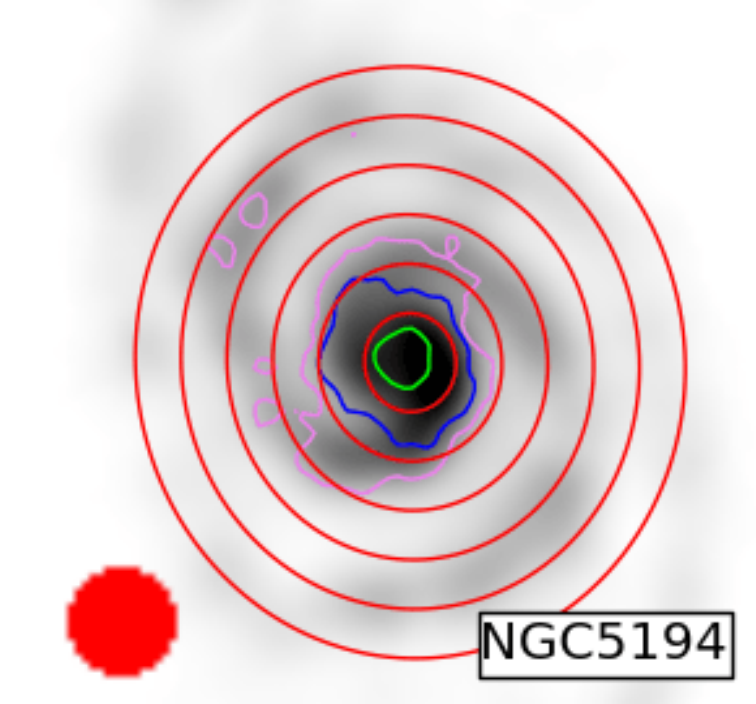}{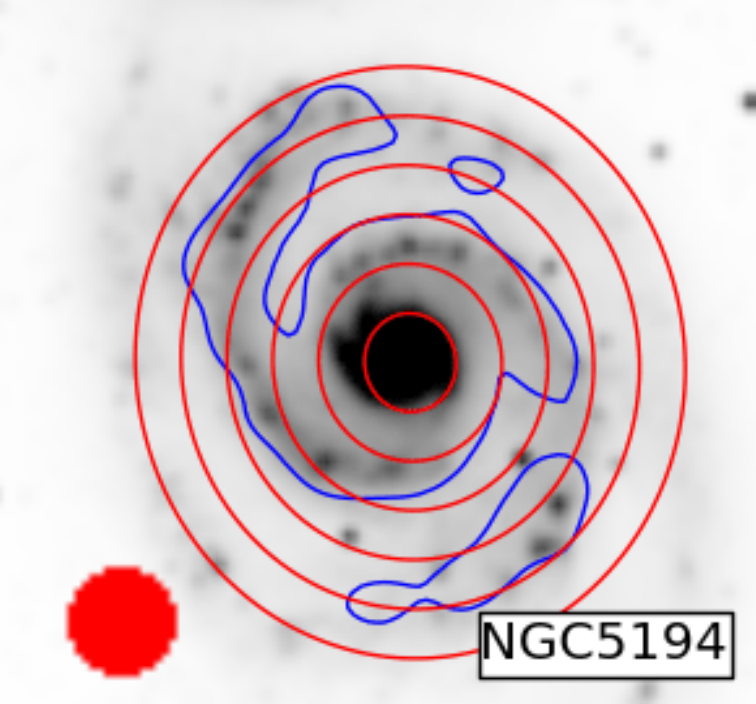}
\plottwo{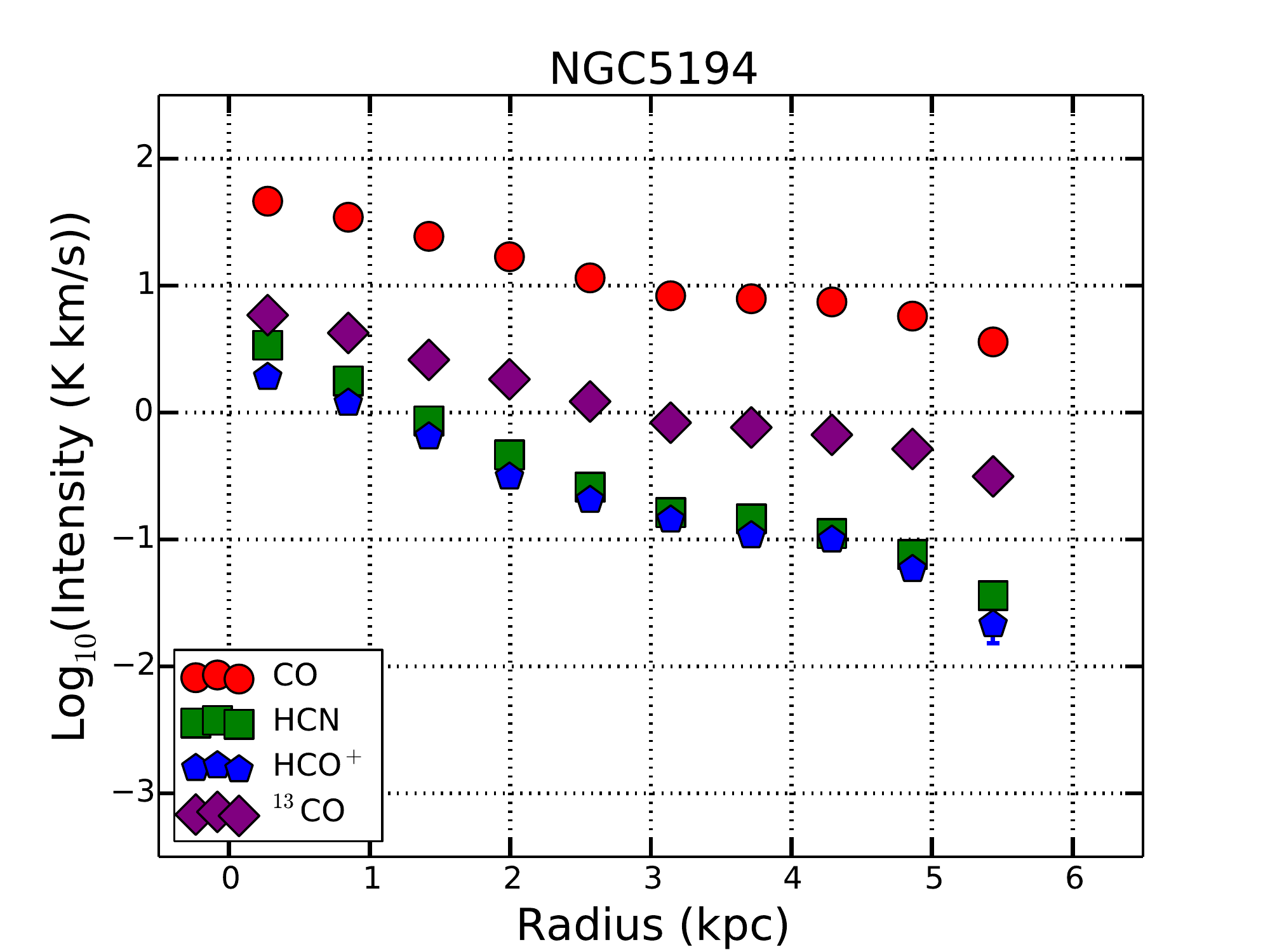}{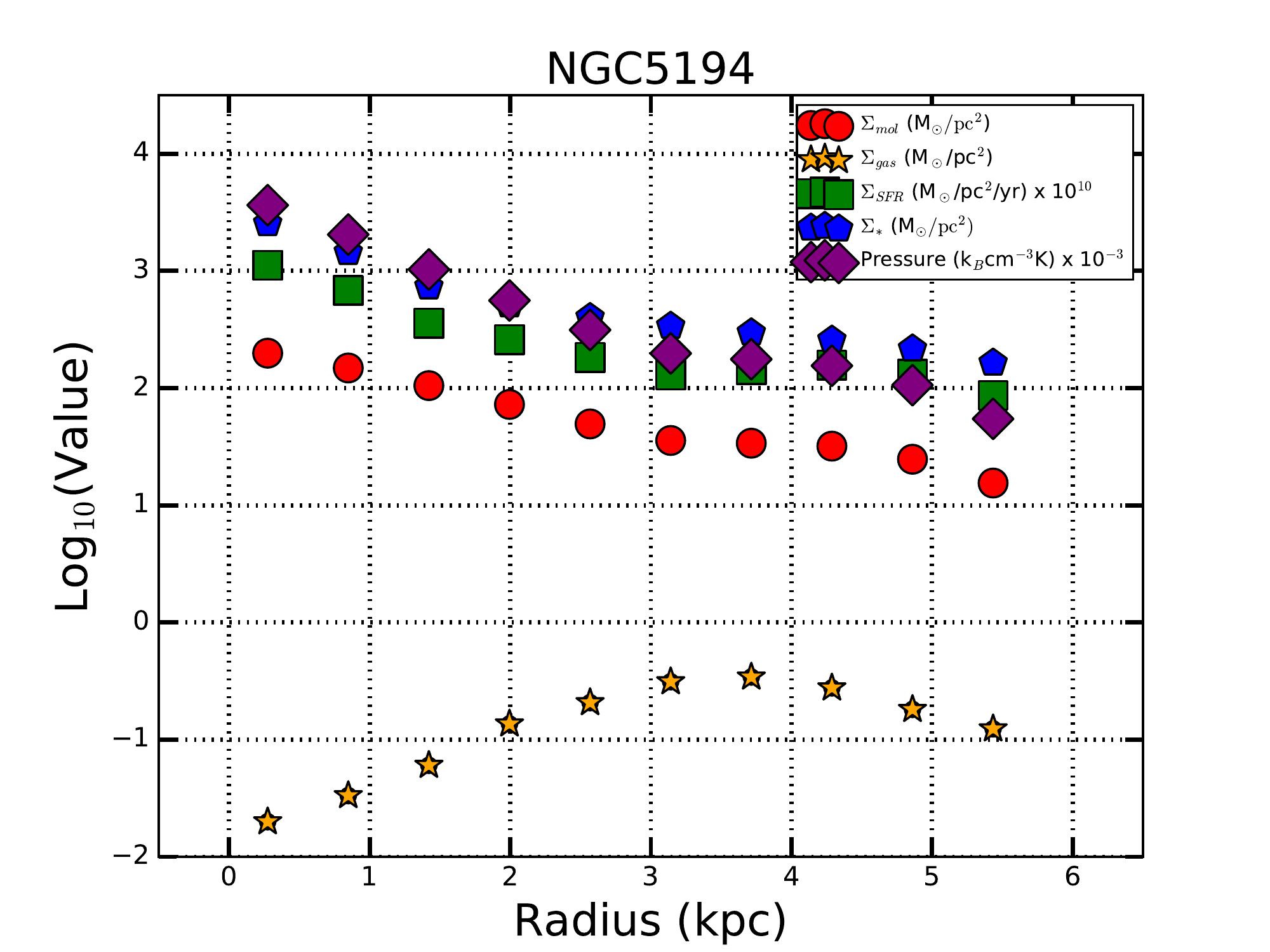}
\plottwo{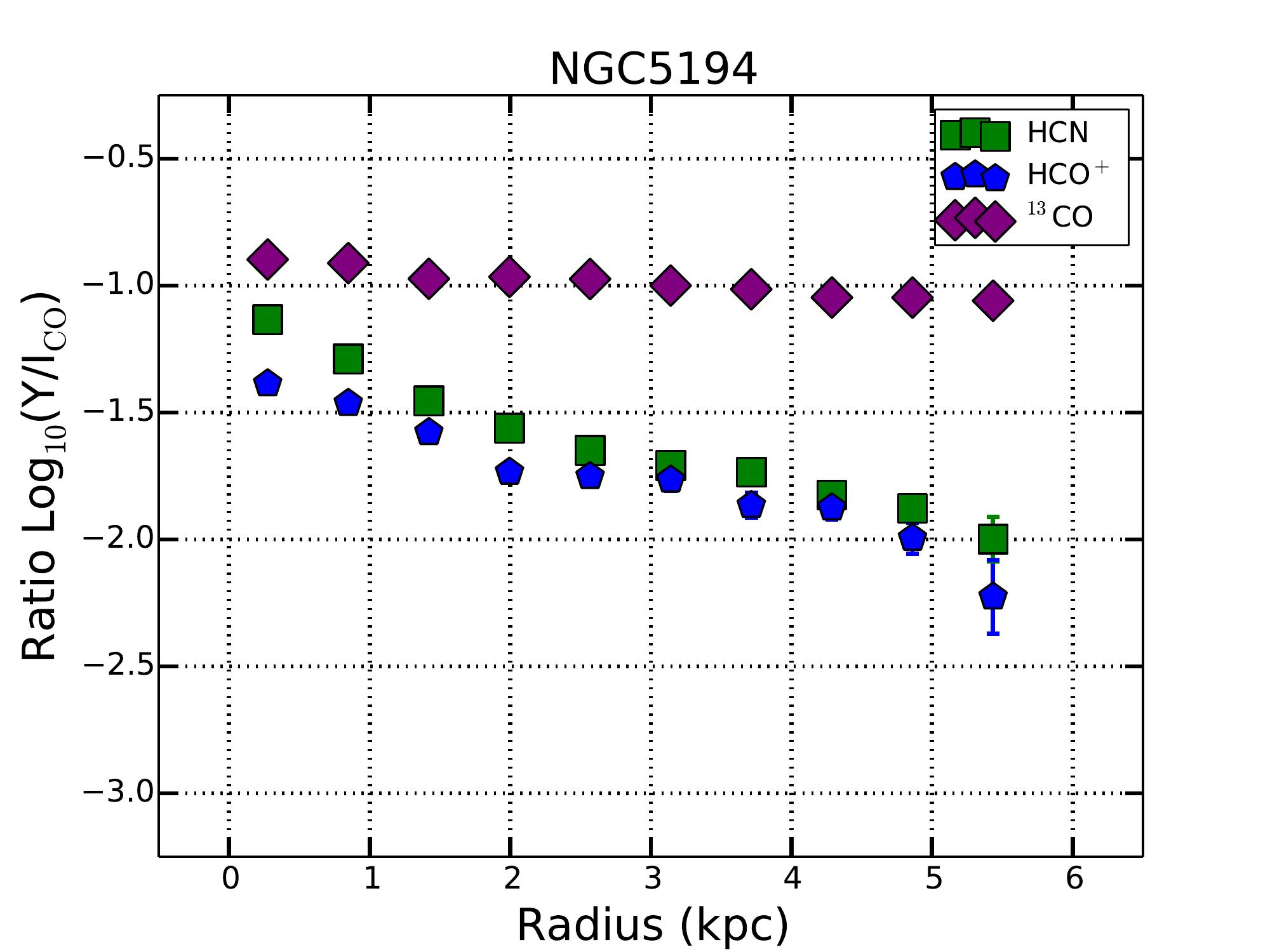}{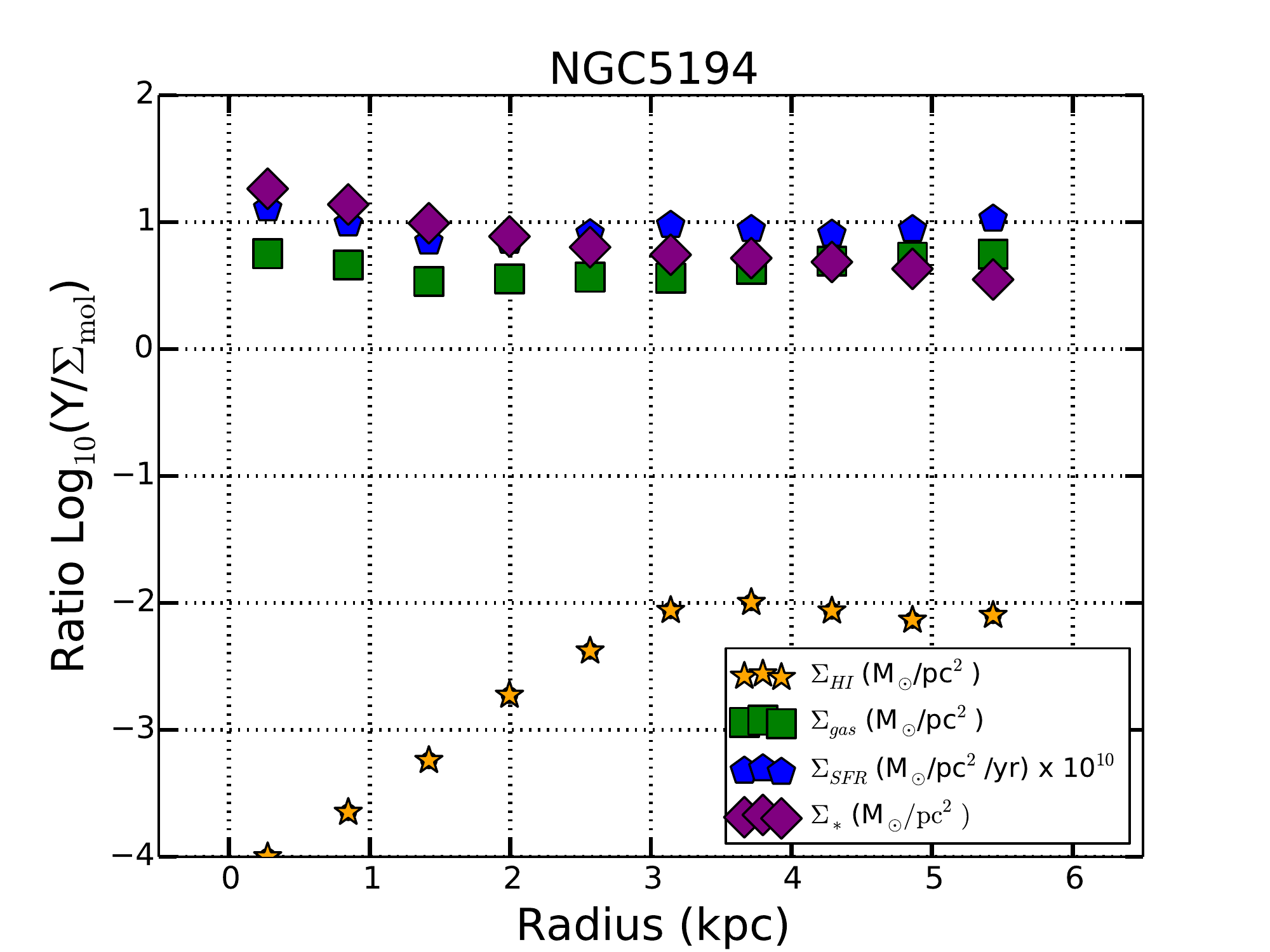}
\caption{As Figure \ref{fig:3351} but for NGC~5194, with data from the IRAM large program EMPIRE \citep{BIGIEL16}. Note that EMPIRE does not cover CS~(2-1). Please note that the y-axis range for the two right panels are different for NGC 5194 than for the other 4 galaxies.}
\label{fig:5194}
\end{figure*}

Figures \ref{fig:3351}-\ref{fig:5194} show CO and HCN emission in context for our four ALMA targets and NGC 5194~(M51). The top left panel shows the distribution of dense gas traced by HCN (colored contours) over the distribution of total molecular gas, traced by CO (grayscale). The top right panel shows the total molecular gas distribution (CO, color contours) and the stellar structure (traced by $3.6\mu$m light in grayscale). In both panels, red ellipses indicate fixed galactocentric radius, allowing one to visually map to the radial profiles below. Our maps extend to $r_{gal} \sim 3$~kpc in NGC 3351 and NGC 3627, to $\sim 5$~kpc in NGC 4254, NGC 4321, and NGC 5194.

We see bright CO emission in the centers of all five galaxies, and fainter CO emission out to many kpc along the bars and spiral arms of all targets except NGC~3351. NGC~3351 shows a bright central disk of molecular gas, but little molecular gas across the stellar bulge. The galaxy has a ring of star formation and molecular material at $\sim 4$~kpc \citep[e.g.,][]{LEROY09}, but this lies near the edge of our mosaic. This outer ring remains undetected in dense gas tracers in our observations and is faint compared to the sensitivity of the BIMA SONG CO~(1-0) map shown here \citep{HELFER03}. Thus NGC~3351 appears as only a compact ($\lesssim 1$~kpc) gas disk in our analysis.

HCN emission follows the CO, with bright HCN intensity ($I_{HCN}$) detected at high significance in the central regions of each galaxy. In the four targets with widespread CO emission, we also detect pervasive HCN emission at the $\sim 2\sigma$ level throughout the bar and spiral arms. In the analysis below, we use the CO emission as a prior to make a clean measurement of $I_{HCN}$. Before doing this, we also inspected the cubes for evidence of HCN and other molecules outside the region of bright CO emission. We do not see evidence for such emission at our sensitivity level. 

These panels illustrate the close connection between molecular gas and stellar structure, which has been shown many times and we see again below \citep[e.g.,][]{YOUNG91,REGAN01,LEROY09}. CO emission is concentrated in the inner regions of the galaxy, where the stellar surface density is high, and follows the arms and bars in the three galaxies that show extended molecular gas.

\subsection{Radial Profiles}
\label{subsec:RPR}

Even with ALMA, emission from dense gas tracers remains hard to detect. To overcome this, we carry out most of our analysis using azimuthally averaged radial profiles (see \S \ref{subsec:RP}).  Note that the masked regions are zeros in this averaging. When we report ratios, these are the ratios among the azimuthal profiles taken after radial profile construction.

The central panels in Figures \ref{fig:3351}-\ref{fig:5194} show these profiles for our target lines ({\em left}) and key measures of the galaxy and ISM structure ({\em right}). Even with azimuthal averaging, we still do not detect the fainter lines in some rings. In these cases we show the $2\sigma$ value as an empty point to indicate an upper limit.

The absolute intensities trace the distribution of material. To capture the changing physical state of the gas, the bottom row of Figures \ref{fig:3351}-\ref{fig:5194} plot ratios among radial profiles. For the lines, we show the intensity of each relative to CO. For the galaxy structure, we show the fractional content of gas (as opposed to stars), the rate of star formation per unit gas, and the balance between atomic and molecular gas. Note that we multiply $\Sigma_{SFR}$ by a factor of 10$^{10}$ to show it on the same scale as the other quantities.

\subsubsection{Galaxy and ISM Structure}

Three of our five galaxies have strong bars \citep{MUNOZMATEOS13} and also show strong nuclear concentrations of gas. NGC 4254, which lacks a strong bar, does not show such a feature. NGC 5194, which also lacks a strong bar, shows a weaker version of the feature. A natural explanation for the difference between the profiles of galaxies with strong bars and those without is that bar-induced streaming motions transport gas inward in the barred galaxies \citep[e.g.,][]{DOWNES96,REGAN99,SHETH05}. This leads to a higher concentration of gas in the inner region, which in turn leads to a higher rate of star formation in the inner $\sim 1$~kpc of these targets. The same mechanism has been used to explain the nuclear region in the Milky Way. 

The radial profiles in Figures \ref{fig:3351}-\ref{fig:5194} show such nuclear enhancements. Within $1{-}2$~kpc, NGC 3351, 3627, 4321, and 5194 all exhibit high $\Sigma_{gas}$, $\Sigma_{mol}$, and $\Sigma_{SFR}$ relative to the surrounding disk. These central gas concentrations are a few hundred pc to a kpc in extent, and result in sharply declining profiles out to $r_{\rm{gal}} \sim$2~kpc. Then they level out, declining more weakly or remaining flat to the edge of our field of view. Recall that these figures show profiles averaging only where CO is detected, so empty space, e.g., between arms is not reflected in the profiles. The stellar surface density, $\Sigma_{*}$, exhibits a similar two-zone behavior, but with a smoother transition between regimes.

The impact of bars is also visible in the images. The top right panels show CO contours over the stellar distribution. Increased gas surface density along an inner bar is visible in NGC 3627 and NGC 4321. In NGC 3351, a strong bar visible in the near-IR links the ring of star formation and molecular gas to the central disk, but our observations detect little gas between the ring and the bar. NGC~5194 shows a comparatively weaker bar than our other targets. Furthermore, the lower resolution for this galaxy makes it hard to compare structural elements between it and the other 4 galaxies. However, the inner parts of the spiral arms may play some of the same role as a bar, funneling material towards the inner galaxy \citep[e.g.,][]{MEIDT13,QUEREJETA16}.

Distinct from our other targets, NGC 4254 shows radial profiles that extend continuously from the central part of the galaxy out into the disk. The central brightness of CO emission is still high in this galaxy, comparable to our other targets. However, the profiles exhibit no strong evidence for a physically discrete nuclear concentration of gas. This galaxy also lacks a strong bar; it is the only galaxy in our sample not identified to have a bar in RC3 or by \citet{MUNOZMATEOS13}. This makes the contrast between NGC 4254 and the other three galaxies a useful point of comparison.

Figures \ref{fig:3351}-\ref{fig:5194} show that emission from the dense gas tracers is also centrally concentrated in our barred targets. Indeed, we will see below that HCN emission appears even more concentrated than CO. Previous work suggests that the gas in these nuclear concentrations also tends to be more excited \citep[e.g., see][]{BRAINE92,LEROY09,LEROY13} and may have a lower CO-to-H$_2$ conversion factor \citep[e.g.,][]{SANDSTROM13} than gas at larger galactocentric radii.

This concentration of gas and star formation resembles a similar feature seen in another strongly barred galaxy: the Milky Way. In the Milky Way, the central molecular zone that covers the inner few hundred pc hosts a large concentration of gas, an overabundance of dense gas, and substantial star formation \citep[though not in a normalized sense, see][]{LONGMORE13}. NGC 3351 may be a particularly striking counterpart, because like the Milky Way, it appears deficient in gas along the bar through the bulge, but rich in gas in the inner region, with a ring of molecular material near the end of the bar \citep[e.g., see review][]{HEYER15}. However, NGC~3351 lacks the bar end starbursts believed to exist in the Milky Way, which do appear, e.g., in NGC~3627 \citep[see also][]{BEUTHER16}.

Radial profiles on the right hand side of Figures \ref{fig:3351}-\ref{fig:5194} show that molecular (and not atomic) gas makes up the overwhelming fraction of the ISM over the fields that we study. We also see that, everywhere we have measurements, stars make up most of the baryonic matter in the disk. The profiles also show variations in the IR-to-CO ratio among and between our targets, tracing the rate at which gas forms stars.

\subsubsection{Line Ratio Profiles}
\label{subsec:LRP}

\capstartfalse
\begin{deluxetable}{lcc}
\tabletypesize{\scriptsize}
\tablecaption{Mean Line Ratios \label{tab:line_ratios}}
\tablewidth{0pt}
\tablehead{
\colhead{Line} & 
\multicolumn{2}{c}{Ratio With CO\tablenotemark{a}} \\
\colhead{} &
\colhead{$r_{gal} \leq 1$~kpc} &
\colhead{$r_{gal} \geq 1$~kpc}
}
\startdata
$^{13}$CO & 0.088 & 0.039\\
HCO$^+$ & 0.026 & 0.008\\
CS\tablenotemark{b} & 0.011 & 0.001 \\
HCN & 0.046 & 0.01 \\
\enddata
\tablenotetext{a}{Mean ratio over our four ALMA targets and NGC 5194. For CS we lack short spacing data and report values only for the four ALMA targets.}
\end{deluxetable}
\capstarttrue

The left central panels show the radial profiles for CO, $^{13}$CO, HCO$^+$, CS, and HCN. Table \ref{tab:line_ratios} reports mean line ratios for our targets for $r_{gal} \leq 1$~kpc and $r_{gal} \geq 1$~kpc. After CO, the order of intensity from brightest to faintest is: $^{13}$CO, HCN, HCO$^+$, and CS. This rank order does not change substantially across the sample, though the relative brightnesses of the lines does vary. The two most common dense gas tracers, HCN and HCO$^+$, have comparable intensities throughout our sample, though their absolute brightness and brightness relative to CO does vary from galaxy to galaxy and with radius (see also \cite{USERO15} and \cite{BIGIEL16}). It is noteworthy that HCN and HCO$^+$ trace each other so closely. We discuss this more in \S \ref{subsec:density}.

This suite of lines traces gas with a range of densities (see Table \ref{tab:line_data}). We utilize the ratios among lines with different $n_{eff}$ to infer changes in the sub-beam density distribution. To first order, the ratios of HCN, HCO$^+$, and CS to CO trace the fraction of dense molecular gas (for more discussion, see \S \ref{subsec:density}). The ratio of $^{13}$CO to CO traces a combination of optical depth, isotopic ratio, and perhaps density of the bulk CO-emitting gas. 

The bottom left panels of Figures \ref{fig:3351}-\ref{fig:5194} show the ratios between the mean intensity of each molecular line and that of CO. Measured line ratios vary within and among our sample, indicating changing physical conditions in the gas.  In the three dense gas ratios, we see similar structure to that of the mean intensity radial profiles. We find inner regions with high ratios of dense gas tracer to CO emission, and this ratio declines as one moves out into the disk.

This decrease in apparent dense gas fraction occurs in all of our targets. In the two galaxies with bars and well-detected disk emission (NGC~3627 and 4321), HCN, HCO$^+$, and CS emission all decrease relative to CO as one moves from the inner region out to the disk. Outside $r_{gal} \sim 1{-2}$kpc, these ratios flatten, remaining approximately constant at values $\sim 2{-}4$ times lower than observed in the inner region of the same galaxy. 

Despite the lack of a distinct inner region in the radial profiles of NGC 4254, $I_{HCN}/I_{CO}$ and $I_{HCO+}/I_{CO}$ also decline with increasing $r_{gal}$. This is consistent with results for NGC 5194 by \cite{BIGIEL16,CHEN15}. Like NGC 4254, NGC 5194 is not strongly barred, but still shows declining density between the inner and outer parts of the galaxy. An anticorrelation between dense gas fraction and galactocentric radius thus appears to be a general feature, not exclusively a product of bar-driven flows. However, when a strong bar is present, its imprint does appear to be visible in the $I_{HCN}/I_{CO}$ profile.

\subsection{Dense Gas and Star Formation}
\label{subsec:DGFASFE}

\begin{figure*}
\plottwo{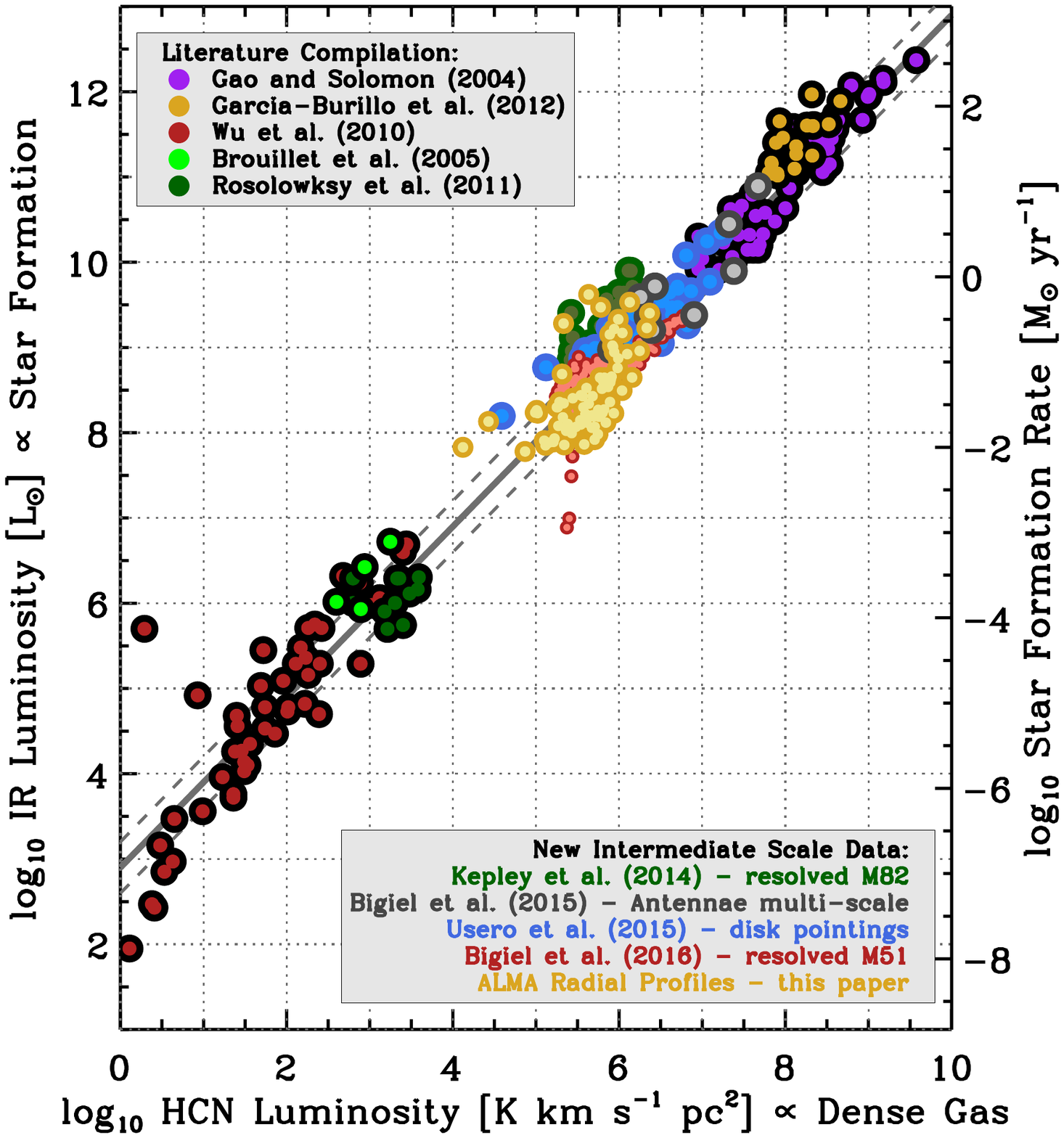}{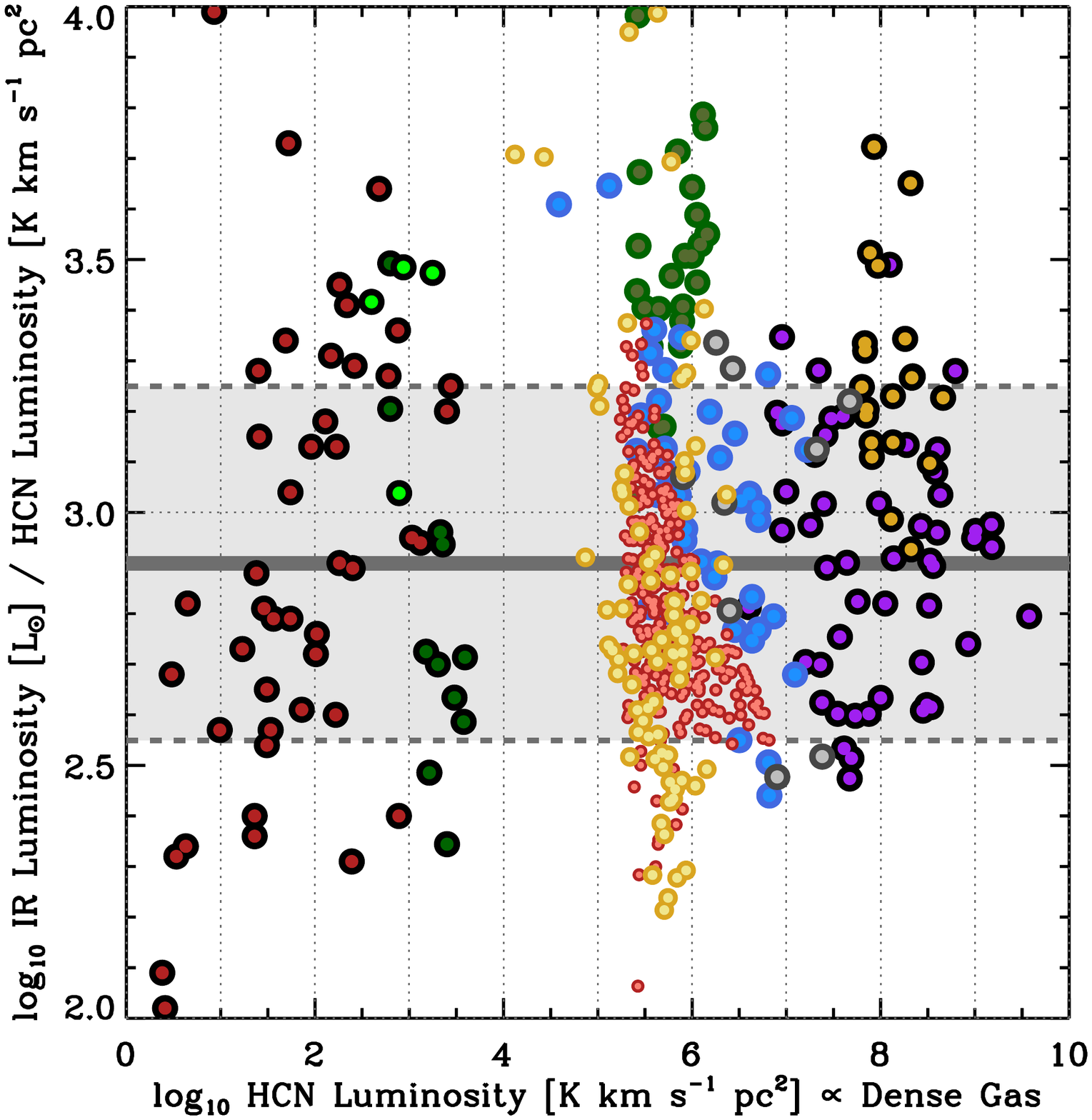}
\caption{IR luminosity, tracing the recent star formation rate, as a function of HCN luminosity, tracing the dense gas mass. Note that to place our rings on to this correlation, we translate our SFRs calculated from H$\alpha$+24$\mu$m to the corresponding TIR luminosity expected from an embedded population with this SFR (using Equation \ref{eq:TIR-to-SFR}). ({\em Left}:) The luminosity-luminosity correlation. ({\em Right}:) The IR-to-HCN ratio as a function of HCN luminosity. The gray lines in both figures show the median IR-to-HCN relation plus or minus $0.35$~dex, the rms scatter across all data. The figure shows our new radial profile data (yellow points) along with a compilation of literature observations. These results span a range of physical scales. {\bf Literature Compilation:} \cite{GAO04} and \cite{GARCIABURILLO12} take luminosities for entire galaxies. \cite{WU10} takes luminosities for individual Milky Way gas clumps.\cite{BROUILLET05} takes luminosities from individual molecular complexes in M31. \cite{ROSOLOWSKY11} takes luminosities from individual GMCs in M33. {\bf New Intermediate Scale Data:} \cite{KEPLEY14}, \cite{BIGIEL15}, \cite{USERO15}, and \cite{BIGIEL16} each target parts of a galaxy from a few hundred pc to a kpc in scale. The $\sim$0.5~kpc-thick rings from this paper falls on the lower end of the extragalactic scale- larger than individual GMCs but smaller than entire galaxies. 
\label{fig:lum_lum}}
\end{figure*}

\subsubsection{Dense Gas-Star Formation Correlation}
\label{sec:dgsf}

We obtained these observations to test the idea that the dense gas fraction determines the star formation efficiency of molecular gas, ${SFE}_{mol} \equiv {SFR}/M_{mol}$. In a simple threshold model, the mass of dense gas determines the star formation rate. Such a relationship would explain the striking correlation between total infrared luminosity (commonly used to trace star formation in the HCN literature) and HCN luminosity (tracing dense gas). This correlation extends from Galactic cores all the way to starburst galaxies (e.g. \citealt{GAO04, GARCIABURILLO12, WU10, BROUILLET05, ROSOLOWSKY11, KEPLEY14, BIGIEL15, USERO15, BIGIEL16}, Figure \ref{fig:lum_lum}). 

To place our rings on to this correlation, we translate our SFRs calculated from H$\alpha$+24$\mu$m to the corresponding TIR luminosity expected from an embedded population with this SFR (using Equation \ref{eq:TIR-to-SFR}). Though indirect, this approach is at least internally consistent with the fiducial SFR estimates in the main text. We use this conversion ONLY for Figure \ref{fig:lum_lum}. An alternative approach, using our best estimates of the local TIR luminosity based on 70$
\mu$m emission (Section \ref{sec:sfr_tir}) yields similar results, and can be carried out with the data provided in the table or profiles.

Figure \ref{fig:lum_lum} shows that our rings do populate the HCN-IR luminosity correlation seen at smaller and larger scales \citep{GAO04,LADA12}. Our new data occupy the intermediate regime between whole galaxies and Milky Way cores or individual GMCs. In this, they resemble results for the disk pointings of \cite{USERO15}, the $\sim30$\arcsec NGC 5194 data of \citealt{BIGIEL16} and \citealt{CHEN15}, and recent studies of M82 and the Antennae galaxies (\citealt{KEPLEY14,BIGIEL15}). Our spatial resolution is finer than that of \cite{BIGIEL16}, but our use of rings raises the luminosity in any individual data point.

Taking all of the literature data together, Figure \ref{fig:lum_lum} shows that that SFR correlates with HCN luminosity, with about a factor of two scatter across almost ten orders of magnitude. Even if the luminosity of a specific ring in a galaxy lacks physical meaning, the narrow spread in the TIR-to-HCN ratio is significant. The right panel shows that over a large range, most data in the literature exhibit TIR-to-HCN ratios within $\sim \pm 0.3$~dex of one another. Although systematic differences are already evident both within and among studies, there is clearly a strong relationship between HCN luminosity and recent star formation.

\subsubsection{Does the Dense Gas Fraction Predict the Star Formation Efficiency of Molecular Gas?}

\capstartfalse
\begin{deluxetable}{lcc}
\tabletypesize{\scriptsize}
\tablecaption{$\Sigma_{SFR}/\Sigma_{mol}$ vs. $\Sigma_{dense}/\Sigma_{mol}$ \label{tab:sfe_dgf}}
\tablewidth{0pt}
\tablehead{
\colhead{Data Set} & 
\colhead{Rank Corr.} &
\colhead{$\log_{10}$~$\Sigma_{SFR}/\Sigma_{dense}$} \\
}
\startdata
Radial Profiles & & \\
$\ldots$NGC~3351 & 0.77 (0.052) & -7.94 ($\pm$0.19) \\
$\ldots$NGC~3627 & -0.01 (0.488) & -7.73 ($\pm$0.34)\\
$\ldots$NGC~4254 & 0.10 (0.342) & -7.89 ($\pm$0.26)\\
$\ldots$NGC~4321 & 0.21 (0.202) &  -8.26 ($\pm$0.23)\\
$\ldots$NGC~5194\tablenotemark{a} & -0.26 (0.236) &  -8.12 ($\pm$0.28) \\
$\ldots$All profiles & 0.13 (0.003) & -7.89 ($\pm$0.33)\\
\citet{USERO15} & 0.42 (0.005)& -7.95 ($\pm$0.25) \\
\citet{GAO04} & 0.65 (0.000)& -9.03 ($\pm$0.24) \\
\citet{GARCIABURILLO12} & 0.35 (0.075) &  -7.84 ($\pm$0.20) \\
\hline
\\
All data & 0.30 (0.000) & -7.96 ($\pm$0.31) \\
\enddata
\tablenotetext{a}{\citet{BIGIEL16}.}
\tablecomments{Rank correlation quotes with $p$ value in parenthesis. $\log_{10}\Sigma_{SFR}/\Sigma_{dense}$ in units of $\log_{10} {\rm yr}^{-1}$. We quote the mean of the logarithm of the ratio and the $\pm1\sigma$ scatter in the log of the ratio. ``All data'' treats all data points with equal weight, regardless of the spatial scale sampled.}
\end{deluxetable}
\capstarttrue

\begin{figure*}
\plotone{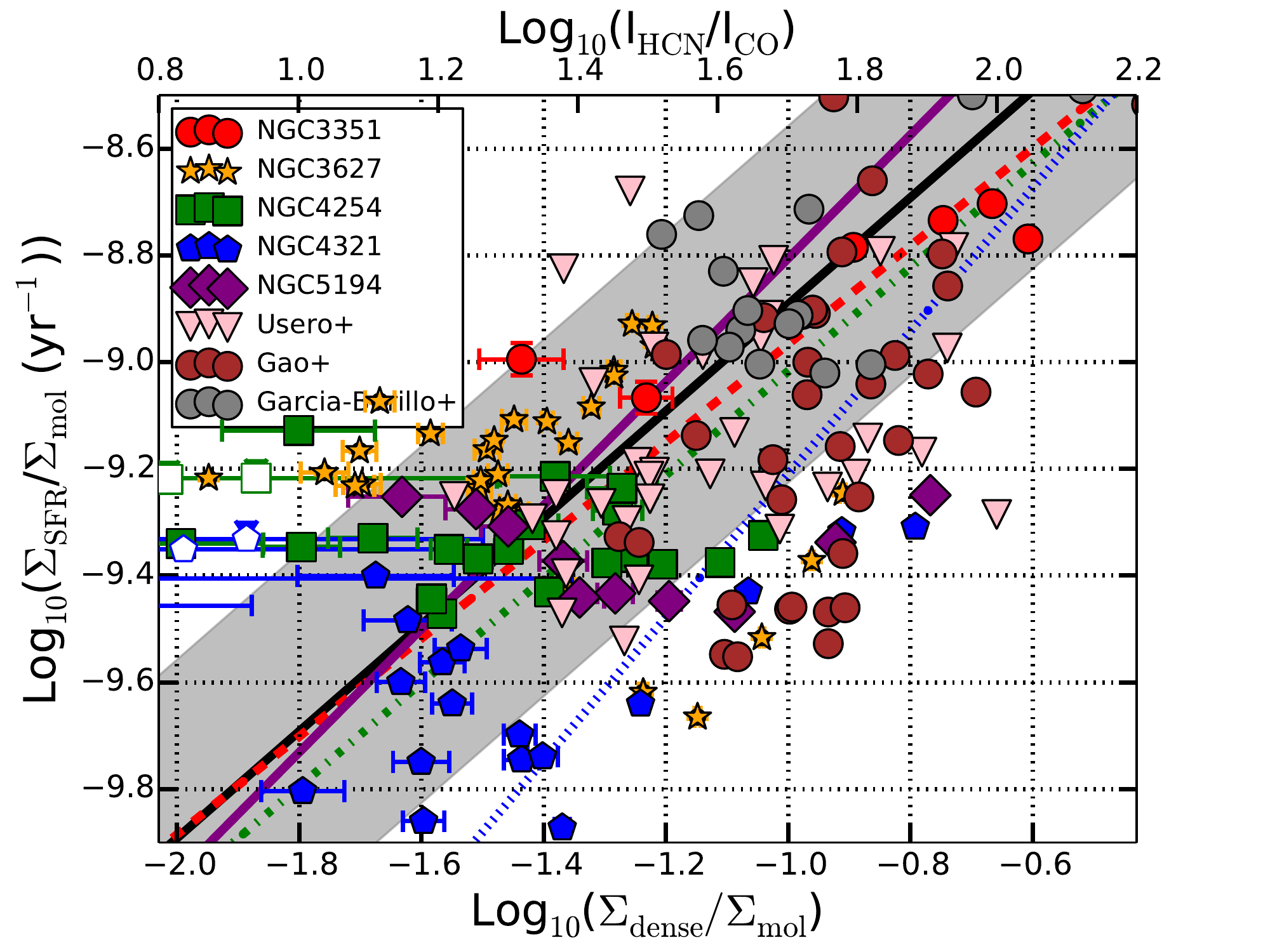}
\caption{$\Sigma_{SFR}/
\Sigma_{mol}$, or the normalized rate at which molecular gas forms stars, as a function of $\Sigma_{dense}/\Sigma_{mol}$, or the dense gas fraction as traced by $I_{HCN}/I_{CO}$. Higher dense gas fractions do correspond to more efficient star formation in the total molecular gas, but with large scatter, which we show below to be physical in nature. Line show show the median, best fit, and scatter for our survey along with the measurements of \citet{GAO04}, \citet{GARCIABURILLO12}, and \citet{USERO15}. Statistically, the HCN/CO ratio is a relatively poor predictor of $\Sigma_{\rm SFR}$/CO when all data are considered together, though it performs better in some individual galaxies or subsamples. \label{fig:dgf_sfe}}
\end{figure*}

One popular approach to thinking about star formation, following, e.g., \citet{GAO04} and \citet{LADA12}, is to imagine a gas density threshold, above which the star formation rate scales with the mass of dense gas. In such a model, the star formation per unit mass of {\em total} molecular gas can and does vary, but star formation will correlate tightly with the mass of {\em dense} gas traced, e.g., by HCN. In such a model, any variation in $SFR_{\rm mol} \equiv \Sigma_{SFR}/\Sigma_{mol}$, should be due to variations in $f_{dense}$ (traced here with $I_{HCN}/I_{CO}$).

We explore this prediction in Figure \ref{fig:dgf_sfe}. We show $\Sigma_{SFR}/\Sigma_{mol}$, or the overall efficiency of star formation, as a function of $\Sigma_{dense}/\Sigma_{mol}$, or $f_{dense}$. We plot results for our ALMA targets, NGC 5194, the \citet{USERO15} pointings, and the \citet{GAO04} survey. 

These data all show a similar picture: a correlation exists between $\Sigma_{SFR}/\Sigma_{mol}$ and $\Sigma_{dense}/\Sigma_{mol}$ with the expected sense, that a higher dense gas fraction does imply a higher star formation efficiency. However, the scatter about this relation is large compared to the dynamic range in $\Sigma_{dense}/\Sigma_{mol}$ or $\Sigma_{SFR}/\Sigma_{mol}$.

The weak correlation does not appear to be an artifact of limited dynamic range. We picked our targets to span a large range of $\Sigma_{SFR}/\Sigma_{mol}$, with the goal of testing the dense gas threshold hypothesis. Indeed, Figure \ref{fig:dgf_sfe} shows that our galaxies exhibit an order of magnitude spread in $\Sigma_{dense}/\Sigma_{mol}$. This is a large fraction of the total variation seen in the local universe \citep[e.g.,][]{SAINTONGE11,LEROY13,LEROY15}. We also see roughly an order of magnitude variation in $\Sigma_{dense}/\Sigma_{mol}$ across our sample. If a sharp correlation between $\Sigma_{SFR}/\Sigma_{mol}$ and $\Sigma_{dense}/\Sigma_{mol}$ was present, we would expect to see it in Figure \ref{fig:dgf_sfe}.

The black line in the left panel of Figure \ref{fig:dgf_sfe} shows the expectation for a fixed star formation rate per unit dense gas. We set the slope to the median $\Sigma_{dense}/\Sigma_{mol}$ value of our sample and illustrate the $\pm 1\sigma$ range as a shaded gray region. This median ratio, $\Sigma_{SFR}/\Sigma_{dense} = 1.10\times10^{-8}$ with $0.25$ dex scatter, is significantly higher than that found by \citet{GAO04}, $9.33\times10^{-10}$, but comparable to those found by \cite{USERO15} and \cite{GARCIABURILLO12}, $1.12\times10^{-8}$ and $1.45\times10^{-8}$ respectively. Recall that each of these works calculate SFR differently and that the \cite{GAO04} and \cite{GARCIABURILLO12} samples focus on starburst systems whereas our sample consists of more normal star forming galaxies.

The red dashed line in Figure \ref{fig:dgf_sfe} shows an ordinary least squares bisector fit to our data. The best fit power law, weighting all significant detections equally, is

\begin{eqnarray}
\label{eq:fit_tirco_hcnco}
\log_{10} \frac{\Sigma_{SFR}}{\Sigma_{mol}} \approx-8.05 + 0.92 \log_{10} \frac{\Sigma_{dense}}{\Sigma_{mol}}.
\end{eqnarray}

\noindent This relation should be taken as indicative of the scaling between $\Sigma_{SFR}/\Sigma_{mol}$ and $\Sigma_{dense}/\Sigma_{mol}$ in our data. As is visible from the plot, the choice of which data to fit (starbursts, whole galaxies, individual regions, etc.), the weighting applied to different data (e.g., by area, luminosity, or galaxy), and the model used (e.g., including astrophysical scatter or not) all have the potential to alter the fit substantially. This would not be true if $\Sigma_{dense}/\Sigma_{mol}$ perfectly predicted $\Sigma_{SFR}/\Sigma_{mol}$ in a universal way across all systems.

The green, blue, and purple lines show similar fits applied to only the \cite{USERO15}, \cite{GAO04}, and \cite{GARCIABURILLO12} data. Considering only the \cite{USERO15} and \cite{GARCIABURILLO12} data, we fit slightly steeper slopes than for our data, but find a consistent normalization near $\log_{10} f_{\rm dense} \sim -1.5$. The fit to the \cite{GAO04} fit has a steeper slope and lower intercept than our data, consistent with the offset in measured $\Sigma_{\rm SFR}/\Sigma_{\rm dense}$ mentioned above. The fits for both \cite{USERO15} and \cite{GARCIABURILLO12} are consistent with the fit to our data within the uncertainties.

Thus, our best fit slope of $\sim 1$ agrees with the schematic picture that  $\Sigma_{dense}/\Sigma_{mol}$ as traced by $\Sigma_{HCN}/\Sigma_{CO}$ should predict $\Sigma_{SFR}/\Sigma_{mol}$ in a simple way. However, our results also show that  $\Sigma_{dense}/\Sigma_{mol}$ is not a highly accurate predictor of $\Sigma_{SFR}/\Sigma_{mol}$. Across our whole sample, the rank correlation between $\Sigma_{SFR}/\Sigma_{mol}$ and $\Sigma_{dense}/\Sigma_{mol}$ is only $0.13$. Meanwhile, the overall scatter in $\Sigma_{SFR}/\Sigma_{mol}$ is 0.33 dex.

Note that individual galaxies do show tighter or looser individual trends in Figure \ref{fig:dgf_sfe}, and we report the rank correlation coefficients and logarithmic fits for all of the scaling relations considered in this work in Tables \ref{tab:sfe_dgf}, \ref{tab:sfe_corr}, \ref{tab:dgf_corr}, and \ref{tab:pressure_corr}. $I_{HCN}/I_{CO}$ can be a good predictor of $\Sigma_{SFR}/\Sigma_{mol}$ for some galaxies. But overall, $I_{HCN}/I_{CO}$ remains only a moderately good predictor of $\Sigma_{SFR}/\Sigma_{mol}$. The relationship appears particularly poor for our radial profiles, which emphasize the contrast between nuclear regions and the surrounding disks.

\cite{MURPHY15} find similar results for NGC 3627. They find that the $SFE_{mol}$ decreases as a function of $f_{dense}$ for the nucleus and two extranuclear star forming regions in the disk of NGC 3627. Furthermore, they find that the velocity dispersion of the dense gas decreases with increasing $SFE_{mol}$. They conclude that the dynamical state of the dense gas, not its abundance, is what sets the star formation rate.

Certainly, stars form in the densest parts of local molecular clouds. Why is $I_{HCN}/I_{CO}$ only a weak predictor of the star formation efficiency of molecular gas? In the next two sections, we consider whether $I_{HCN}/I_{CO}$ indeed traces the dense gas mass fraction (or at least indicates the mean ISM density) and then present evidence for a context-dependent role for dense gas in star formation \citep[building on][]{GARCIABURILLO12,USERO15,BIGIEL16}.

\subsection{Is HCN-to-CO Tracing Density?}
\label{subsec:density}

\begin{figure*}
\plottwo{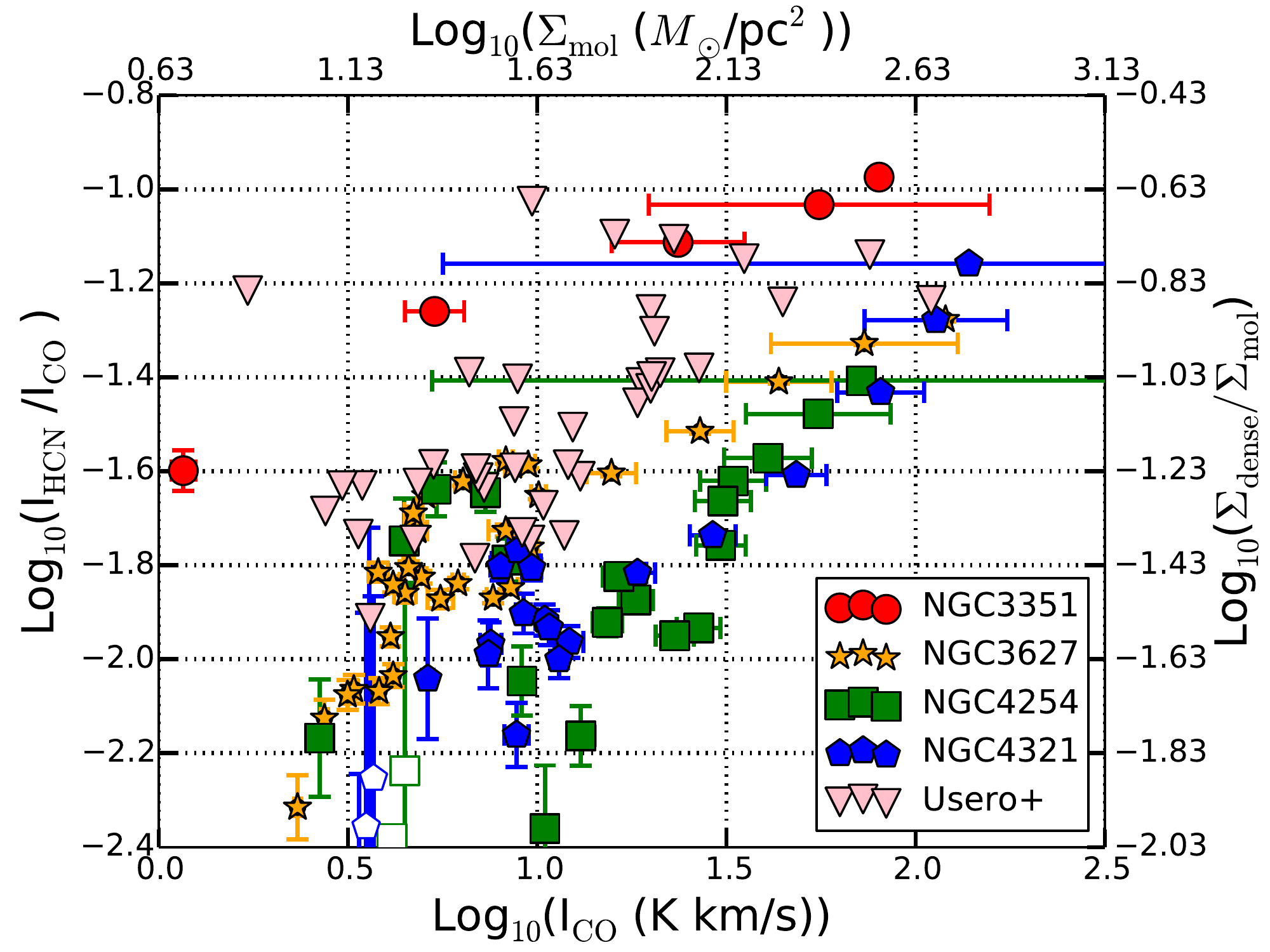}{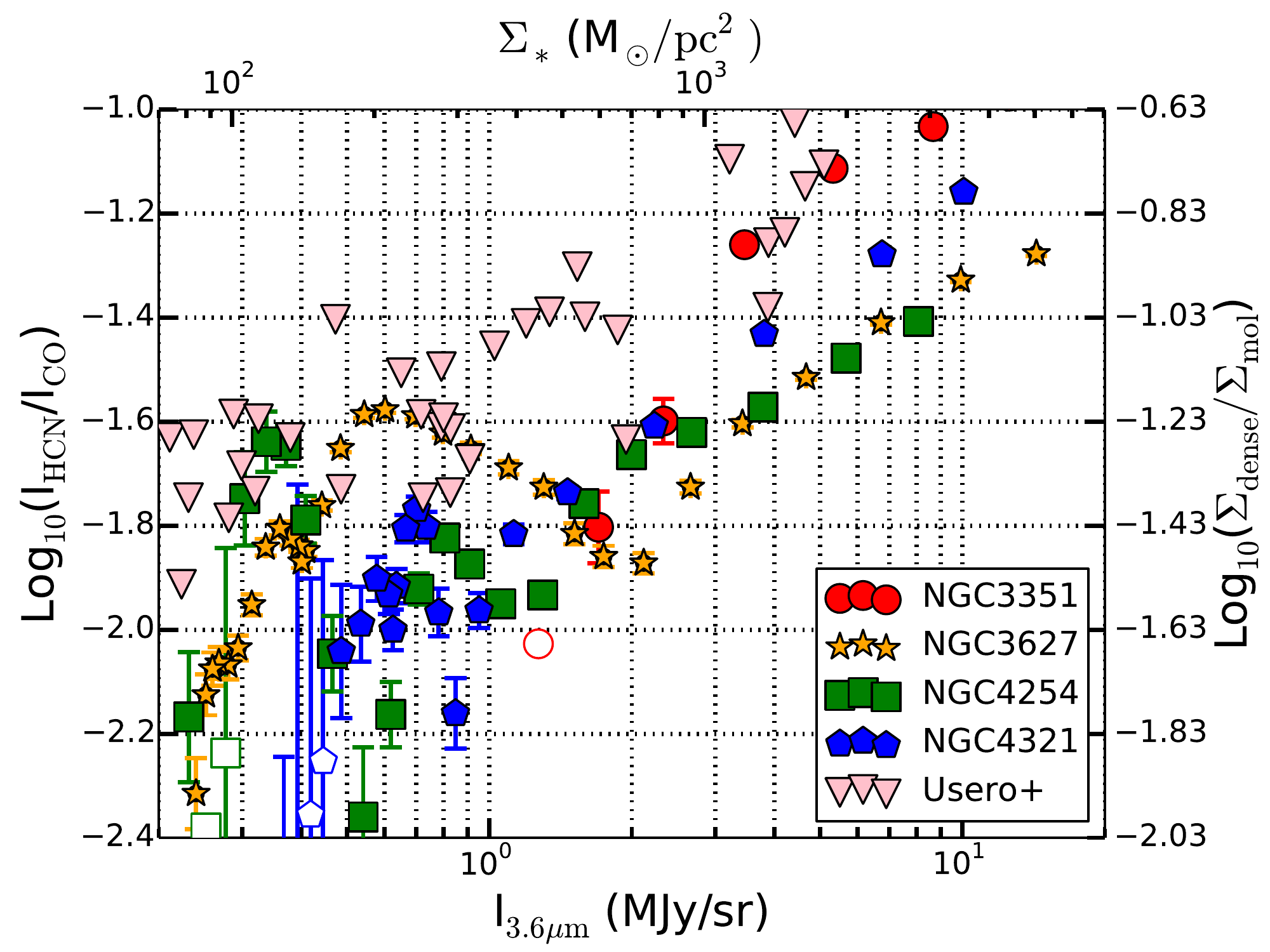}
\caption{HCN/CO, tracing the dense gas fraction, as a function observable cognates of ({\em left}) molecular gas and ({\em right}) stellar surface density. The left panel shows HCN/CO as a function of mean molecular gas surface density, estimated from $I_{\rm CO}$ using a fixed conversion factor. The right panel shows HCN/CO as a function of stellar surface density, estimated from contaminant-corrected $3.6\mu$m maps. In both cases, a higher overall surface density corresponds to a larger apparent dense gas fraction. The correlation with stellar surface density is particularly striking \citep[as in][]{USERO15}.
\label{fig:co_dgf}}
\end{figure*}

\begin{figure*}
\plottwo{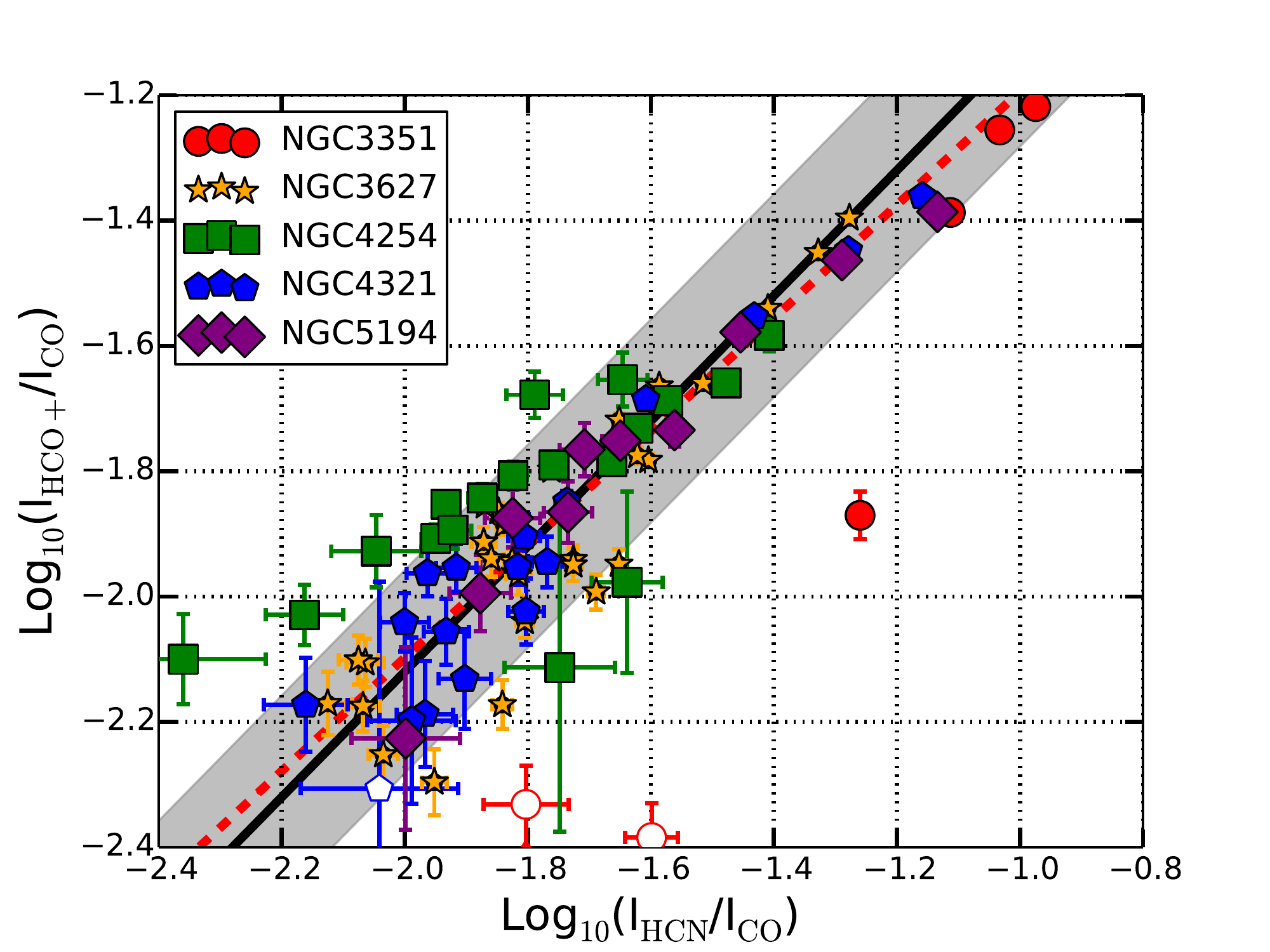}{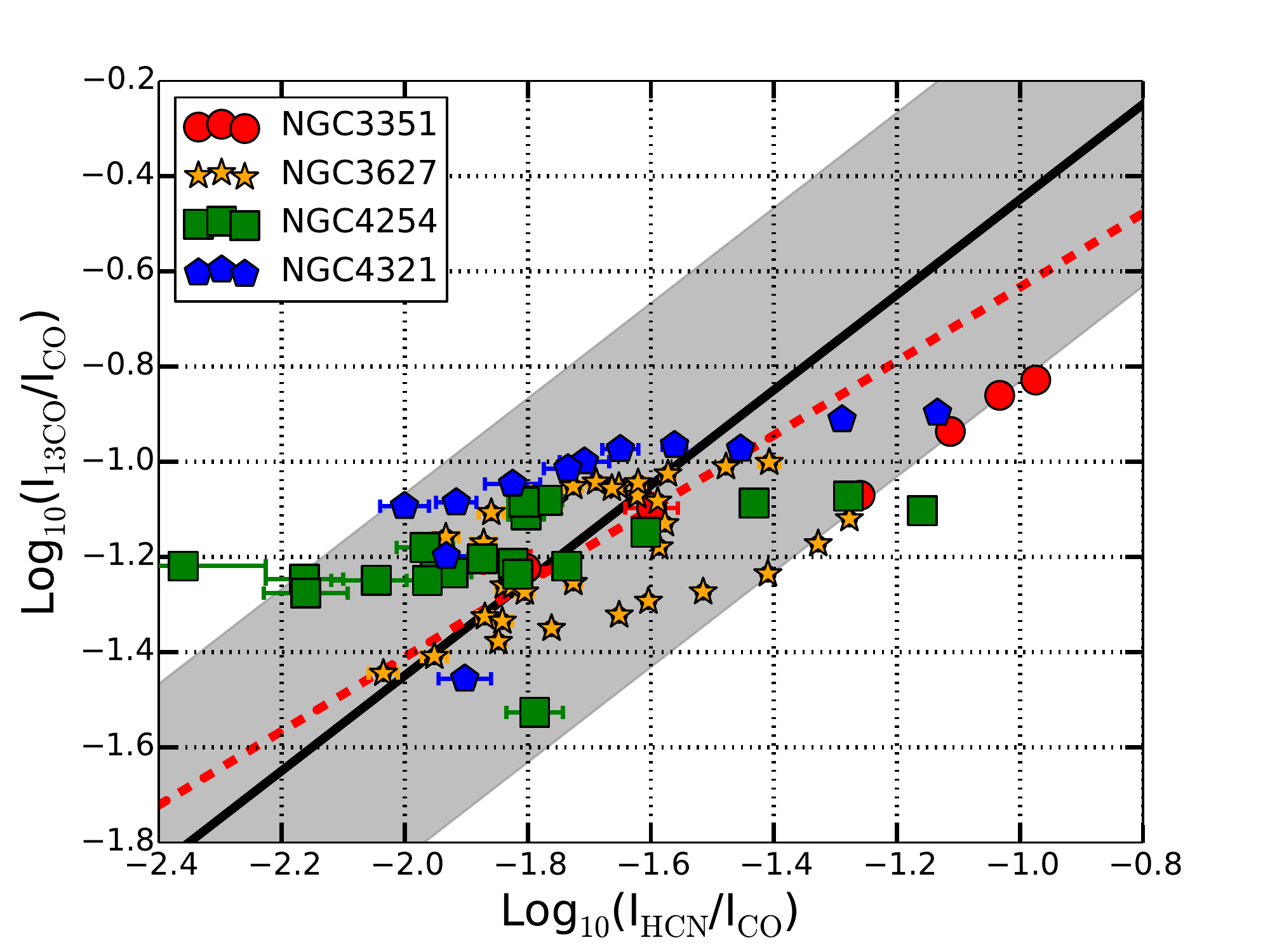}
\caption{Line ratios for our radial profiles. ({\em Left:}) $I_{HCO+}/I_{CO}$ as a function of $I_{HCN}/I_{CO}$. The black line shows the median HCO$^+$/HCN ratio and its $\pm 1\sigma$ range in our data. The gray shaded region illustrates the $\pm 1\sigma$ range found by \cite{PRIVON15} for luminous infrared galaxies. The two line ratios track one another, reinforcing that our results are not specific to HCN. There is some indication of a slightly sublinear slope to the $I_{HCO+}/I_{CO}$ vs $I_{HCN}/I_{CO}$ ralation, which would be expected for a lognormal distribution give the lower critical density of HCO$^+$. ({\em Right:}) $^{13}$CO/CO as a function of $I_{HCN}/I_{CO}$. The black line shows the median $^{13}$CO/HCN in our sample. The sublinear slope can be expected if the effective density for $^{13}$CO emission lies below the median density by mass while HCN traces comparatively dense gas.
\label{fig:hcn_hcop_13co}}
\end{figure*}

Does the surprising lack of correlation between $\Sigma_{TIR}/I_{CO}$ and $I_{HCN}/I_{CO}$ reflect a problem with our interpretation of $I_{HCN}/I_{CO}$? We take this ratio to trace the fraction of dense gas, $f_{dense}$. In this picture, CO traces the bulk of the molecular gas, while HCN traces material with $n_{H2} \gtrsim 10^{4}$cm$^{-3}$ \citep[e.g.,][]{GAO04,KRUMHOLZ07}. This is the standard interpretation for this ratio, because the most effective density for HCN emission, $n_{eff}$, is much higher than both the $n_{eff}$ for CO and the mean density of a molecular cloud in the Solar Neighborhood. Meanwhile, CO traces the overall molecular gas mass, though not without caveats \citep[see][]{BOLATTO13}.

Reality may be more complicated. We expect a wide distribution of densities in each beam, and the shape of that distribution can affect our interpretation of the line ratio. Variations in the optical depth, temperature, and chemical abundance of one or both species can also produce line ratio variations that mimic those expected from changing density. Although we lack a ``smoking gun,'' our observations provide several indirect ways to test whether $I_{HCN}/I_{CO}$ indeed reflects the density distribution in our data.

First, in the left panel of Figure \ref{fig:co_dgf} we compare $I_{HCN}/I_{CO}$ to the average surface density on much larger scales, traced by the average surface brightness of CO in a ring. This large-scale surface density convolves the distribution of clouds and the internal density of each cloud, but still provides a coarse measure of how dense the molecular ISM is in a given region of our galaxy.  The figure shows that $I_{HCN}/I_{CO}$ indeed correlates with $I_{CO}$, suggesting that high dense gas fractions occur in regions of high mean density. 

This correlation appears stronger within individual galaxies than taking all of our targets together (see Table \ref{tab:dgf_corr}). Conversion factor variations \citep{SANDSTROM13} and differences in the sub-beam ISM structure \citep[e.g.,][]{LEROY16} likely contribute to these galaxy-to-galaxy variations. In that sense, our $8\arcsec \sim 500$~pc resolution still limits this test. A stronger version of this test will be to correlate $I_{HCN}/I_{CO}$ with cloud-scale (10s of pc) surface and volume density estimates based on high resolution CO imaging. We will present such a comparison in M. Gallagher et al. (in preparation).

Our line suite allows us to construct several ratios that constrain the shape of the density distribution and the potential impact of abundance variations. We show these in Figure \ref{fig:hcn_hcop_13co}. There we plot $I_{HCO+}/I_{CO}$ and $I_{13CO}/I_{CO}$, each as a function of $I_{HCN}/I_{CO}$. These ratios offer a check that other dense gas tracers show the same results as HCN and offer some constraint on the sub-beam density distribution.

What do we see? $I_{HCN}/I_{CO}$ and $I_{HCO+}/I_{CO}$ correlate throughout our data, with a median ratio of $I_{HCO+}/I_{HCN}= 0.76$ and a scatter of 0.24. The solid black line illustrates a constant $I_{HCO+}/I_{HCN}$ equal to this median ratio, with the black dashed lines showing the scatter in our data. We would expect abundance variations unrelated to density to induce scatter in this relation. The observed correlation suggests that our results would be qualitatively similar if we used $I_{HCO+}/I_{CO}$ to trace dense gas instead of $I_{HCN}/I_{CO}$. 

Correlations between the $I_{HCO+}/I_{HCN}$ ratio and environment have been studied in great detail in luminous infrared galaxies (LIRGs) \citep[e.g.,][]{GARCIACARPIO06,IMANISHI06,GARCIACARPIO08,PRIVON15}.  These studies reveal some source-to-source variations. But consistent with our results here, starburst dominated systems show a mean ratio $I_{HCO+}/I_{HCN}$ $\approx 0.77 \pm 0.24$ \citep{PRIVON15} and an overall good correlation between $I_{HCO+}$ and $I_{HCN}$. The gray line and region in Figure \ref{fig:hcn_hcop_13co} show this ratio and scatter from \cite{PRIVON15}.

$I_{HCO+}/I_{CO}$ variations appear modestly weaker than $I_{HCN}/I_{CO}$ variations. That is, the slope in the left panel of Figure \ref{fig:hcn_hcop_13co} appears slightly sublinear. Meanwhile, the right panel shows that the magnitude of $I_{13CO}/I_{CO}$ variations are much smaller than those in $I_{HCN}/I_{CO}$. As discussed in \citet{LEROY17}, this is the expected behavior for a distribution of gas densities similar to a lognormal. The $n_{eff}$ of optically thin $^{13}$CO is higher than that of CO, which is usually significantly optically thick. But both $n_{eff}$ are lower than mean density of the medium. Thus the fractional change in the gas above these two densities will vary only weakly as the density distribution shifts. Meanwhile, HCO$^+$ has a slightly lower $n_{eff}$ than HCN. For a curving distribution like a lognormal, we expect variations in HCN emission to be stronger than those in HCO$^+$ as the mean of the density distribution shifts \citep{LEROY17}. Conversely, the modest changes in $I_{13CO}/I_{CO}$ indicates that a power law or other self-similar distribution would be a poor description of the entire density distribution.

Thus, our data offer support for the idea that $I_{HCN}/I_{CO}$ traces the dense gas fraction: high $I_{HCN}/I_{CO}$ occurs in regions of high surface brightness, $I_{HCN}/I_{CO}$ tracks with $I_{HCO+}/I_{CO}$, and the sense of line ratio variations is that expected from a lognormal-like distribution sub-resolution. However, variations in the quantitative translation of $I_{HCN}/I_{CO}$ to $f_{dense}$ remain likely. Temperature, abundance, and optical depth variations of both lines remain plausible and hard to quantify \citep[e.g.,][]{JIMENEZDONAIRE17}. These could affect both $\alpha_{HCN}$ and $\alpha_{CO}$. If both lines are optically thick, then temperature variations might be expected to cancel out to some degree, but this is far from certain. We proceed quoting results in terms of the observable $I_{HCN}/I_{CO}$ and noting the corresponding $f_{dense}$ for our fiducial $\alpha_{HCN}$ and $\alpha_{CO}$, but note the latter as uncertain.

Wide field mapping of local molecular clouds do indicate $\alpha_{HCN}$ variations within, e.g., Orion \citep[see][]{PETY17,KAUFFMANN17}. \citet{USERO15} considered a wide range of possible $\alpha_{HCN}$. They show that such variations are highly unlikely to conspire in a way to return a fixed star formation efficiency of dense gas. Still, the quantiative translation of line ratios in to density variations remains uncertain. More multi-line wide area maps of the Milky Way will be key in this regard. Meanwhile, synthetic multi-line modeling of the full line suite \citep[e.g., building on][]{LEROY17}, comparison to small scale ISM structure measurements (M. Gallagher et al. in prep.), and improved constraints on optical depth \citep{JIMENEZDONAIRE17} are next goals of this project and the EMPIRE survey.

\subsection{What Sets the SFE of Dense Gas?}
\label{subsec:sfedense}

\capstartfalse
\begin{deluxetable}{lcc}
\tabletypesize{\scriptsize}
\tablecaption{$\Sigma_{SFR}/\Sigma_{dense}$ vs Other \label{tab:sfe_corr}}
\tablewidth{0pt}
\tablehead{
\colhead{Data Set} & 
\colhead{Rank Corr. (vs $\Sigma_*$)} &
\colhead{Rank Corr. (vs $\Sigma_{mol}$)} \\
}
\startdata
Radial Profiles & & \\
$\ldots$NGC~3351 & -1.0 (0.002) & -1.0 (0.001) \\
$\ldots$NGC~3627 & -0.74 (0.000) & -0.76 (0.000) \\
$\ldots$NGC~4254 & -0.69 (0.001) & -0.69 (0.001)\\
$\ldots$NGC~4321 & -0.90 (0.000) & -0.74 (0.000) \\
$\ldots$NGC~5194\tablenotemark{a} & -1.0  (0.000) & -1.0  (0.000) \\
$\ldots$All profiles & -0.71 (0.000) & -0.73 (0.000)\\
\citet{USERO15} & -0.60 (0.000) & -0.55 (0.000)  \\
\hline
\\
All data & -0.66 (0.000) & -0.72 (0.000)  \\
\enddata
\tablenotetext{a}{\citet{BIGIEL16}.}
\tablecomments{Rank correlation quotes with $p$ value in parenthesis. We quote the mean of the logarithm of the ratio and the $\pm1\sigma$ scatter in the log of the ratio. ``All data'' treats all data points with equal weight, regardless of the spatial scale sampled.}
\end{deluxetable}
\capstarttrue

\begin{figure*}
\plottwo{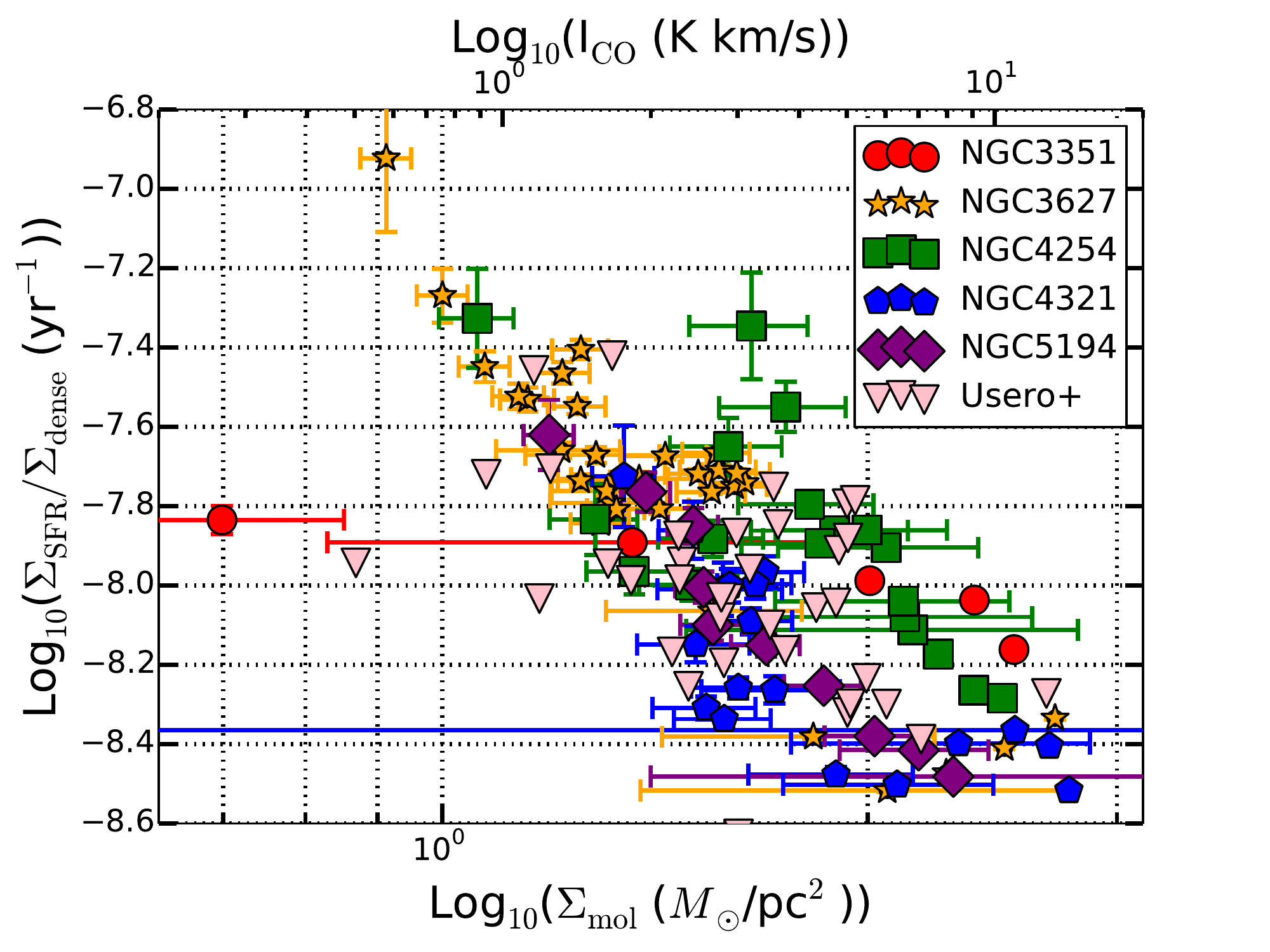}{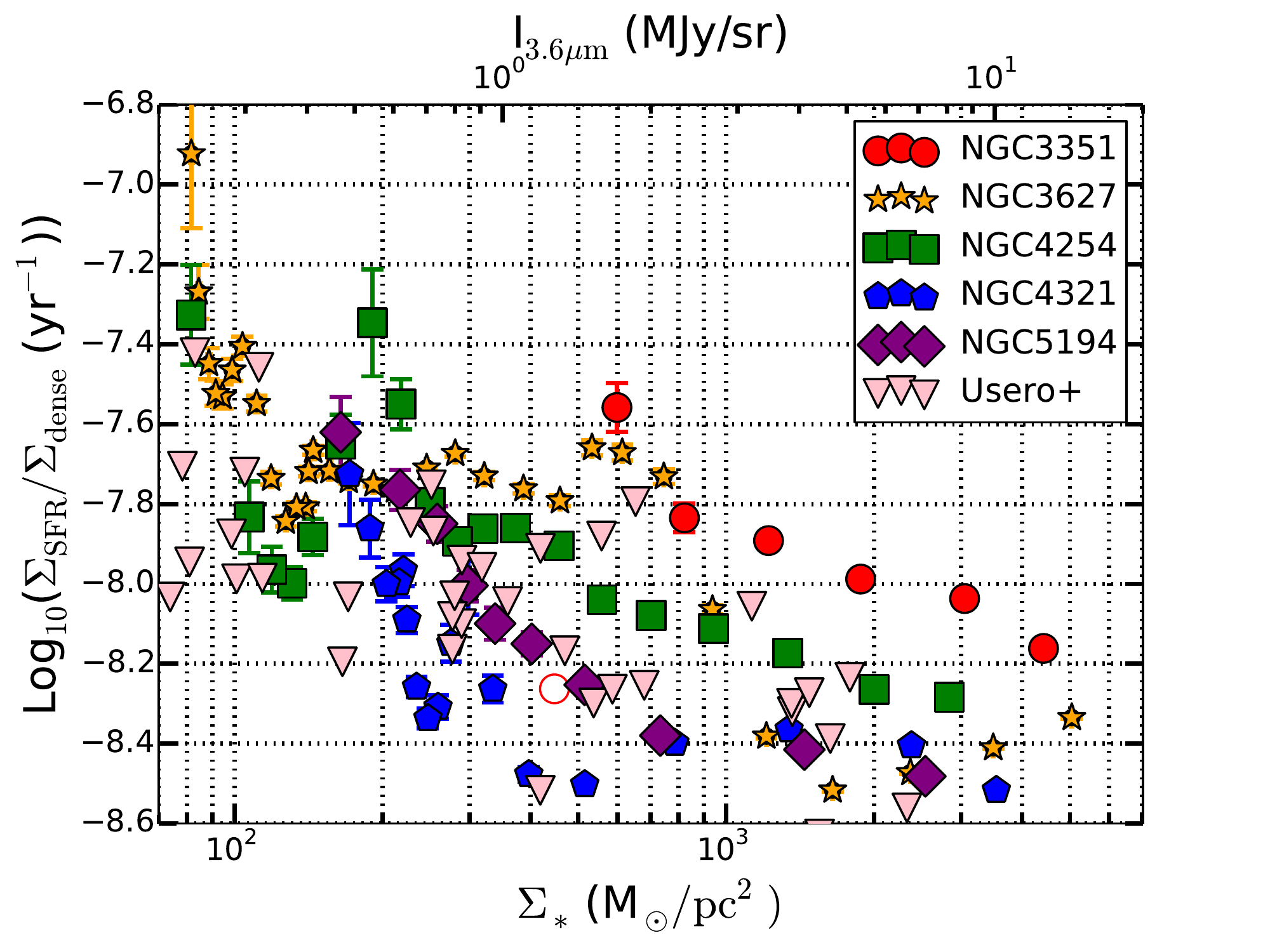}
\caption{$\Sigma_{SFR}/\Sigma_{dense}$, or the star formation efficiency of dense gas, as a function of tracers of the molecular gas ({\em left}) and stellar ({\em right}) surface density. $\Sigma_{SFR}/\Sigma_{dense}$ anti-correlates with $\Sigma_{mol}$, traced by $I_{CO}$, and with $\Sigma_*$, as traced contaminant-corrected $3.6\mu$m intensity. Thus dense gas appears less efficient at star formation at high surface densities. The primary axes adopt our fiducial translations to physical quantities and report the molecular gas or stellar surface density ($x$) and the star formation efficiency of dense gas (SFR/$M_{dense}$). The alternate axes present the observational data. \label{fig:co_sfe}}
\end{figure*}

Our data follow the broad correlation between SFR and HCN luminosity, but we see clear evidence of variations in $\Sigma_{SFR}/\Sigma_{dense}$, the apparent efficiency with which dense gas forms stars.

\subsubsection{Implied Depletion Time and Efficiency Per Free Fall Time}

For our fiducial conversion factors, our measured HCN/SFR ratios imply a median dense gas depletion time, $\tau_{dep}^{dense} \equiv M_{dense}$/SFR, of $\tau_{dep}^{dense} \sim 78$~Myr. This value is lower than the median $\tau_{dep}^{dense} \sim 110$~Myr from \cite{USERO15} and $\tau_{dep}^{dense} \sim 140$~Myr from \cite{GAO04}. 

Our CO/SFR ratios a mean molecular gas depletion time, $\tau_{dep}^{mol} \equiv M_{mol}$/SFR, of 2.0~Gyr. This median value is comparable to the median $\tau_{dep}^{mol}=1.7$~Gyr from \cite{USERO15} and somewhat higher than to the median $\tau_{dep}^{mol}=1.1$~Gyr from \cite{GAO04}.

When comparing these numbers, note that the regions, galaxies, and spatial scales sampled differ among our work, the bright disk pointings of \citet{USERO15}, and the starburst-heavy sample of \citet{GAO04}. Also recall that our sample was constructed to capture variations in CO/SFR, not to be representative of the local galaxy population. With this in mind, these values agree reasonably well with one another and with the $\tau_{dep}^{mol} = 1{-2}$~Gyr found from observations targeting large areas and many galaxies \citep[e.g.,][]{SAINTONGE11,LEROY13,BOLATTO17}.

If we naively take all CO and HCN emitting gas to lie at the effective critical density for the line, then we can estimate the fraction of total and dense gas converted into stars per free fall time, e.g., following \citet{KRUMHOLZTAN07}. We take the density of all molecular gas to be $n_{H2} \sim 10^2$cm$^{-3}$ ($\tau_{ff} \sim 3$~Myr) and the density of the dense gas producing HCN emission to be $n_{H2} \sim 10^5$cm$^{-3}$ ($\tau_{ff} \sim 0.1$~Myr). Then both HCN and CO suggest an efficiency per free fall time of $\sim 0.2{-}0.3$\%. This value agrees well with direct estimates of the efficiency per free fall time based on high resolution imaging in NGC 5194 \citep{LEROY17B} and on previous estimates based on dense gas spectroscopy \citep{GARCIABURILLO12,USERO15}.

\subsubsection{Variation of Dense Gas SFE With Environment}

\citet{USERO15} and \citet{BIGIEL16} showed that stellar surface density ($\Sigma_*$) and the molecular-to-atomic gas ratio act as ``third parameters'' in the relationship between $f_{dense}$  and $SFE_{dense}$. In their studies, regions with high stellar surface density or high molecular gas fraction also have high dense gas fractions. However, in these same high stellar surface density regions dense gas appears less efficient at forming stars. 

Such behavior might be expected if clouds with a high mean density ($n_0$) still only form stars in local overdensities within the cloud. This would be true if these cloud have a typical, roughly virialized, dynamical state at large scales despite their high $n_0$. Then, where $n_{0}$ is high, the immediately star forming gas engaged in direct collapse and star formation may lie at still higher densities. Similar arguments have been proposed to explain the behavior of our own Galactic center \citep[e.g.,][]{KAUFFMANN13,KRUIJSSEN14,RATHBORNE15}.

In this scenario, $SFE_{dense}$ for ``dense'' gas defined by some fixed density threshold will anti-correlate with $n_0$. Thus, e.g., we would expect $\Sigma_{SFR}/I_{HCN}$ to go down as the mean density of the ISM goes up and HCN traces more ``normal'' and less directly star-forming gas. In this case $SFE_{dense}$ should anti-correlate with environmental quantities related to the mean density of the ISM.

To explore this, we plot $\Sigma_{SFR}/\Sigma_{dense}$ where $\Sigma_{dense}$ is a scaled $I_{HCN}$, or $SFE_{HCN}$, as a function of the surface densities of stars and $I_{CO}$, tracing molecular gas surface density. Figure \ref{fig:co_sfe} shows that $\Sigma_{SFR}/\Sigma_{dense}$ indeed decreases with increasing $\Sigma_{mol}$. A reasonable fit to the relationship is

\begin{equation}
\label{eq:fit_tirhcn_co}
\log_{10} \frac{\Sigma_{SFR}}{\Sigma_{dense}} = -6.72 - 0.72 \log_{10}\Sigma_{mol}~.
\end{equation}

\noindent Here $\Sigma_{mol}$ in our sample is the azimuthally averaged intensity of CO emission multiplied by $\alpha_{CO}$. This traces molecular gas surface density on large scales (recall that our resolution is already 100s of pc). A more rigorous test of this idea using higher resolution imaging will be presented in M. Gallagher et al. (in preparation).

Stellar surface densities exceed the gas surface density over the whole area of our survey, and the stars play a key role in setting the gravitational potential within which the gas exists. If the gas is in some semblance of  vertical dynamic equilibrium, then one expects high $\Sigma_*$ to also lead to higher average midplane pressure and so a higher average gas density. The right panel of Figure \ref{fig:co_sfe} shows $\Sigma_{SFR}/\Sigma_{dense}$ as a function of $\Sigma_*$. As in \citet{USERO15} and \citet{BIGIEL16}, we observe a clear anti-correlation between $\Sigma_{SFR}/\Sigma_{dense}$ and $\Sigma_*$, so that in regions with high stellar surface density, dense gas appears worse at forming stars. 

$\Sigma_{SFR}/\Sigma_{dense}$ and $\Sigma_*$ have a rank correlation coefficient of $-0.71$, indicating a stronger relationship than between $\Sigma_{SFR}/\Sigma_{dense}$ and $\Sigma_{dense}/\Sigma_{mol}$. Using the bisector method, we fit a power law of:

\begin{equation}
\label{eq:fit_tirhcn_stellar}
\log_{10} \frac{\Sigma_{SFR}}{\Sigma_{dense}} = -6.08  -0.72 \log_{10}\Sigma_*~.
\end{equation}

\noindent Both Equation \ref{eq:fit_tirhcn_co} and Equation \ref{eq:fit_tirhcn_stellar} should be taken as indicative. We use all of the data available to us, but note that our sample is in no way statistically representative of the whole galaxy population.

Both panels in Figure \ref{fig:co_sfe} support the hypothesis that in a high mean density ISM, the gas that emits HCN is less effective at forming stars than in regions of low mean density. Caveats related to the ability of HCN to trace dense gas apply; we cannot rule out that instead this gas may simply be much better at emitting HCN. However, as discussed in \S \ref{subsec:density}, indirect evidence in our data set, and the discussion in \citet{USERO15}, give us some confidence that we do capture density effects. And as discussed by \citet{USERO15}, HCN emission has so far been widely used to trace dense gas in the extragalactic literature \citep{GAO04}. These results complicate any threshold-style interpretation based on those observations.

We focus on $I_{CO} \sim \Sigma_{mol}$ and $\Sigma_*$ because they relate to the mean density of the gas. Assuming that the weight of the gas is balanced by its random motions, then this relationship may be captured by considering the dynamical equilibrium pressure, $P_{tot}\sim P_{DE}$. $P_{DE}$ will depend on both $\Sigma_*$ and $\Sigma_{mol}$, and may in turn relate to the mean density of a cloud, $n_0$. We return to this point in \S \ref{subsec:pressure_disc}.

\subsection{What Sets the Dense Gas Fraction?}
\label{subsec:DG}

\capstartfalse
\begin{deluxetable}{lcc}
\tabletypesize{\scriptsize}
\tablecaption{$I_{HCN}/I_{CO}$ vs Other \label{tab:dgf_corr}}
\tablewidth{0pt}
\tablehead{
\colhead{Data Set} & 
\colhead{Rank Corr. (vs $\Sigma_*$)} &
\colhead{Rank Corr. (vs $I_{CO}$)} \\
}
\startdata
Radial Profiles & & \\
$\ldots$NGC~3351 & 1.0 (0.001) & 1.0 (0.001) \\
$\ldots$NGC~3627 & 0.81 (0.000) & 0.86 (0.000) \\
$\ldots$NGC~4254 & 0.43 (0.037) & 0.43 (0.037)\\
$\ldots$NGC~4321 & 0.71 (0.001) & 0.72 (0.001) \\
$\ldots$NGC~5194\tablenotemark{a} & 1.0 (0.000)  & 1.0 (0.000) \\
$\ldots$All profiles & 0.67 (0.000)  & 0.56 (0.000) \\
\citet{USERO15} & 0.78 (0.000) & 0.72 (0.000) \\
\hline
\\
All data & 0.67 (0.000) & 0.56 (0.000)  \\
\enddata
\tablenotetext{a}{\citet{BIGIEL16}.}
\tablecomments{Rank correlation quotes with $p$ value in parenthesis. We quote the mean of the logarithm of the ratio and the $\pm1\sigma$ scatter in the log of the ratio. ``All data'' treats all data points with equal weight, regardless of the spatial scale sampled.}
\end{deluxetable}
\capstarttrue

We also observe $I_{HCN}/I_{CO}$, our observational tracer of $f_{dense}$, to vary significantly across our sample. Adopting our fiducial conversion factors, our observed $I_{HCN}/I_{CO}$ imply $f_{dense}$  between $\sim 0.5$\% and $\sim 25$\% with a median $f_{dense} \sim 4$\%.

This median value is lower than the median value of $f_{dense} \sim 8$\% from \cite{USERO15} and the median value of $f_{dense} \sim 12$\% from \cite{GAO04}. This deviation could be expected based on the construction of samples, as $f_{dense}$ is not expected to remain constant. \cite{GAO04} study gas rich starburst galaxies. Meanwhile the pointings of \cite{USERO15} were chosen to be bright in CO, and so rich in gas. We target a wider area with less of a selection bias, and thus recover lower dense gas fractions. Still wider area maps covering whole galaxy disks may return even lower median $f_{dense}$ (a main goal of EMPIRE, see M. Jimenez Donaire et al. in preparation).

Our median $f_{dense} \sim 4$\% is lower than the typical $f_{dense} \sim 10$\% found for local clouds by \cite{LADA12}. This may reflect different ways of tracing dense gas, they use extinction rather than HCN emission, and adopt a different nominal density. It may also reflect more sensitivity of our observations to an extended molecular component.

Following the logic laid out in the previous Section, we would expect the $I_{HCN}/I_{CO}$, tracing $f_{dense}$, to correlate with both $\Sigma_*$ and $I_{CO}$, tracing $\Sigma_{mol}$. In Figure \ref{fig:co_dgf}, we plot $I_{HCN}/I_{CO}$ as a function of both quantities. The figure shows that $I_{HCN}/I_{CO}$ indeed increases with increasing $I_{CO}$. Using the bisector method, we find a power law fit for $I_{HCN}/I_{CO}$ vs. $I_{CO}$ of:

\begin{equation}
\label{eq:fit_hcnco_co}
\log_{10} \frac{HCN}{CO} = -2.42 +0.72  \log_{10}I_{CO}~.
\end{equation}

\noindent Again, we emphasize that Equation \ref{eq:fit_hcnco_co} is specific to our data and that $I_{CO}$ traces the molecular gas surface density averaged on moderately large (few hundred pc to $\sim$kpc) scales.

We also expect to find a higher mean density, and so perhaps more dense gas where the gravitational potential well is deeper.  We compare $I_{HCN}/I_{CO}$ to $\Sigma_*$ in the right panel of Figure \ref{fig:co_dgf}. We observe a clear correlation between $I_{HCN}/I_{CO}$ and $\Sigma_*$. This has the sense that in regions with high stellar surface density, there is more gas dense enough to produce HCN emission. $I_{HCN}/I_{CO}$ and $\Sigma_*$ exhibit a rank correlation coefficient of $0.74$, indicating a stronger relationship between $f_{dense}$ and $\Sigma_*$ than between $\Sigma_{TIR}/I_{CO}$ and $I_{HCN}/I_{CO}$. 

A reasonable fit to our data, including NGC 5194 and the data of \citet{USERO15} is:

\begin{equation}
\label{eq:fit_hcnco_stellar}
\log_{10} \frac{HCN}{CO} = -3.28  +0.62  \log_{10}\Sigma_*~.
\end{equation}

\noindent Which is moderately steeper than the relation found by \cite{USERO15}:

\begin{equation}
\label{eq:fit_hcnco_stellar_usero}
\log_{10} \frac{HCN}{CO}_{\rm{Usero}} = -2.58+0.42\log_{10}\Sigma_* ~.
\end{equation}

Thus, following \citet{USERO15} and \citet{BIGIEL16}, but now for maps of the disk and central regions of five galaxies, our observations reveal strong correlations between disk structure and $I_{HCN}/I_{CO}$. These correlations extend over roughly an order of magnitude in $I_{HCN}/I_{CO}$ and span two orders of magnitude in stellar surface density and $I_{CO}$. We interpret this as evidence for a strong link between the dense gas fraction and local disk structure.

\subsection{Dynamical Equilibrium Pressure, Dense Gas, and Star Formation}
\label{subsec:pressure_disc}

\capstartfalse
\begin{deluxetable}{lcc}
\tabletypesize{\scriptsize}
\tablecaption{Other vs. Pressure \label{tab:pressure_corr}}
\tablewidth{0pt}
\tablehead{
\colhead{Data Set} & 
\colhead{Rank Corr.} &
\colhead{Rank Corr.} \\
\colhead{} & 
\colhead{($\Sigma_{SFR}/\Sigma_{dense}$ vs)} &
\colhead{($\Sigma_{dense}/\Sigma_{mol}$ vs)} \\
}
\startdata
Radial Profiles & & \\
$\ldots$NGC~3351 & -1.0 (0.001) & 1.0 (0.001) \\
$\ldots$NGC~3627 & -0.65 (0.000) & 0.79 (0.000) \\
$\ldots$NGC~4254 & -0.69 (0.001) & 0.43 (0.037)\\
$\ldots$NGC~4321 & -0.70 (0.001) & 0.67 (0.002) \\
$\ldots$NGC~5194\tablenotemark{a} & -1.0 (0.000) & 1.0 (0.000) \\
$\ldots$All profiles & -0.71 (0.000) & 0.57 (0.000)\\
\citet{USERO15} & -0.47 (0.001)& 0.67 (0.000) \\
\hline
\\
All data & -0.54 (0.000) & 0.36 (0.000) \\
\enddata
\tablenotetext{a}{\citet{BIGIEL16}.}
\tablecomments{Rank correlation coefficients with $p$ value in parenthesis. We quote the mean of the logarithm of the ratio and the $\pm1\sigma$ scatter in the log of the ratio. ``All data'' treats all data points with equal weight, regardless of the spatial scale sampled.}
\end{deluxetable}
\capstarttrue

\begin{figure*}
\plottwo{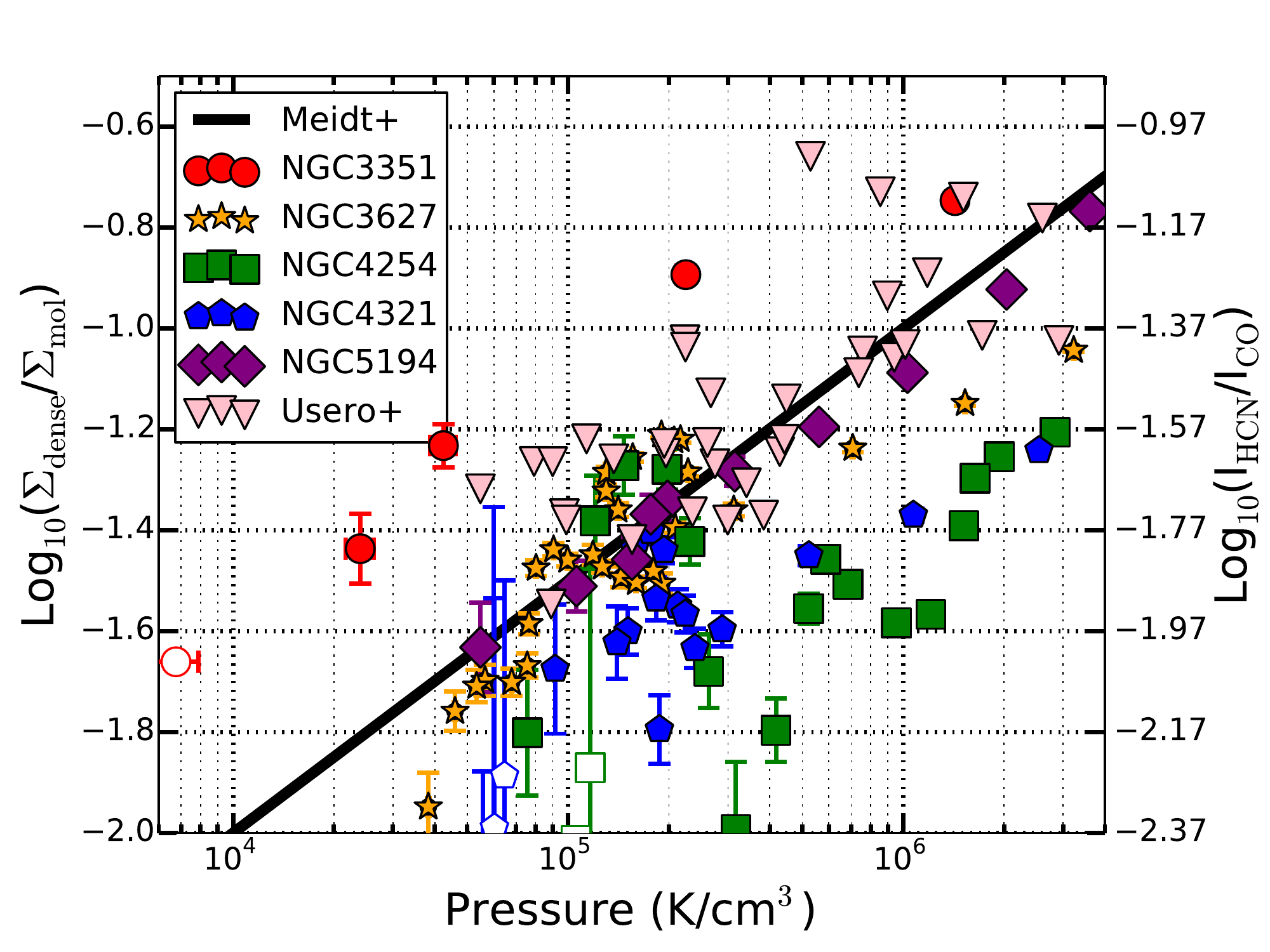}{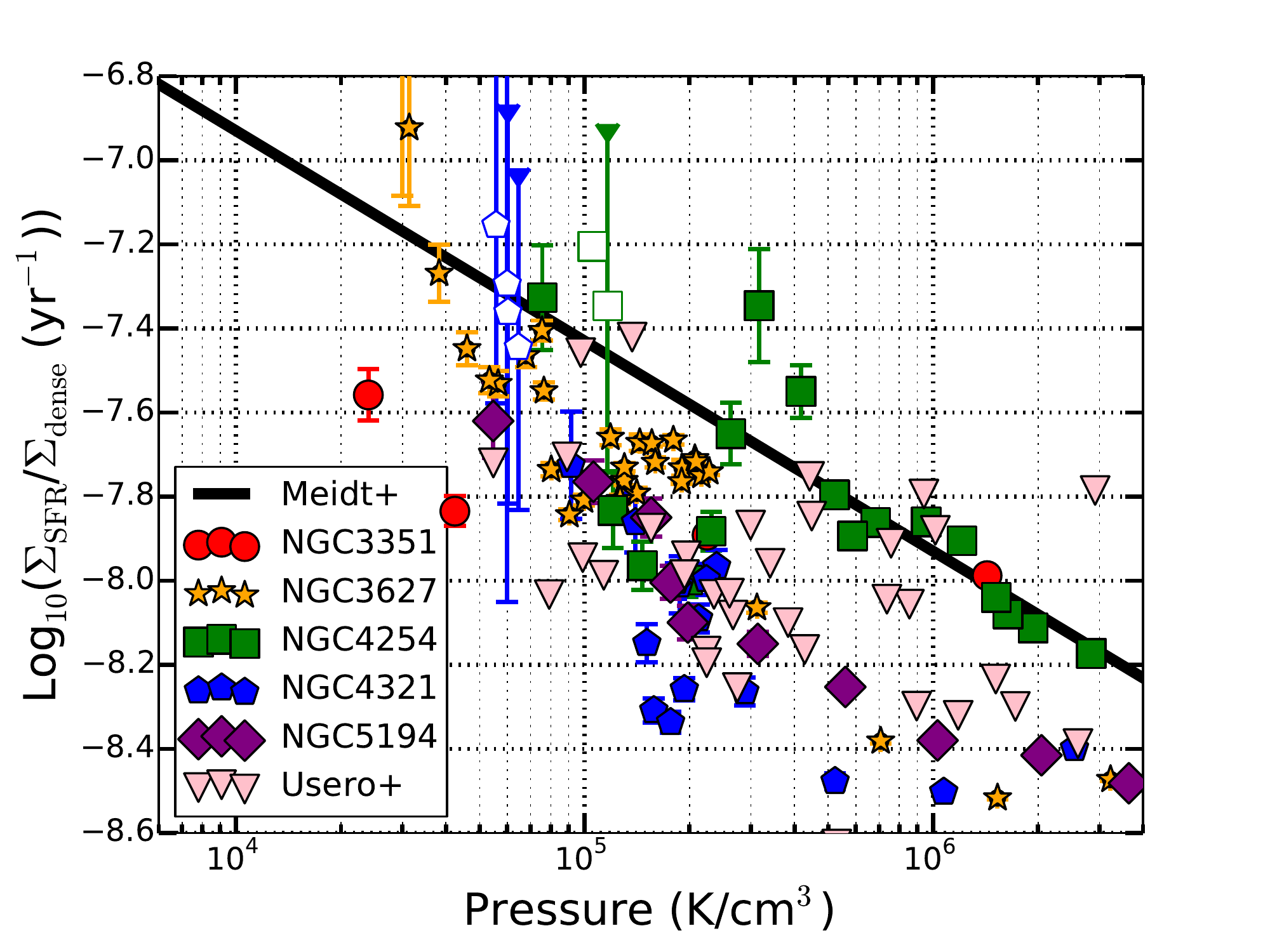}
\caption{({\em Left:}) $I_{HCN}/I_{CO}$, tracing dense gas fraction, and ({\em right}) $\Sigma_{SFR}/\Sigma_{dense}$, or the star formation efficiency of dense gas, both as a function of dynamical equilibrium pressure calculated following Equation \ref{eq:ph}. Regions where high $P_{DE}$ is needed to sustain the weight of the gas have high dense gas fractions but apparently low dense gas star formation efficiencies. This is consistent with the idea that the mean cloud density reflects the large scale environment and that the role of any particular density in star formation (here the density needed to emit HCN) depends on the environment. Gas of a particular density appears to be less efficient in regions of high mean density. The black lines are model predictions from \cite{MEIDT16}.
\label{fig:pressure}}
\end{figure*}

The large scale gas surface density, traced by $I_{CO}$, and the stellar surface density, $\Sigma_*$, appear to affect both $SFE_{dense}$ and $f_{dense}$. Above, we hypothesized, that these relationships emerge because the mean density of molecular clouds reflect their large scale environment. Specifically, following \citet{ELMEGREEN89}, \citet{OSTRIKER10}, \citet{KIM11}, and others, we noted that the state of the ISM should self-regulate so that the pressure averaged over some relatively large time and length scales, will approach the ``dynamical equilibrium'' value. In this case, molecular clouds will represent over-pressured regions against a background pressure that reflects the environment. \citet{HELFER97} argued for a version of this idea to explain the variable dense gas fractions observed in nearby galaxies, and both \citet{USERO15} and \citet{BIGIEL16} noted this as one plausible family of explanations for their observations. Galactic center studies have considered similar scenarios.

To explore the role of pressure directly, we estimate the dynamical equilibrium pressure, $P_{DE}$, for each ring in our sample (see \S \ref{subsec:Ph}). In Figure \ref{fig:pressure}, we plot $\Sigma_{dense}/\Sigma_{mol} (\propto I_{HCN}/I_{CO}$) and $\Sigma_{SFR}/\Sigma_{dense} (\propto \Sigma_{SFR}/I_{HCN}$) as a function of $P_{DE}$. 

A reasonable fit relating $P_{DE}$ and $\Sigma_{dense}/\Sigma_{mol}$ is:

\begin{equation}
\label{eq:fit_dgf_pressure}
\log_{10} \frac{\Sigma_{dense}}{\Sigma_{mol}} = -4.01+0.48 \log_{10}P_{DE}~.
\end{equation}

\noindent $P_{DE}$ and $\Sigma_{dense}/\Sigma_{mol}$ relate via a rank correlation coefficient of 0.60. The sense of this correlation, as expected, is that regions with high $P_{DE}$ exhibit high apparent dense gas fractions. This is the same as the sense of the correlation between $I_{CO}$ and $I_{HCN}/I_{CO}$ and $\Sigma_*$ and $I_{HCN}/I_{CO}$. The strength of the correlations does not recommend one of these as more fundamental, but we note that the correlation with $P_{DE}$ has a physical explanation that would be expected to yield the other two.

High $P_{DE}$ corresponds to a higher dense gas fraction, but also a lower star formation efficiency in the dense gas. In our data, a bisector fit between $P_{DE}$ and $\Sigma_{SFR}/\Sigma_{dense}$ yields:

\begin{equation}
\label{eq:fit_sfe_pressure}
\log_{10} \frac{\Sigma_{SFR}}{\Sigma_{dense}} = -5.15 - 0.50  \log_{10}P_{DE}~.
\end{equation}

\noindent The two quantities show a rank correlation coefficient of $-0.71$, indicating a reasonably strong anti-correlation. Again, this goes in the same direction as the anti-correlation between $\Sigma_*$ and $\Sigma_{SFR}/\Sigma_{dense}$ and between $I_{CO}$ and $\Sigma_{SFR}/\Sigma_{dense}$. In detail, the slopes differ, with somewhat shallower slopes for the $P_{DE}$ relations than those treating $\Sigma_*$ or $I_{CO}$ as the independent variable.

{\em Should $P_{DE}$ affect the mean cloud density?} Almost all regions that we consider have $P_{DE}/k_{B} \gtrsim 10^5$~cm$^{-3}$~K. We expect that the average pressure of the ISM must approach these values. Compare this to the typical internal pressure for a local molecular cloud with $\Sigma_{mol} \sim 100$~M$_\odot$~pc$^{-2}$. Taking $P_{int} \sim G\pi\Sigma_{mol}^2/2 $ we see $P_{in}/k_{B} \sim 3 \times 10^{5}$~cm$^{-3}$~K. In the dynamical equilibrium picture, we would expect the internal pressure of molecular clouds to be somewhat higher than $P_{DE}$, as molecular clouds represent overpressured regions in a dynamical ISM.  In this sense, almost all of our regions have $P_{DE}$ high enough that $P_{int}$ cannot remain fixed for clouds across our survey while still balancing the weight of the gas in the stellar potential. 

Our result for the dense gas fraction agrees well with the prediction by \cite{MEIDT16}, shown by the black line in Figure \ref{fig:pressure}. They consider isothermal clouds bounded by interstellar pressure with $\rho \sim r^{-2}$ density profiles. We show their predicted $f_{dense} = \Sigma_{t, HCN}/\Sigma_{t, mol}$. Here $\Sigma_{t, HNC}$ is the column density above the critical density of HCN and $\Sigma_{t, mol}$ is the column density of the molecular material at the interface with the ambient ISM and is characterized by the ISM pressure. In this view, our calculated dynamical equilibrium pressure is interpreted as an ambient interstellar pressure; this assumption should work as long as the cloud cores modeled by \citet{MEIDT16} do not make up too much of the mass. Figure \ref{fig:pressure} shows good agreement with most of our data and with the data of \citet{USERO15}. The corresponding prediction for $SFE_{dense}$ also captures our measured slope and intecept reasonably well. 

{\em Is there a threshold pressure for environment to affect dense gas?} Conversely, our data do not yield much insight into the question of whether the apparent dense gas fraction and $SFE_{dense}$ continue to correlate with $P_{DE}$ at the lower values of $P_{DE}$ expected over most of the area in galaxy disks. One scenario would be that when $P_{DE}$ falls far below $P_{int}$ of a typical molecular cloud and {\sc Hi} dominates the ISM, the clouds decouple and converge to a fairly universal population. In this case, the threshold picture of \citet{LADA12} might still hold for analogs of the Solar Neighborhood.

Our study, which is optimized to study the contrast between inner molecular disks and the bright central regions of our targets, does not test this scenario. The full EMPIRE survey (Jimenez Donaire et al. in preparation) and full disk, sensitive ALMA follow up of our targets should offer more insight.

\section{Discussion}
\label{sec:discussion}

We observed the inner few kpc of four star forming galaxies, which show a range of apparent molecular gas depletion times, traced by CO/SFR. We analyze these data in conjunction with the first full-galaxy map from EMPIRE \citep{BIGIEL16} and the pointed observations targeting $\sim 30$ galaxy disks from \citet{USERO15}. Extending \citet{USERO15} and \citet{BIGIEL16} and following \citet{GARCIABURILLO12}, these data provide strong evidence that the ratios $I_{HCN}/I_{CO}$ and $\Sigma_{SFR}/I_{HCN}$ depend systematically on galactic environment. Assuming these to trace the dense gas fraction and the star formation efficiency of dense gas, then this provides strong evidence for a dependence of dense gas fraction on environment and context-dependent role for density in star formation.

With maps at ALMA's resolution, we are able to distinguish the relative structure of SFR, HCN emission, and CO emission. Our results suggest strong changes in the ratios among these quantities between the dense, central regions of some galaxies and the lower surface density disk regions. In our sample, this contrast appears strongest for strongly barred galaxies (NGC 3351, NGC 3627, and NGC 4321), though NGC 5194 (M51) shows some of the same signal despite the lack of a strong bar. A logical scenario to explain this behavior is that the bar drives gas to the center of the galaxy, creating bright, dense concentrations of molecular gas \citep[e.g.,][]{SHETH05,KRUMHOLZKRUIJSSEN15}. In these regions, dense gas is common but the star formation per unit dense appears low. NGC 5194 may exhibit a similar phenomenon thanks in part to the role of the arms in enabling gas streaming towards the center of the galaxy \citep[e.g., see][]{MEIDT13,QUEREJETA16}.

This behavior is reminiscent of that seen in our own Milky Way galaxy. The Milky Way also has a strong bar, a massive concentration of dense molecular gas in the inner few hundred pc, and an apparently low ratio of star formation to dense gas compared to local clouds. \citet{LONGMORE13} showed that in the galactic center the dense gas integrated intensity is $\sim$100 times that of non-galactic center regions and that the star formation efficiency in the galactic center is an order-of-magnitude lower than predicted.

We interpret our results as strong evidence that the behavior seen in the Milky Way is a general feature of strongly barred galaxies, with (at least apparent) central concentrations of dense gas but comparatively low (apparent) star formation rate per unit dense gas. More statistics are needed, especially to constrain any possible duty cycle \citep{KRUMHOLZKRUIJSSEN15} and to understand how often such phenomena also occur in unbarred galaxies. But it seems clear that the Milky Way is far from a unique case in this regard.

\subsection{A Context Dependent Role for Density}
\label{subsec:theory}

Our main observational result is that $\Sigma_{dense}/\Sigma_{mol}$ and $\Sigma_{SFR}/\Sigma_{dense}$ show correlations with environment that are stronger than the correlation between $\Sigma_{SFR}/\Sigma_{mol}$ and $\Sigma_{dense}/\Sigma_{mol}$. A natural interpretation for this is that the density distribution is a strong function of environment while the role of any particular density in star formation is context-dependent.

{\em Environment and Density:} We find that the fraction of dense gas depends on environment. In this paper, $f_{dense}$ is traced mainly by $I_{HCN}/I_{CO}$, but HCO$^+$ and CS show similar results. Here ``environment'' means that $f_{dense}$ appears to increase as $\Sigma_*$ and $I_{mol} \sim \Sigma_{mol}$ increase.

Stellar and gas surface density both relate to the mean pressure, $P_{DE}$, needed for the ISM to support its own weight in the (mostly stellar) potential well. Inferring $P_{DE}$ from observables, we show that $f_{dense}$ does correlate with $P_{DE}$. As $P_{DE}$ increases, $f_{dense}$ appears to increase. This follows closely on the results of \cite{HELFER97} that gas becomes denser in a deeper gravitational potential well where the ISM pressure is higher.

We suggest the following sketch to explain this correlation. In order for approximate equilibrium to hold, a disk must self-regulate to have $\sim P_{DE}$ over a moderate length or time scale \citep{OSTRIKER10,OSTRIKER11,KIM11,KIM13,KIM15}. This dynamical equilibrium captures an average behavior. Molecular clouds will represent overdensities relative to this average. $P_{DE}$ over large parts of our sample already corresponds to the typical internal pressure of a standard local cloud, $P_{int} / k_{B} \sim \pi G \Sigma_{gas}^2 / 2 k_{B} \sim 3 \times 10^5$~cm$^{-3}$~K. For molecular clouds to represent over-pressured regions (relative to the time and space average) in our data, their internal pressure must vary, increasing with $P_{DE}$. 

Our view is that the mean internal pressure of molecular clouds tracks $P_{DE}$, and that as $P_{DE}$ shifts so does the density distribution. That is, in high pressure regions, the mean density is higher and the overall density distribution also shifts to higher values. We would also expect these regions to have higher line widths, and likely higher Mach numbers, which could also yield a larger dense gas fraction \citep[e.g., see][]{PADOAN02,KRUMHOLZ07}.

Our own observations, along with \citet{USERO15} and \citet{BIGIEL16} support a scenario where the mean density of molecular clouds varies with the surface density of the disk. Viewing such a scenario through the lens of self-regulation to some mean $P_{DE}$ based on disk structure and gas content seems reasonable. This explanation is not unique, but does offer a logical scenario to explain what we see.

{\em Mean Density, Dense Gas Fraction, and Star Formation:} While $I_{HCN}/I_{CO}$ increases with $\Sigma_*$, $I_{CO} \sim \Sigma_{mol}$, and $P_{DE}$, $\Sigma_{SFR}/I_{HCN}$ decreases. That is, as the dense gas fraction, and so likely the mean density, increases, the apparent rate of star formation per unit dense gas decreases. Taken at face value, this implies that any particular density --- here specific $n_{\rm eff}$ of HCN --- has a context dependent role in star formation.

A context-dependent role for density in star formation could be expected if clouds are always marginally bound or virialized but vary in their mean density. Following, e.g., \citet{KRUMHOLZKRUIJSSEN15} and many others, overdensities will collapse out of the turbulent cascade and form stars. This immediately star forming material seems likely to exhibit a power law distribution of densities \citep[e.g., see][and many others]{KAINULAINEN09,FEDERRATH13}. In clouds with high mean density, the immediate condition for these star forming structures may be shifted to higher densities \citep[e.g.,][]{KRUIJSSEN14}. That is, stars should form out of overdensities in turbulent molecular clouds.

If HCN always traces the same gas density, $n_{HCN}$, and if that density does not always capture the collapsing, power law tail of the density distribution, then we could expect that the ratio of star formation per unit HCN will change as the density distribution changes.
In regions of low $P_{DE}$ and low mean cloud density, HCN traces high density gas more immediately related to star formation. In high $P_{DE}$ regions with high mean density, HCN may trace a large fraction of the bulk mass in the cloud.

As above, this is not a unique interpretation. The mapping between HCN emission and dense gas may change with environment. And this interpretation requires that HCN not always sample a self-similar power law tail of the density distribution that behaves the same in any environment \citep[e.g., see][]{LEROY17}. There is support for a scenario like what we describe from the Milky Way center \citep[][]{KEPLEY14,RATHBORNE15}, but much more work remains to be done.

{\em Turbulence or Pressure?} We have emphasized mean density in this section, but note that the Mach number also plays a key role in the density distribution for a turbulent cloud \citep{PADOAN02,KRUMHOLZ05}. \citet{USERO15} showed that using the model of \citet{KRUMHOLZ07}, plausible variations in the Mach number could explain many of their observations along with those of \citet{GARCIABURILLO12} and \citet{GAO04}. Observations suggest that as the mean density changes, the turbulent velocity dispersion will also change \citep[e.g., see][and J. Sun et al. ApJ submitted]{LEROY16}, so many of the same considerations described above will apply. We intend to return to this topic using high resolution CO observations of our targets.

\subsection{Caveats and Next Steps}
\label{subsec:caveats}

Resolved maps of density sensitive line ratios across galaxies are new. We have focused on scaling relations that relate closely to observables. More, we placed particular emphasis on HCN, the brightest high critical density mm-wave line, and we use our other lines to investigate the robustness of our physical interpretation. We also target a limited area and set of targets designed to highlight the contrast between galaxy disks and central starburst regions. In this section, we highlight some of the caveats and logical next steps related to these choices.

First, our survey does not yet offer a representative sample. It was designed to capture variations in $\Sigma_{SFR}/\Sigma_{mol}$ rather than to span whole galaxy disks or the local galaxy population. Thus, our scaling relations should be taking as indicative rather than authoritative. In particular, we do not yet know how they would extend to the low surface density, outer parts of galactic disks. We also lack sizable samples barred and unbarred galaxies to quantitatively estimate the impact of a strong bar. We expect the full EMPIRE sample to help with this, as well as future deeper, wide area mapping by the GBT and ALMA.

Second, we adopt a simple translation from HCN and CO to dense and total molecular gas mass. Given the novelty of wide area HCN mapping, we emphasize reporting the basic scaling relations. We also checked that our other dense gas tracers yield qualitatively similar results. Moving forward, a next direction for this field is clearly to combine a large suite of lines to model the density distribution; see \citet{LEROY17} for a first step in this direction following \citet{KRUMHOLZ07,NARAYANAN08}. Just as important, higher $J$ transitions are needed to understand the impact of excitation effect. Optically isotopologue observations are needed to constrain the optical depth of the dense gas tracers, which directly impacts their $n_{\rm eff}$ \citep[see][]{JIMENEZDONAIRE17}. Extragalactic observations lack an external constrain on the dense gas mass (e.g., like dust observations can provide in local clouds). As a result, wide area Milky Way mapping of diverse environments \citep{PETY17} has a crucial role to play in informing the translation from line ratios to the density distribution.

We also adopt a simplified treatment of $\alpha_{CO}$ in order to remain close to observables. Based on \citet{SANDSTROM13}, who studied some of our targets, we know that this represents an oversimplification. $\alpha_{CO}$ often drops towards the bright centers of barred galaxies. However, as discussed by \citet{USERO15}, implementing $\alpha_{CO}$ corrections with no knowledge of $\alpha_{HCN}$ variations may obscure the basic scaling relations. Still, our targets have also been mapped by ALMA with the explicit goal of deriving $\alpha_{CO}$ from resolved multi-line modeling and dust comparison. We expect to revisit this topic in detail in future work.

Other physical parameter estimates may also induce uncertainty. We adopt a combination of H$\alpha$ and 24$\mu$m to trace star formation. Much of the starburst literature \citep{GAO04,GARCIABURILLO12} has focused on bolometric infrared emission as a star formation tracer, motivating us to explore the impact of adopting other star formation rate tracers (e.g., IR tracer) in the appendix. While there are subtle differences between using TIR, UV and 24$\mu$m, or H$\alpha$ and 24$\mu$m, none of them appear to change our core results.

Finally, we make frequent reference to the structure of turbulent clouds, discussing the relationship between mean density, dynamical state (i.e., virial parameter), and the density distribution probed by our spectroscopy. Our galaxies are also targets of high resolution ($\theta \sim 1\arcsec$) ALMA CO~(2-1) imaging (``PHANGS-ALMA,'' A. K. Leroy, E. Schinnerer, et al. in prep.). A next main direction for these kind of observations will be to directly compare the turbulent structure of clouds at $\sim 50{-}100$ pc scales from high resolution imaging to the density distribution inferred from spectroscopy. The same data set will allow detailed comparison of the structure of the nuclear region and gas flow along the bars to the density distribution.

\section{Summary}
\label{sec:summary}

Our key findings are:

\begin{enumerate}

\item We find strong central enhancements in all molecular lines, including tracers of dense gas, and tracers of star formation in the central regions of our three targets with strong bars (see Figures \ref{fig:3351}-\ref{fig:5194}). These inner regions show high ratios of dense gas tracers to CO emission (e.g., $I_{HCN}/I_{CO}$), but also signatures of lower $\Sigma_{SFR}/I_{HCN}$. These findings closely resemble those for our own Galaxy \citep[e.g.,][]{LONGMORE13,KRUIJSSEN14}, where bar-induced streaming motions transport gas to the center of the galaxy, creating a wealth of dense gas, but that dense gas does not appear particularly good at forming stars. This appears to be a general phenomenon in strongly barred galaxies (see \S \ref{subsec:RPR}). 

\item In contrast to some current theories, the apparent efficiency of star formation in molecular gas ($\Sigma_{SFR}/I_{CO}$) is only weakly predicted by the apparent dense gas fraction ($I_{HCN}/I_{CO}$). Our observations do fall on the previously observed TIR-HCN scaling, but real physical scatter in $\Sigma_{SFR}/I_{HCN}$ in our sample and the current literature renders the relationship weaker than is sometimes claimed. We find a Spearman rank correlation coefficient of $\sim 0.13$ relating the two quantities and a scatter of a factor of $\sim 2$ in SFR-to-HCN. How this result extends to the wider, fainter regions of whole galaxy disks remains to be seen (\S \ref{sec:dgsf}).

\item $I_{HCO+}/I_{CO}$ correlates well with $I_{HCN}/I_{CO}$ with a slightly sublinear slope, while $^{13}$CO shows a flatter correlation. These are the correlations expected for a density distribution with a mean above the effective density for $^{13}$CO emission but below that of HCO$^+$ and HCN. These results suggest give some confidence that $I_{HCN}/I_{CO}$ indeed traces the density distribution, but the suite of tracers available to extragalactic studies remains limited and much more work needs to be done to understand the translation of observed line ratios to gas density distributions (see \S \ref{subsec:density}).

\item $\Sigma_{SFR}/I_{HCN}$, an observational tracer of the star formation efficiency of dense gas, anti-correlates with the apparent surface densities of stars, traced by $3.6\mu$m emission, and molecular gas, traced by $I_{CO}$ (see Figure \ref{fig:co_sfe}). Indeed, these correlations are stronger than that between $I_{HCN}/I_{CO}$ and $\Sigma_{SFR}/I_{CO}$. This provides additional support for a context dependent role for density in star formation, building on the results of \citet{GARCIABURILLO12}, \citet{USERO15}, \cite{BIGIEL16} and the Milky Way results mentioned above (\S \ref{subsec:sfedense} and \S \ref{subsec:theory}).

\item On the other hand, $I_{HCN}/I_{CO}$, an observational tracer of the dense gas fraction, positively correlates with the apparent surface densities of stars, traced by $3.6\mu$m emission, and molecular gas, traced by $I_{CO}$ (see Figure \ref{fig:co_dgf}). Thus in the high surface density, deep potential well regions of galaxies, the dense gas fraction increases.  There is a higher fraction of dense gas where there is more gas, and the density distribution appears to be tied to the gravitational potential well in which it sits (see \S \ref{subsec:DG}). This result shows that trends seen among galaxies or in individual galaxies \citep[e.g.,][]{GAO04,GARCIABURILLO12,USERO15,CHEN15,BIGIEL16} also hold in resolved profiles of a sample of galaxies.

\item Both of these correlations can be re-expressed in terms of the dynamical equilibrium pressure, $P_{DE}$, that must be maintained, on average, to support the weight of gas in a part of galaxy \citep[e.g.,][]{ELMEGREEN89,OSTRIKER10,KIM11}. $I_{HCN}/I_{CO}$ correlates with $P_{DE}$, supporting with a large amount of data the suggestion by \citet{HELFER97} of a link between ISM pressure and dense gas fraction. $\Sigma_{SFR}/I_{HCN}$ anti-correlates with $P_{DE}$, suggesting a picture in which the mean internal pressure of a cloud reflects its environment, but only the high density tail of the density distribution of the density distribution directly participates in star formation.

\end{enumerate}
  
\acknowledgments{This paper makes use of the following ALMA data: ADS/JAO.ALMA\#2013.1.00634.S and ADS/JAO.ALMA\#2011.0.00004.SV. ALMA is a partnership of ESO (representing its member states), NSF (USA) and NINS (Japan), together with NRC (Canada), NSC and ASIAA (Taiwan), and KASI (Republic of Korea), in cooperation with the Republic of Chile. The Joint ALMA Observatory is operated by ESO, AUI/NRAO and NAOJ. The National Radio Astronomy Observatory is a facility of the National Science Foundation operated under cooperative agreement by Associated Universities, Inc. This work is partially based on observations carried out with the IRAM IRAM 30-m telescope. IRAM is supported by INSU/CNRS (France), MPG (Germany) and IGN (Spain). The work of MG and AKL is partially supported by the National Science Foundation under Grants No. 1615105, 1615109, and 1653300. ES acknowledge financial support to the DAGAL network from the People Programme (Marie Curie Actions) of the European Union's Seventh Framework Programme FP7/2007- 2013/ under REA grant agreement number PITN-GA-2011-289313. AH acknowledges support from the Centre National d'Etudes Spatiales (CNES). AU acknowledges support from Spanish MINECO grants ESP2015-68964 and AYA2016-79006. ER is supported by a Discovery Grant from NSERC Canada. FB acknowledges funding from the European Union’s Horizon 2020 research and innovation programme (grant agreement No 726384 - EMPIRE).The work of ADB is supported in part by the NSF under grant AST-1412419. S.G.B. acknowledges support from the Spanish MINECO grant
AYA2016-76682-C3-2-P. This paper made use of the NASA Astrophysics Data Abstract service, AstroPy, the IDL Astronomy User's Library, CASA, and the NASA Extragalactic Database. MG and AKL gratefully acknowledge helpful discussions with Todd Thompson and the OSU galaxy and ISM discussion groups. Diane Cormier is supported by the European Union's Horizon 2020 research and innovation programme under the Marie Sk\l{}odowska-Curie grant agreement No 702622. Mar\'{i}a J. Jim\'{e}nez-Donaire and Frank Bigiel acknowledge support from DFG grant BI 1546/1-1. ECO is supported by grant AST- 1713949 from the National Science Foundation.
}

\begin{appendix}
\section{70 Micron Emission, Total Infrared Emission, and SFR Estimation in Our Targets}
\label{sec:70tTIR}

As discussed in \S \ref{sec:sfr_tir}, we are interested in estimating the TIR luminosity surface density, $\Sigma_{\rm TIR}$ for each ring. The bolometric infrared emission has been a standard measure of recent star formation in the HCN and ``dense gas'' literature, and it offers an alternative to and a check on our fiducial H$\alpha+24\mu$m approach.

To calculate $\Sigma_{\rm TIR}$, we use maps of 70, 100, 160, and 250$\mu$m emission from {\em Herschel} and 24$\mu$m emission from {\em Spitzer}. We used the kernels provided by \citet{ANIANO11} to convolve all of these maps to share a Gaussian $30\arcsec$ (FWHM) beam. At this coarse resolution, we calculate $\Sigma_{\rm TIR}$ using all bands and following the prescription of \citet{GALAMETZ13}.

Ideally, we would simply use all of these data to estimate $\Sigma_{\rm TIR}$ ring-by-ring. Unfortunately, the 160 and 250$\mu$m data, which constrain the long wavelength part of the IR spectral energy distribution, have coarse angular resolution compared to our our 8\arcsec working resolution. Therefore, we use Herschel PACS 70$\mu$m emission ($I_{70}$) as a proxy for the TIR surface density ($\Sigma_{TIR}$).

To do this, we adopt an empirical approach calibrated galaxy-by-galaxy. For each target, we fit a relation between $\Sigma_{\rm TIR}$ to 70$\mu$m intensity, $I_{70}$ at $30\arcsec$ resolution and assume that relation to hold at the higher $8\arcsec$ resolution of the data. The relation has the form:

\begin{equation} 
\log_{10} \Sigma_{TIR}=a\log_{10} I_{70} + b~.
\label{eq:70-to-TIR}
\end{equation}

\noindent Here $I_{70}$ is the surface brightness at 70$\mu$m, $\Sigma_{TIR}$ is total infrared luminosity surface density, and $a$ and $b$ are the best fit constants. \cite{GALAMETZ13} already calculated galaxy-specific $a$ and $b$ for all four of our ALMA targets. We modify their values slightly based on our own fits to the data for the specific fields of view that we study and choice of resolution. We do not consider these prescriptions as corrections to \citet{GALAMETZ13}, only more appropriate the precise data that we study. Table \ref{tab:tir_corr} reports our adopted conversion for each galaxy. For M51, we adopt the prescription by \citet{LEROY17B}, who carried out a similar exercise in that galaxy in an overlapping region.

\begin{figure*}
\plottwo{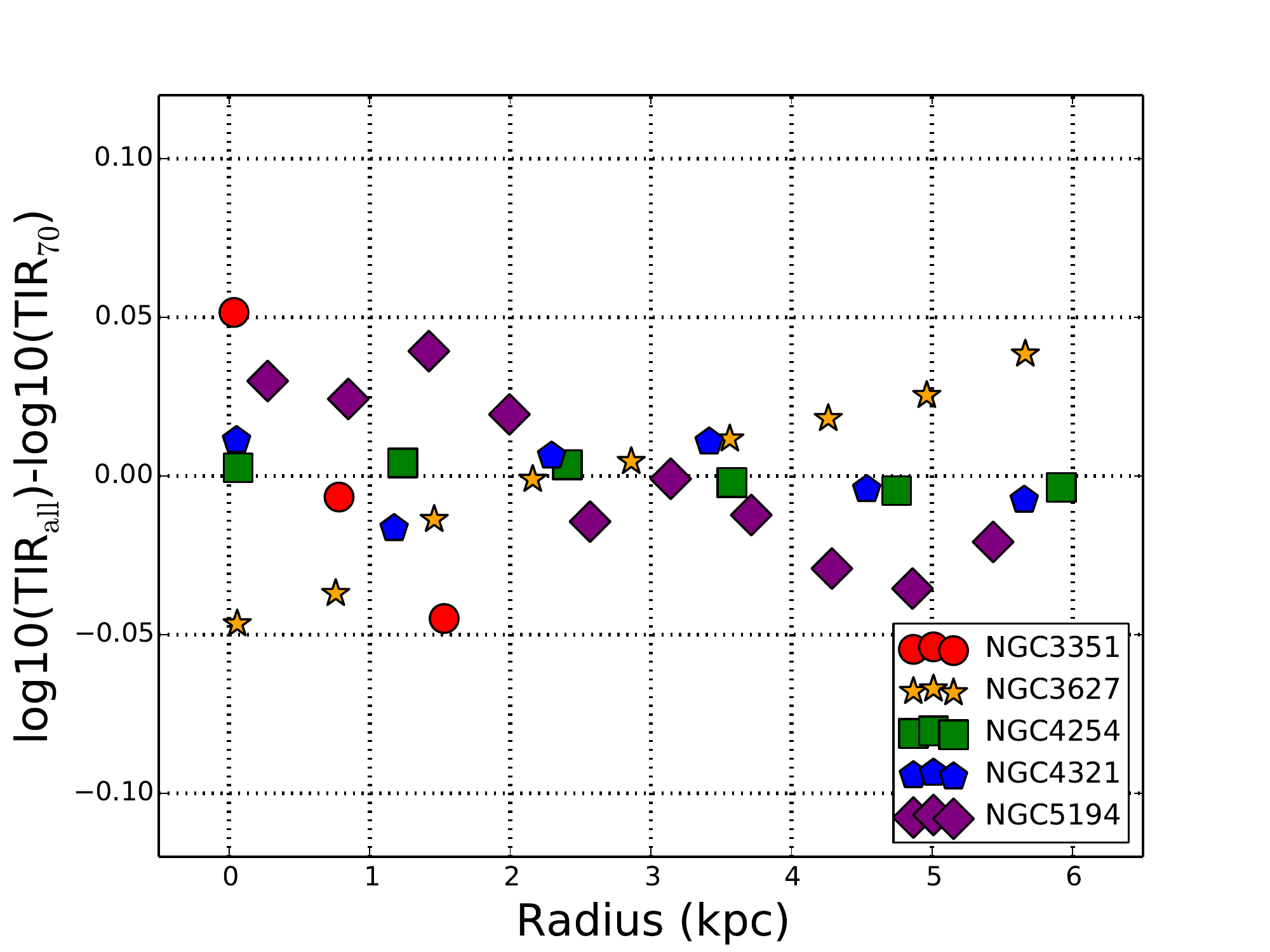}{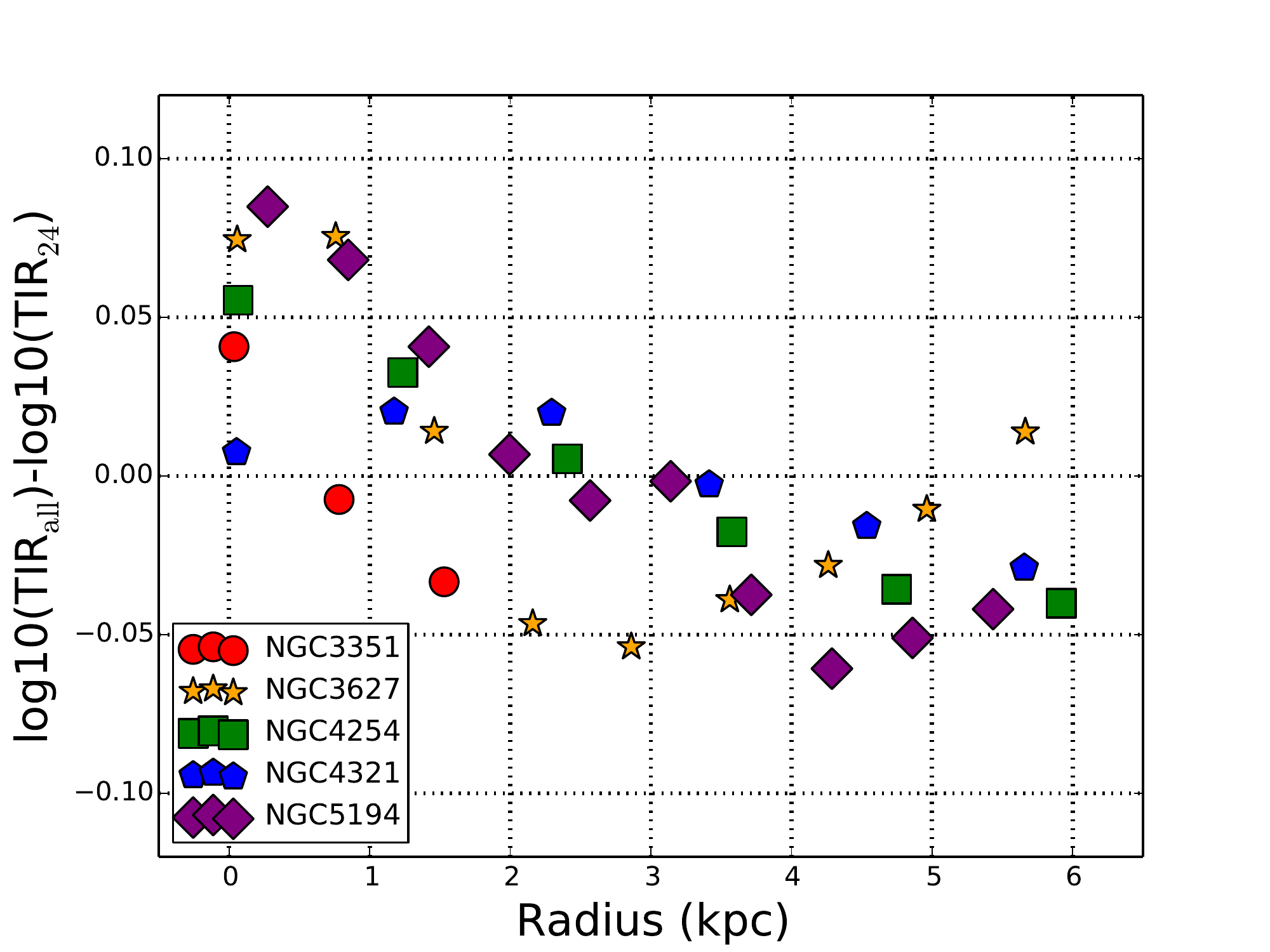}
\caption{Residuals between $\Sigma_{\rm TIR}$ (TIR) calculated using all IR data and $\Sigma_{\rm TIR}$ predicted using only $70\mu$m (TIR$_{70}$, left) or 24$\mu$m (TIR$_{24}$, right) following Equation \ref{eq:70-to-TIR} and Table \ref{tab:tir_corr}. We show the logarithmic residual between the predicted and best-estimate TIR for each ring at $30\arcsec$ resolution. The 70$\mu$m-based prediction recovers the best-estimate TIR with $\sim 0.05$~dex ($\sim 12\%$) scatter and only modest residuals.
\label{fig:tir_comp}}
\end{figure*}

We calculate a similar conversion using 24$\mu$m emission to estimate $\Sigma_{\rm TIR}$. These are also reported in Table \ref{tab:tir_corr}. We consider the 70$\mu$m-based $\Sigma_{\rm TIR}$ estimates more reliable than the $24\mu$m-based ones, which agrees with results from \citet{CALZETTI10} and \citet{GALAMETZ13}. The $70\mu$m band lies nearer to the peak of the IR SED, and as a result our $70\mu$m-based fits recover $\Sigma_{\rm TIR}$ at $30\arcsec$ resolution with less residual scatter than the $24\mu$m based fits. We plot the residuals about the best fit relation in Figure \ref{fig:tir_comp}. The figure shows the difference between the TIR predicted using a single band ($70\mu$m or $24\mu$m) with Table \ref{tab:tir_corr} and the calculation using all bands. The $70\mu$m-based approach predicts $\Sigma_{\rm TIR}$ with better than $\pm 0.05$~dex ($\sim 12$\%) accuracy and only weak residual trends. The $24\mu$m based approach (right panel) shows larger residuals and systematic trends --- a power law is an inadequate description of the $24\mu$m-to-TIR relationship. Coupled with the cleaner and more centrally peaked PSF, this leads us to prefer a $70\mu$m-based $\Sigma_{\rm TIR}$ estimate at $8\arcsec$ resolution.

We present these estimates in the table described in Appendix \ref{sec:table} and recommend them for those interested in using the bolometric luminosity of young stars to capture recent star formation. They provide a useful check when compared with the recombination line-based estimates used in the main paper. 

 Note that these bolometric IR estimates carry the usual uncertainties related to semi-resolved SFR estimates. The star formation history over the last $\sim 100$~Myr may vary from place-to-place and some fraction of the emitted starlight may not be reprocessed by dust (though see below). Dust properties may also vary, perhaps driving some of the observed scatter in the $70\mu$m-to-TIR ratio. \citet{CALZETTI13} and \citet{KENNICUTT12} review key systematics for both this approach and our fiducial H$\alpha+24\mu$m tracer.

Also note that \citet{CALZETTI10} has calibrated the monochromatic $70\mu$m as an SFR indicator for whole galaxies. The difference in scales between that study and this one will likely introduces a difference in the mean age of the population sampled. As a result, we do not expect their prescription to apply exactly to our case. \citet{LI13} observe exactly such a scale dependence in a study using Br$\gamma$ to calibrate 70$\mu$m emission as a star formation rate tracer at several scales in two galaxies. We refer readers interested in more details on the performance of $70\mu$m-based tracers of recent star formation to those papers and the reviews listed above.

Dust heating by an older stellar population may also contaminate our IR-based SFR estimates, though we expect this to be sub-dominant in our sample. Studying $24\mu$m emission from our targets, \citet{LEROY12} estimated that $7\%$ (NGC 3351), $12\%$ (NGC 3627), $16\%$ (NGC 4254), $23\%$ (NGC 4321), and $35\%$ (NGC 5194) of the $24\mu$m light within $r_{\rm gal} < 4$~kpc arises from dust heated by an older stellar population. Though these estimates have substantial uncertainty and apply to $24\mu$m rather than 70$\mu$m, they give an idea of the likely magnitude of a ``cirrus'' correction in these targets \citep[see also][]{LIU11}.

\capstartfalse
\begin{deluxetable}{lccc}
\tabletypesize{\scriptsize}
\tablecaption{{\bf 70$\mu$m and 24$\mu$m to TIR Conversion} \label{tab:tir_corr}}
\tablewidth{0pt}
\tablehead{
\colhead{Galaxy} & 
\colhead{$log_{10}(TIR_{24})$} &
\colhead{$log_{10}(TIR_{70})$}  \\
}
\startdata
NGC~3351 & $\log_{10}(I_{24}) \ \times 0.853 + 5.772$  & $\log_{10}(I_{70}) \ \times 0.852 + 5.452$ \\ 
NGC~3627 & $\log_{10}(I_{24}) \ \times 0.984 + 1.504$ & $\log_{10}(I_{70}) \ \times 0.885 + 4.285$  \\
NGC~4254 & $\log_{10}(I_{24}) \ \times 0.948 + 2.779$ & $\log_{10}(I_{70}) \ \times 0.893 + 4.081$  \\
NGC~4321 & $\log_{10}(I_{24}) \ \times 0.923 + 3.550$ & $\log_{10}(I_{70}) \ \times 0.862 + 5.092$ \\
NGC~5194 ($r < 2.5$~kpc) & $\log_{10}(I_{24}) \ \times 0.919 + 3.803$ & $\log_{10}(1.1807 \times 10^6(1.93+0.01r+0.28r^2-0.048r^3)I_{70}\times L_{\odot}$[Watts])\\
NGC~5194 ($r \geq 2.5\rm{kpc}$) & $\log_{10}(I_{24}) \ \times 0.919 + 3.803$ & $\log_{10}(3.495I_{70}\times L_{\odot}[Watts]\times10^6$) \\
\enddata
\tablenotetext{a}{Both TIR$_{24}$ and TIR$_{70}$ should have units of W kpc$^{-2}$}
\end{deluxetable}
\capstarttrue

\section{Different Star Formation Tracers}
\label{sec:TIRSF}

\begin{figure*}
\plottwo{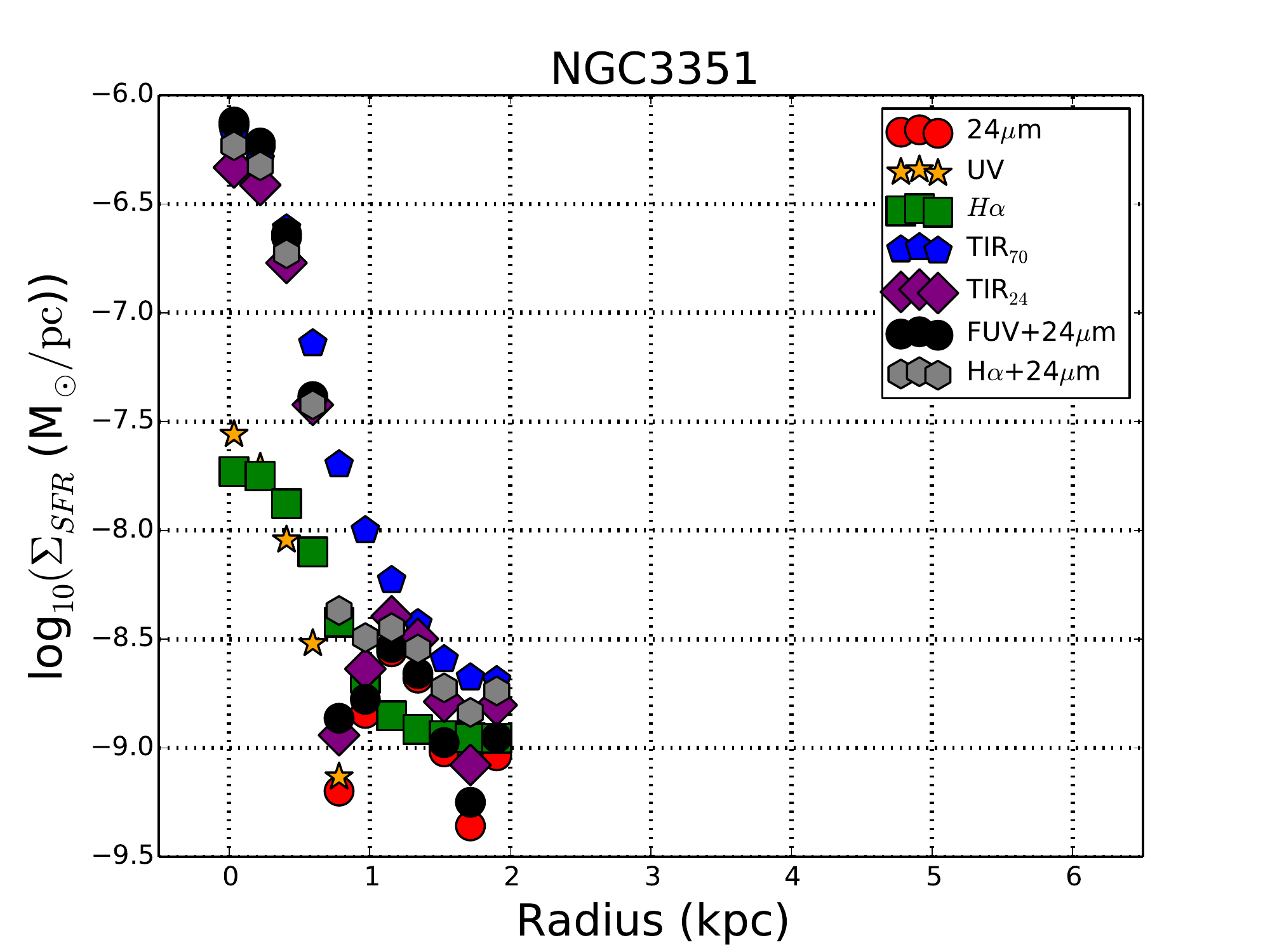}{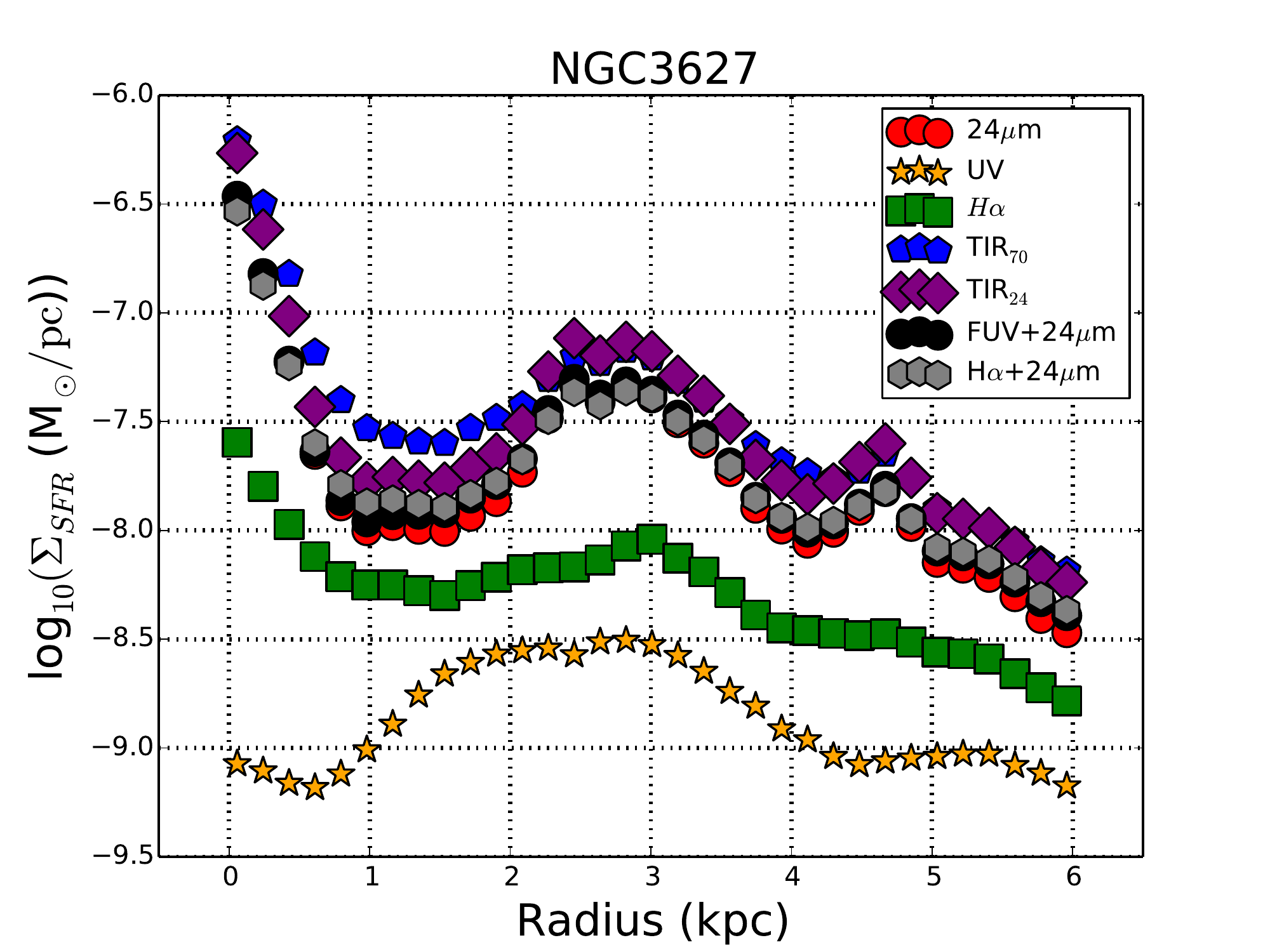}
\plottwo{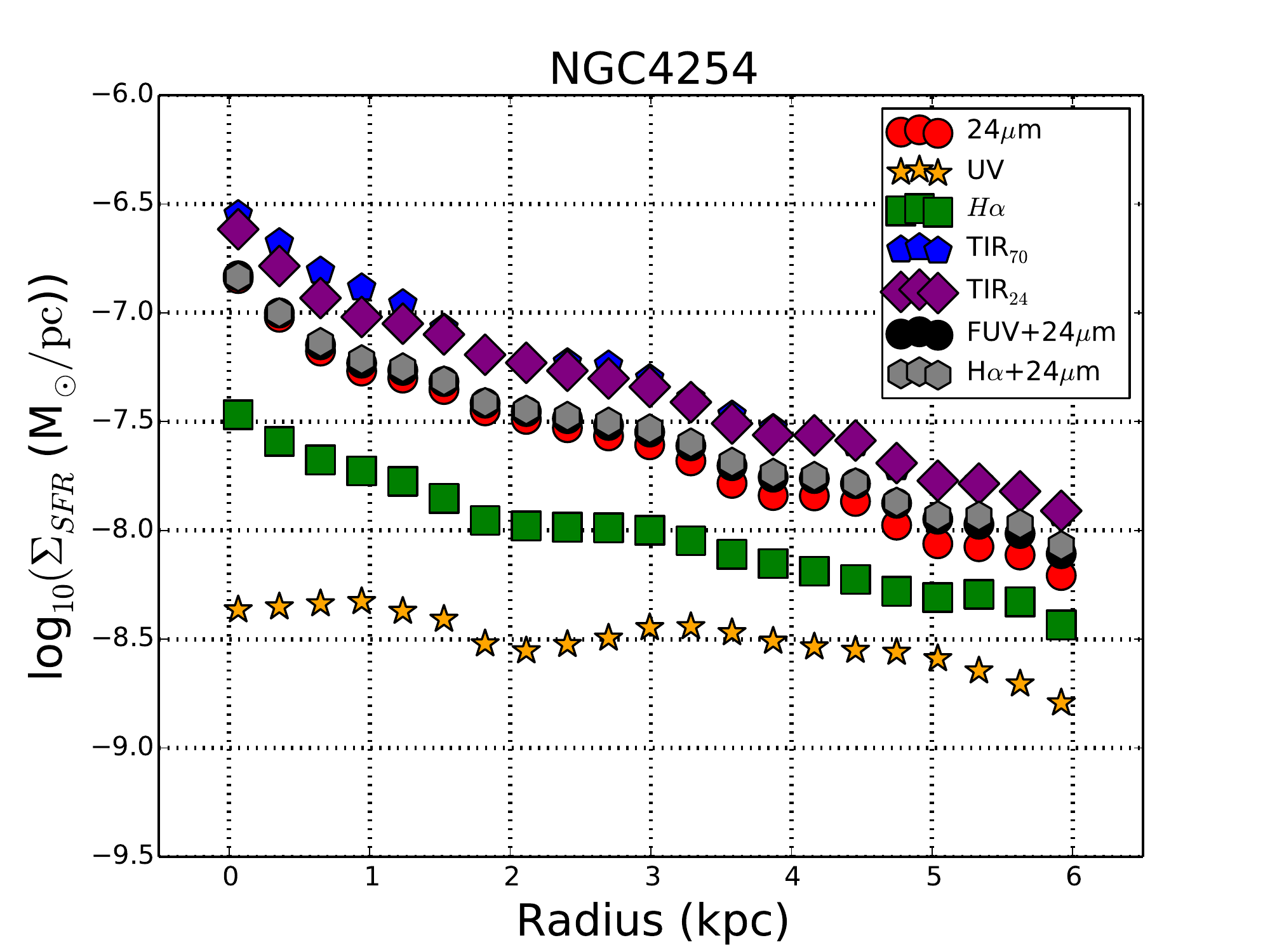}{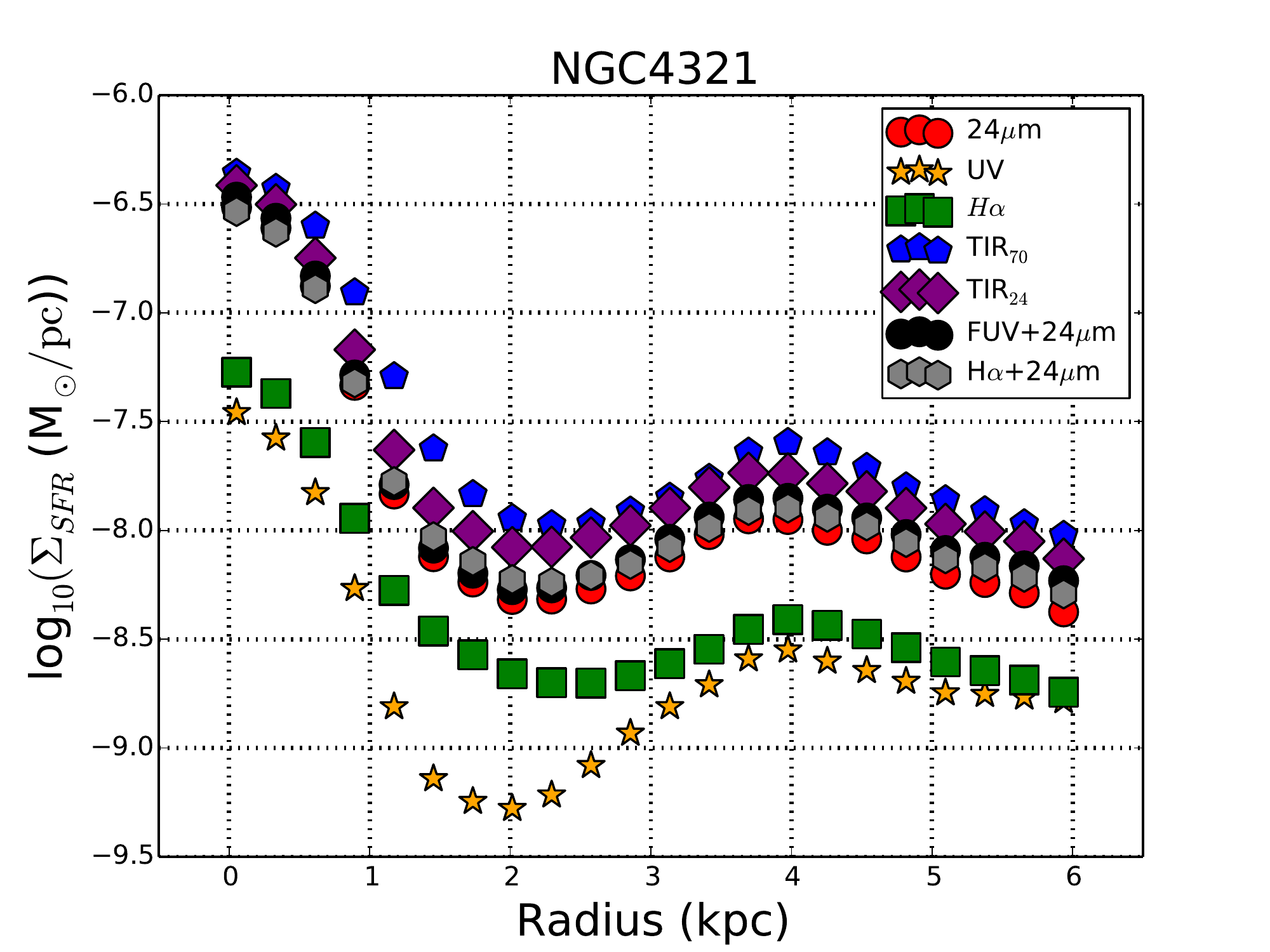}
\caption{A comparison of the SFR as determined using only 24$\mu$m emission, only FUV emission, only H$\alpha$ emission, only TIR emission, a combination of FUV and 24$\mu$m emission, and a combination of H$\alpha$ and 24$\mu$m emission. We find that IR emission dominates significantly over FUV or H$\alpha$ emission, indication a high level of extinction.
   \label{fig:sfr_compare}}
\end{figure*}

In the main text, we use a combination of 24$\mu$m and H$\alpha$ emission to estimate $\Sigma_{\rm SFR}$. Here, we explore the consequences of adopting other SFR tracers. We estimate $\Sigma_{\rm SFR}$ based on the 70$\mu$m- and $24\mu$m-based $\Sigma_{\rm TIR}$ described in the previous section, H$\alpha$ alone, and FUV emission. We compare these to our fiducial H$\alpha+24\mu$m estimates and look at the relative contribution of different terms. We also show that our basic conclusions are robust to the choice of SFR tracer.

For this comparison, we calculate $\Sigma_{\rm SFR}$ from:

\begin{enumerate}
\item $\Sigma_{\rm TIR}$ estimated from $I_{70}$ and $I_{24}$ (separately) following Table \ref{tab:tir_corr} and using Equation \ref{eq:TIR-to-SFR}.
\item FUV only using the coefficient quoted in \citet{LEROY12}, consistent with the one in \citet{KENNICUTT12}.
\item A combination of FUV and $24\mu$m emission following \citet{LEROY12}.
\item H$\alpha$ only using that part of Equation \ref{eq:HA24-to-SFR}.
\item 24$\mu$m only using that part of Equation \ref{eq:HA24-to-SFR}, which is our estimate for extinguished H$\alpha$ emission.
\end{enumerate}

Figure \ref{fig:sfr_compare} shows the radial profiles of $\Sigma_{\rm SFR}$ for each of these estimates. In all cases, the IR-based tracers imply higher $\Sigma_{\rm SFR}$ than either H$\alpha$ or FUV alone. In fact, the figure shows that over the area of interest in all of our targets, the contribution from the unobscured FUV term (yellow stars) to any $\Sigma_{\rm SFR}$ estimate is negligible. Estimates of $\Sigma_{\rm SFR}$ using only H$\alpha$ (green squares) are also lower everywhere than any estimate involving IR emission. This could be expected given that we focus on gas-rich inner regions of actively star-forming galaxies. As a result, over most our target area the ``hybrid'' tracers combining 24$\mu$m emission and FUV or H$\alpha$ yield results that are mostly indistinguishable from using only the 24$\mu$m part of the coefficient. 

As a result of this bright IR emission compared to other tracers, Figure \ref{fig:sfr_compare}, also supports the utility of our alternative, pure infrared estimator. Though the caveats listed in the previous section apply, the plots suggest that across our sample, most of the light from young stars is reprocessed by dust. We still cannot rule out important ``cirrus'' contributions to the IR emission from these regions (see the estimates noted in the previous section). But we do see that our targets are high $\Sigma_{\rm SFR}$ in all tracers, giving us some confidence that the cirrus contribution is likely to be subdominant.

One of our main results is that only the measured HCN/CO ratio in a ring is not a perfect, or even necessarily very good, predictor of the ratio $\Sigma_{\rm SFR}$/CO. We test whether that conclusion changes if we substitute a different SFR tracer for that which we use in the main paper. Equations \ref{eq:fit_sfrmol_densemol_TIR}-\ref{eq:fit_sfrmol_densemol_halpha} and Table \ref{tab:sfr_tracers} compare the effects of using the three different types of SFR tracers on our main result. We see that our result, that the dense gas fraction is not a good tracer of the molecular gas star formation efficiency, does not depend on our choice of SFR tracer. For this comparison, we convert HCN and CO emission into dense gas and molecular gas surface densities, respectively, in order to compare physical quantities. For a more detailed description of these conversions, see \S \ref{subsec:CF}. Here, we make no correction for IR cirrus. This is an area for future improvement (but see \cite{USERO15} for a comparison that does check the effect of IR cirrus estimates on similar results).

These four SFR calculations yield the following fitted relations:

\begin{equation}
\label{eq:fit_sfrmol_densemol_TIR}
\log_{10} \frac{ \Sigma_{SFR,TIR, 70}}{\Sigma_{mol}} = -7.89 +0.85 \log_{10} \frac{\Sigma_{dense}}{\Sigma_{mol}}
\end{equation}

\begin{equation}
\label{eq:fit_sfrmol_densemol_24}
\log_{10} \frac{ \Sigma_{SFR,TIR, 24}}{\Sigma_{mol}} = -7.95 +0.88 \log_{10} \frac{\Sigma_{dense}}{\Sigma_{mol}}
\end{equation}

\begin{equation}
\label{eq:fit_sfrmol_densemol_FUV}
\log_{10} \frac{\Sigma_{SFR,UV+24}}{\Sigma_{mol}} = -8.02 +0.94 \log_{10} \frac{\Sigma_{dense}}{\Sigma_{mol}}
\end{equation}

\begin{equation}
\label{eq:fit_sfrmol_densemol_halpha}
\log_{10} \frac{\Sigma_{SFR,H\alpha+24}}{\Sigma_{mol}} = -8.05 + 0.92 \log_{10} \frac{\Sigma_{dense}}{\Sigma_{mol}} ~.
\end{equation}

The slopes and intercepts of all three relationships are not identical, indicating that choice of SFR tracer does have some impact on our results. However, no matter the choice of tracer, the rank correlation coefficient relating $\Sigma_{SFR}/\Sigma_{mol}$ to $\Sigma_{dense}/\Sigma_{mol}$  is low when we consider all galaxies together. These values appear in Table \ref{tab:sfr_tracers}. None of these SFR tracers yield strong correlations with dense gas fraction, and neither the fit nor the correlation change significantly with SFR tracer.

\cite{USERO15} explore the issue of different SFR tracers, using TIR , H$\alpha$, 24$\mu$m data, and a combination of H$\alpha$ and 24$\mu$m. They find that all SFR tracers yield similar trends, albeit with different levels of intrinsic scatter. \cite{LEROY12} also find good agreement between the results using different SFR tracers. They use H$\alpha$ with one magnitude of extinction, FUV and 24$\mu$m, H$\alpha$ and 24$\mu$m, and H$\alpha$ and 24$\mu$m with a correction for cirrus.

Our analysis indicates that, while there is some variation, the choice of SFR tracer does not change our qualitative results. Quantitatively, a more rigorous examination of what SFR tracers are valid in these different galactic environments is needed. Such an examination is beyond the scope of this paper.

\capstartfalse
\begin{deluxetable}{lcc}
\tabletypesize{\scriptsize}
\tablecaption{$\Sigma_{SFR}/\Sigma_{mol}$ vs. $\Sigma_{dense}/\Sigma_{mol}$ \label{tab:sfr_tracers}}
\tablewidth{0pt}
\tablehead{
\colhead{SFR Tracer} & 
\colhead{Rank Corr.}  \\
}
\startdata
TIR$_{70}$ & 0.28 (0.000)  \\
TIR$_{24}$ & 0.16 (0.002) \\
FUV + 24$\mu$m & 0.13 (0.003) \\
H$\alpha$ + 24$\mu$m & 0.13 (0.003) \\
\enddata
\tablecomments{Rank correlation quotes with $p$ value in parenthesis. Each data point represents the result using all 5 data profiles, including NGC~5194 (\citealt{BIGIEL16}).}
\end{deluxetable}
\capstarttrue

\section{Radial Profile Data Table}
\label{sec:table}

We provide the radial profiles used in this paper as a machine readable table. In this table, each row reports our measurements for one radial ring in one galaxy. The contents of each row are:

\begin{enumerate}
\item The NGC number of the galaxy, and the inner radius of the ring, in kpc. 
\item The average intensity and rms uncertainty on the average intensity for the CO (1-0), HCN (1-0), CS (2-1), HCO$^+$ (1-0), $^{13}$CO (1-0), and C$^{18}$O (1-0) emission lines.
\item The average intensity and rms uncertainty for 3.6$\mu$m, 24$\mu$m, 70$\mu$m, FUV, and H$\alpha$ emission.
\item Quantities derived from the intensities above and the assumptions described in \S \ref{sec:data}: dynamical equilibrium pressure, stellar surface density, TIR luminosity surface density, total (HI+molecular), molecular, and dense gas surface density.
\item Estimates of the star formation rate surface density constructed using several approaches, as well as individual terms in the hybrid estimators discussed in \S \ref{sec:data} and Appendix \ref{sec:TIRSF}.
\end{enumerate}

\end{appendix}

\nocite{*}
\bibliography{main}{}

\end{document}